\documentclass[11pt,a4paper]{article}
\pdfoutput=1

\usepackage{jheppub}
\usepackage{mathrsfs}
\usepackage{tensor,mathtools,bbm}

\newcommand{\cA}{\mathcal{A}}
\newcommand{\cB}{\mathcal{B}}

\newcommand{\A}{\underline{\mathsf{a}}}
\newcommand{\K}{\mathsf{U}}
\newcommand{\N}{\mathcal{N}}
\newcommand{\twista}{\mathsf{a}}
\renewcommand{\d}{\mathrm{d}}
\newcommand{\tr}{\,\mathrm{tr}\,}
\newcommand{\C}{\mathbb{C}}
\renewcommand{\H}{\mathsf{H}}
\newcommand{\CP}{\mathbb{CP}}
\newcommand{\PT}{\mathbb{PT}}
\newcommand{\R}{\mathbb{R}}

\newcommand{\scri}{\mathscr{I}}
\newcommand{\im}{\mathrm{i}}
\newcommand{\M}{\mathbb{M}}
\newcommand{\p}{\partial}
\newcommand{\dbar}{\bar\p}
\newcommand{\dbarx}{\left.\bar\partial\right|_X}
\newcommand{\D}{\mathrm{D}}
\newcommand{\SL}{\mathrm{SL}}
\newcommand{\SU}{\, \mathrm{SU}}
\newcommand{\GL}{\mathrm{GL}}

\newcommand{\uA}{\underline{A}}
\newcommand{\rD}{\mathcal{D}}

\newtheorem{thm}{Theorem}

\newcommand{\unabla}{\underline{\nabla}}
\newcommand{\uF}{\underline{F}}
\newcommand{\utF}{\underline{\tilde F}}
\newcommand{\upsi}{\underline{\psi}}
\newcommand{\utpsi}{\underline{\tilde{\psi}}}
\newcommand{\uphi}{\underline{\phi}}
\newcommand{\uB}{\underline{B}}
\newcommand{\usquare}{\underline{\square}}

\title{Yang-Mills form factors on self-dual backgrounds}

\author{Giuseppe Bogna,}
\author{Lionel Mason\footnote{On leave at the Institut des Haut \'Etudes Scientifique, 35 Route de Chartres, 91440 Bures-Sur-Yvette, and the Laboratoire de Physiques, ENS, Rue L'Homnond, Paris 75005, France}}

\affiliation{The Mathematical Institute, University of Oxford, Woodstock Road, Oxford OX2 6GG, United Kingdom}

\emailAdd{giuseppe.bogna@maths.ox.ac.uk}
\emailAdd{lmason@maths.ox.ac.uk}

\abstract{The construction of perturbative quantities on non-linear backgrounds leads  to the possibility of incorporating strong field effects in perturbation theory. We continue a programme to construct QFT observables on self-dual backgrounds.  The approach works with asymptotic data for fields defined at null infinity $\scri$, extending earlier work on Yang-Mills amplitudes on self-dual backgrounds to form factors and incorporating supersymmetry. Since our analysis is based on reconstruction from data at null infinity, it naturally ties into work on celestial and twisted holography.
We study form factors both in pure Yang-Mills and their supersymmetric counterparts in $\N=4$ SYM, giving a full treatment of $\N=4$ super-Yang-Mills at null infinity and their self-dual nonlinear  backgrounds. 
We obtain tree-level MHV form factors around these backgrounds using new formulae for lifting operators to twistor space leading to simple dressings of the corresponding form factors around the vacuum. We give brief indications on how to go beyond the MHV sector by introducing dressed versions of the MHV diagram propagator.  We discuss generating functionals of the MHV all plus 1-loop amplitude in this context together with its various dual conformal representations.
}

\begin{document}
\maketitle
\flushbottom
\section{Introduction}
Integrability is a powerful tool for the study of non-linear problems, and, although it doesn't apply directly to generic gauge and gravity theories, in four dimensions  such theories possess self-dual sectors that are integrable \cite{Penrose:1976js, Ward:1977ta, Mason:1991rf}.  There is by now a long tradition of exploiting the integrability of the self-dual sector to provide non-perturbative results such as the construction of instantons and monopoles \cite{Ward:1990vs}.  
These structures are also  intimately related to the rich structures discovered in scattering amplitudes, from the famous Parke-Taylor formula for the tree-level MHV scattering of gluons \cite{Parke:1986gb, Nair:1988bq},  to the more general constructions of  \cite{ Witten:2003nn, Berkovits:2004hg, Roiban:2004yf, Cachazo:2004kj, Britto:2005fq, Brandhuber:2008pf, Bern:2010ue, ArkaniHamed:2010kv, Cachazo:2013hca, Cachazo:2014xea, Geyer:2014fka}, 
see \cite{Dixon:1996wi, Bern:2007dw,  Elvang:2013cua, Travaglini:2022uwo, Geyer:2022cey} for reviews.  
The role of integrability was made explicit in studies of scattering amplitudes using twistor actions defined on twistor space that allow the  direct exploitation of the  integrability of the self-dual sectors of Yang-Mills and gravity theories.  
Moreover, the non-self-dual theory can be formulated as a perturbation around the self-dual sector \cite{Chalmers:1996rq, Mason:2005zm, Boels:2006ir, Boels:2007qn, Mason:2010yk, Adamo:2011cb} on twistor space, so many interesting results for the full theory can be readily obtained via perturbation theory; see the reviews \cite{Adamo:2011pv,  Adamo:2013cra} for further details. From a different perspective, the existence of an integrable self-dual sector was at the heart of recent, exciting developments in flat-space holography, most notably celestial \cite{Strominger:2017zoo} and twisted \cite{Costello:2022wso} holography. These sectors host chiral symmetry algebras -- whose existence is underpinned by infinitely many soft symmetries \cite{Adamo:2019ipt, Strominger:2021lvk, Ball:2021tmb, Strominger:2021mtt} -- that can be used to significantly constrain celestial correlators. The non-local nature of twistor constructions means that they can be naturally formulated at null infinity \cite{Newman:1976gc, Hansen:1978jz, Sparling:1990, Newman:1978ze}, so these features become apparent if one adopts a twistorial description of flat holography; for example, twistor methods give a nice way to understand the symmetries of self-dual gravity \cite{Adamo:2021lrv}  and can be used to derive the gluon celestial OPE at all orders \cite{Adamo:2022wjo}.

These approaches have by now been extended to obtain formulae for form factors. Form factors are expectation values of local composite operators between the vacuum and an $n$-particle on-shell state and therefore represent intermediate observables between on-shell amplitudes and off-shell correlators. They have important physical applications as well, for example, arising as scattering amplitudes after additional fields have been integrated out in effective field theories and, since form factors are only partially off-shell, many amplitudes techniques have been extended to the construction of these observables. This includes the MHV formalism, recursion relations, and methods inspired by twistor theory in both $\mathcal{N}=4$ super-Yang-Mills and pure Yang-Mills, both at tree and at loop level, see \cite{Brandhuber:2010ad, Brandhuber:2011tv, Brandhuber:2012vm, Penante:2014sza, Brandhuber:2016xue, He:2016jdg,  Brandhuber:2017bkg, Brandhuber:2018kqb, Brandhuber:2018xzk, Bianchi:2018peu, Bianchi:2018rrj, AccettulliHuber:2019abj}. The twistor action approach has been pursued also for form factors  leading to many further results \cite{Koster:2016ebi, Koster:2016loo, Koster:2016fna, Chicherin:2016soh, Chicherin:2016ybl, Chicherin:2016fac, Chicherin:2016fbj, Chicherin:2014uca}, building on work on correlation functions in twistor space \cite{Adamo:2011dq, Adamo:2011pv, Eden:2017fow}. Form factors have also played an interesting role in celestial and twisted holography where they were recently used to resolve singularities in the celestial amplitudes \cite{Casali:2022fro}, to deform the soft symmetry algebras \cite{Melton:2022fsf}, as well as to compute both amplitudes and form factors in terms of correlators of a 2d chiral algebra reminiscent of the celestial soft algebra \cite{Costello:2022wso}. 

The above works concern observables  around a flat background.  In this paper, we address the question of extending the computations of form factors around non-trivial backgrounds. Although work already exists in this direction in the case of scattering amplitudes \cite{Furry:1951zz, DeWitt:1967ub, tHooft:1975uxh, Abbott:1981ke}, these face difficulties as soon as symmetries are lost and explicit formulae for background-coupled fields are no longer easily available; standard techniques and constraints based on momentum space and hence the rationality of tree diagrams, such as BCFW recursion and unitarity, cannot be applied. However, progress can be made on simple backgrounds:  it is for example possible to extract much information about scattering around plane wave backgrounds \cite{Adamo:2017nia, Adamo:2017sze, Adamo:2018mpq, Adamo:2020qru, Adamo:2021hno}. In this work, we work around \emph{self-dual} and \emph{radiative} backgrounds, continuing the programme established in \cite{Adamo:2020syc, Adamo:2020yzi}.  In these works, all-multiplicity expressions for tree-level MHV gluon scattering amplitude around such backgrounds were obtained for the first time, together with conjectural formulae for the N$^k$MHV amplitudes.  Analogous gravitational formulae were found in  \cite{Adamo:2022mev}. 

The class of backgrounds we consider are  \emph{radiative} in the sense of being determined by their data at $\scri$,  but they are otherwise generic.  They could be taken to be a sum of plane waves at infinity, to make contact with higher-point formulae, but there is no particular reason to do so and one can choose data for more general backgrounds, such as instantons.  We use the complete integrability of the self-dual sector to construct background-coupled fields from their asymptotic data at null infinity and  study their interactions.  Without local symmetries of the background, there will now be no straightforward local definition of plane waves in the interior, but fields can be taken to be plane waves at infinity; they can be characterized by the same data at null infinity that leads to momentum eigenstates on the trivial background. Complete integrability implies that these solutions do not themselves scatter as they pass through the background field, so their values from past null infinity $\scri^-$ can be identified with their values in the future at $\scri^+$ with no ambiguity, thus preserving crossing symmetry. Moreover, the backgrounds themselves are by assumption determined by their data at null infinity, so the whole setting is intrinsically holographic and fits well into the celestial holography programme.

As we explain below, the argument in \cite{Adamo:2020yzi} that leads to the MHV amplitude is closely related to the construction of an MHV form factor: the generating functional for the MHV amplitude can be viewed as the $q\to 0$ limit of the generating functional for the form factor of the operator $\tr B^2$, where $B_{\alpha\beta}$ is the anti-self-dual component of the field strength in the chiral Yang-Mills formulation of Chalmers and Siegel \cite{Chalmers:1996rq, Chalmers:1997sg} and $q$ is the momentum associated to the local operator $\tr B^2$  in the expectation value.  Similar reasoning was also the basis of the twistor action approach of \cite{Boels:2007qn} to the construction of the MHV formalism of \cite{Cachazo:2004kj}, as well as  recent works in celestial holography on a trivial background \cite{Costello:2022wso, Costello:2022upu, Bu:2022dis}, but  see also \cite{Costello:2022jpg} for an intriguing example on Burns space.
Even away from $q=0$, the tree-level, colour-ordered MHV form factor for $\tr B^2$ around a Cartan-valued self-dual radiative background is extremely simple 
\begin{equation}
    \mathscr{F}_{\tr B^2}(1^+,\ldots,i^-,\ldots,j^-,\ldots n^+;q)=\frac{\langle ij\rangle^4}{\langle 12\rangle\ldots\langle n1\rangle}\int_{\M}\d^4x\,e^{\im (Q-q)\cdot x+\sum_je_jg(x,\kappa_j)}\,.
\end{equation}
Here $Q=k_1+\ldots +k_n$ is the  sum of the gluon momenta measured at null infinity, $g$ is a function that depends on the background field\footnote{For example, for a self-dual plane wave background, $g$ is known as \emph{Volkov exponent} \cite{Wolkow:1935zz, Seipt:2017ckc}.} and $e_j$ is the set of charges of each gluon with respect to the background, with values in the Cartan subalgebra of the gauge group -- relatively simple formulae are available for generic backgrounds as well. The corresponding  expression for the form factor for the operator $\tr\tilde F^2$ is quite different due to the chirality in our focus on MHV form factors.  We show that the tree-level MHV form factor around a Cartan-valued self-dual radiative background has the compact form
\begin{equation}
    \mathscr{F}_{\tr\tilde F^2}(1^+,\ldots,n^+;q)=\frac{(q\cdot Q)^2}{\langle 12\rangle\ldots\langle n1\rangle}\int_\M\d^4x\,e^{\im (Q-q)\cdot x+\sum_je_jg(x,\kappa_j)}\,.
\end{equation}
In both these examples, the space-time integral represents a simple dressing of the form factor around the flat background; in the limit $g\to 0$ (corresponding to the flat-background limit) we recover momentum conservation in the presence of the local operators, i.e. the momentum of either $\tr B^2$ or $\tr\tilde F^2$ is constrained to be equal to the sum of the gluon momenta, as a consequence of the translational invariance of the trivial background. In that  limit, the expression of the  $\tr\tilde F^2$ form factor reduces to the well-known formula for the tree-level scattering of a massive Higgs and arbitrarily many positive-helicity gluons \cite{Dixon:2004za}. The form factor around a general background thus retains much of the simplicity observed in the trivial background case, although it involves a single residual space-time integral because the background is not translation invariant, but the kinematical prefactors  coincide on the support of momentum conservation. Similar considerations apply to other tree-level form factors in the MHV sector, see Equations \eqref{eq:trf^3formfactor} and \eqref{eq:trb^kformfactor} for $k=3$ below for explicit examples of $\tr\tilde F^3$ and $\tr B^3$, as well as for tree-level MHV form factors in $\N=4$ SYM, see Equation \eqref{eq:phi_super_form_factor}.

This work is organized as follows. We review some elementary results in twistor theory and develop the necessary tools to describe self-dual radiative gauge fields in Section \ref{sec:sd_fields}, as well as the extension to self-dual radiative backgrounds in $\N=4$ SYM in \S\ref{sec:sd_fields_N=4}, paying special regard to the fermionic expansion for the asymptotic data near $\scri$.  The key technique is introduced in \S\ref{sec:integral}, where we introduce new  explicit integral representations for the background fields in terms of their radiative data and for the linear fields propagating around these backgrounds. These are then used to lift  for example $\tr\tilde F^2$ and $\tr\tilde{F}^3$ to twistor space in Section \ref{sec:form_factors}.  We use these to show how expressions for the tree-level MHV form factors around non-trivial backgrounds in pure Yang-Mills can be readily obtained by their expressions around the trivial background. We conclude by showing that similar results hold for tree-level MHV super form factors in $\N=4$ SYM around gluonic self-dual radiative backgrounds. In the discussion section \ref{n-kmhv} we briefly explain how the MHV-diagram propagator can be dressed; this can in principle be used to compute higher MHV degree and loop-level expressions on backgrounds.   The machinery developed in the text is used in \ref{one-loop} to discuss generating functions for the MHV all plus one-loop amplitude and its  extension to backgrounds, and we show how the formulation can be naturally  used to obtain dual conformal invariant region momentum formulae for the amplitude such as those obtained by \cite{Henn:2019mvc,Chicherin:2022bov}.   In Appendix \ref{app:N=4_constraints} we show an equivalence between the equations of motion for $\N=4$ SYM and constraint equations for super-connections on chiral superspace. Finally, we defer some more computational details of our construction to Appendix \ref{app:computation}.

\section{Self-dual radiative backgrounds in pure Yang-Mills}\label{sec:sd_fields}
In this section, we review self-dual radiative Yang-Mills backgrounds in four-dimensional complexified Minkowski space-time and their twistor theory; 
see also \cite{Penrose:1986uia, Mason:1991rf, Adamo:2017qyl, Adamo:2020yzi} for further details. Working on complex Minkowski space-time $\M\cong\C^4$ with coordinates $x^\mu$, $\mu=0,1,2,3$, it's useful to recall the local isomorphism between $\mathrm{SO}(4,\C)$ and $\SL(2,\C)\times\SL(2,\C)$ and to introduce the 2-spinor notation, trading tensor indices with pairs of spinor indices with opposite chirality
\begin{equation}
	x^{\alpha\dot\alpha}=\frac{1}{\sqrt{2}}\left(\begin{array}{c c}
		x^0+x^3&x^1-\im x^2\\x^1+\im x^2&x^0-x^3
	\end{array}\right)\,.
\end{equation}
The $\SL(2,\C)$-invariant Levi-Civita symbols $\varepsilon_{\alpha\beta}$ and $\varepsilon_{\dot \alpha\dot \beta}$ are used to raise and lower spinor indices.  Following standard spinor-helicity notation, we denote the contractions between spinors by  $
\langle ab\rangle\coloneqq a^\alpha b_\alpha=\varepsilon^{\alpha\beta}a_\beta b_\alpha$ and $[\tilde a\tilde b]\coloneqq\tilde a^{\dot\alpha}\tilde b_{\dot\alpha}=\varepsilon^{\dot\alpha\dot\beta}\tilde{a}_{\dot\beta}\tilde{b}_{\dot\beta}$.
Given a gauge field on $\M$, the field strength $F_{ab}$ admits the decomposition
\begin{equation}
F_{\alpha\dot\alpha\beta\dot\beta}=\varepsilon_{\alpha\beta}\tilde{F}_{\dot\alpha\dot\beta}+\varepsilon_{\dot\alpha\dot\beta}F_{\alpha\beta}\,,
\end{equation}
where $F_{\alpha\beta}$ and $\tilde{F}_{\dot\alpha\dot\beta}$ are symmetric spinors and represent the anti-self-dual (ASD) and self-dual (SD) parts of the field strength, respectively.

A source-free gauge field will be said to be radiative if it extends to null infinity inside the conformal compactification of Minkowski space-time and is completely determined by its free characteristic data at either past or future null infinity. Here we will in fact assume that we are working with complex fields on the conformal compactification of $\M$ that includes $\scri_\C=\scri^+_\C\cup \scri ^-_\C$. This  is a partial complexification of standard real null infinity $\scri=\scri^+\cup\scri^-$ of $\R^{1,3}$ where advanced and retarded fields are allowed to be complex. Focusing on future null infinity, recall that in the real case, $\scri^+\cong\R\times S^2$ can be understood as the inversion of the light-cone of the origin of $\R^{1,3}$; $\scri^+_\C$ is obtained by complexifying the $\R$ factor to $\C$ while keeping the $S^2$ base.\footnote{This will guarantee that each $\alpha$-plane representing a twistor in $\M$ will intersect $\scri_\C^+$ in a unique point.} In order to connect with homogeneous coordinates on twistor space, we will use a homogeneous version $(u,\lambda_\alpha,\bar\lambda_{\dot\alpha})$ of Bondi coordinates subject to the equivalence relation
\begin{equation}
(u,\lambda_\alpha,\bar\lambda_{\dot\alpha})\sim(b\bar bu,b\lambda_\alpha,\bar b\bar\lambda_{\dot\alpha})\,, 
\end{equation}
for any $b\in\C^*$. $u$ is a complexification of the standard Bondi retarded time, while $(\lambda_\alpha,\bar\lambda_{\dot\alpha})$ are homogeneous coordinates on the celestial sphere thought of as the complex projective line $\CP^1$. The homogeneous coordinates  allow us to encode spin and conformal weights in terms of homogenous line bundles $\mathcal{O}(p,q)\to\scri^+_\C$, where a section of $\mathcal{O}(p,q)$ is represented as a function of $f_{p,q}(u,\lambda,\bar\lambda)$ with weights $(p,q)$ under rescaling of the homogeneous coordinates
\begin{eqnarray}
f_{p,q}(|b|^2u,b\lambda,\bar b\bar\lambda)=b^p\bar b^qf_{p,q}(u,\lambda,\bar\lambda)\,.
\end{eqnarray} 
The conformal and spin weights $(h,s)$ are then given by $h=(p+q)/2$ and $s=(p-q)/2$, respectively. A similar description can be set up on $\scri^-_\C$ by replacing the retarded time $u$ with the advanced time $v$.

Within this projective formalism, the restriction of a gauge field $A$ on $\M$ to $\scri^+_\C$ 
 in temporal gauge $A_u=0$ is \cite{vanderBurg:1969, Newman:1978ze, Strominger:2013lka, Barnich:2013sxa}
\begin{equation}
	\left.A\right|_{\scri^+}=A_-(u,\lambda,\bar\lambda)\D\lambda+A_+(u,\lambda,\bar\lambda)\D\bar\lambda\,,\label{eq:radiative-data-pure-yang-mills}
\end{equation}
where $\D\lambda\coloneqq\langle\lambda\,\d\lambda\rangle$, $\D\bar\lambda\coloneqq[\bar\lambda\,\d\bar\lambda]$.  The restriction of the leading components of the SD and ASD parts of the curvature are
\begin{subequations}
    \begin{eqnarray}
         F^0_+
         &=&\p_u A_+\,\d u \wedge\D\bar\lambda\,,\\
         F^0_-
         &=&\p_u A_-\,\d u\wedge\D \lambda\,,
    \end{eqnarray}
    \end{subequations}
thus  $A_+$, $A_-$ are  the free data for the SD and ASD components of the field strength, respectively.  In terms of the line bundles above,  $A_+$ is a section of $\mathcal{O}(-2,0)\otimes\mathfrak{g}$, while $A_-$ is a section of $\mathcal{O}(0,-2)\otimes\mathfrak{g}$, where $\mathfrak{g}$ is the Lie algebra of the gauge group. A self-dual, radiative gauge field is a gauge field completely characterized by the free data $\left.A\right|_{\scri^+_\C}= A_+\,\D\bar\lambda$, while $A_-=0$.

\paragraph{Radiative fields from their asymptotic data.}
We can understand the previous discussion in terms of the peeling properties of the gauge fields and by means of the Kirchhoff-d'Adhémar integral formula \cite{Penrose:1962ij, Penrose:1980yx, Penrose:1984uia}; these can in turn be regarded as twistor integral formulae using twistor representatives built from asymptotic data at future null infinity. This perspective will also provide a natural way to introduce radiative data for scalars and fermions in $\N=4$ SYM. 

Recall that the radiative data for a  field $\Phi_{\alpha_1\ldots\alpha_{2|h|}}$ of helicity $h\leq0$ consist of a function $\Phi_{2h}=\Phi_{2h}(u,\lambda,\bar\lambda)$ of weight $(-2|h|-1,-1))$ on (future) null infinity; it is well-known that the characteristic data $\Phi_{2h}$ is the leading-order component of the field that decays as $r^{-1}$ as $r\rightarrow \infty$, where $r$ is an affine parameter; the $2|h|+1$ components of $\Phi_{\alpha_1\ldots\alpha_{2|h|}}$ \emph{peel} at different rates  as $r\rightarrow \infty$ \cite{Penrose:1984uia}. In terms of $\Phi_{2h}$, the bulk field can be reconstructed using the Kirchhoff-d'Adhémar formula
\begin{equation}
    \Phi_{\alpha_1\ldots\alpha_{2|h|}}(x)=\int_{\CP^1}\frac{\D\lambda\wedge\D\bar\lambda}{2\pi \im}\ \lambda_{\alpha_1}\ldots\lambda_{\alpha_{2|h|}}\left.\frac{\p\Phi_{2h}}{\p u}\right|_{u=\langle \lambda|x|\bar\lambda]}\,, \label{K-dA} 
\end{equation}
where the integral is evaluated on the \emph{light-cone cut} of $x$, that is the intersection of the null cone with apex $x$ and future null infinity
\begin{equation}
    u=x^{\alpha\dot\alpha}\lambda_\alpha\bar \lambda_{\dot\alpha} \,.    
\end{equation} 
In particular, for $h=-1$, we identify the radiative data $\Phi_{-2}$  for an ASD gauge field $\Tilde{\mathcal{A}}$ by
\begin{equation}    
\Phi_{-2}(u,\lambda,\bar\lambda)= \p_u A_-(u,\lambda,\bar\lambda)
    \,,\label{eq:spin-1-peeling}
\end{equation}
that is, $\Phi_{-2}$ is precisely the leading part at $\scri^+_\C$ of the ASD curvature. For positive helicities, we can take the conjugate of the above and the radiative data $\Phi_{2h}$ are valued in $\mathcal{O}(-1,2h-1)$ (in particular, we identify $\Phi_2=\p_uA_+$), but the corresponding Kirchhoff-d'Adhémar will have a less direct connection with twistor representatives. 

\paragraph{The $J$ and  $K$ potentials for self-dual gauge fields.}
Self-dual gauge fields can be described also in terms of scalar second  potentials. Given a reference spinor $\iota^\alpha$, the vanishing of the ASD curvature component $\iota^\alpha \iota^\beta F_{\alpha\beta}=0$ implies flatness in the two-plane  tangent to $\iota^\alpha \beta^{\dot\alpha}$ for all $\beta^{\dot\alpha}$, so we can work with the ansatz 
\begin{equation}
    A_{\alpha\dot\alpha}=\iota_\alpha A_{\dot\alpha}
    \,,
\end{equation}
for the gauge connection. In particular, the gauge field is in light-cone gauge with respect to any null vector of the form $n^{\alpha\dot\alpha}=\iota^\alpha\beta^{\dot\alpha}$. The equation $\iota^\alpha F_{\alpha\beta}=0$ then implies the existence of a matrix-valued scalar potential $K$, the $K$-matrix, so that
\begin{equation}
A_{\dot\alpha}=\iota^\alpha\partial_{\alpha\dot\alpha}K\,.\label{eq:kmatrix}
\end{equation}
The gauge field is automatically in Lorentz gauge as well and the curvature can now be written in terms of $K$ as
\begin{subequations}
\begin{eqnarray}
    F_{\alpha\beta}&=& -\frac{1}{2}\iota_\alpha \iota_\beta(\square K+\im[\tilde\d_{\dot\alpha}K,\tilde\d^{\dot\alpha}K])\, , \\ 
\tilde F_{\dot\alpha\dot\beta}&=&  \d_{\dot\alpha}\d_{\dot\beta}K \,,\label{K-curv}
\end{eqnarray}
\end{subequations}
where we introduced the notation $\d_{\dot\alpha}\coloneqq \iota^\alpha\partial_{\alpha\dot\alpha}$.
The self-duality equation is therefore  
\begin{equation}
\square K+\im[\d_{\dot\alpha}K,\d^{\dot\alpha}K]=0\,. \label{eq:kmatrixeom}
\end{equation}

The $J$-matrix potential requires the choice of a second  spinor $o_\alpha$, which we normalize by $\langle \iota o\rangle=1$. Using  $o^\alpha o^\beta F_{\alpha\beta}=0$, we can deduce the existence of a matrix function $J$ so that 
\begin{equation}
    A_{\alpha\dot\alpha}=-\im\, \iota_\alpha J^{-1}o^\beta\p_{\beta\dot\alpha}J\,.\label{J-gauge}
\end{equation} Defining $\tilde\d_{\dot\alpha}\coloneqq o^\alpha\p_{\alpha\dot\alpha}$, the curvatures become
\begin{subequations}
\begin{eqnarray}
    F_{\alpha\beta}&=&-\im\, o_{(\alpha}\iota_{\beta)} \d^{\dot\alpha}(J^{-1}\tilde\d_{\dot\alpha}J)\,, \\
    \tilde F_{\dot\alpha\dot\beta}&=&-\im\, \d_{(\dot\alpha}(J^{-1}\tilde\d_{\dot\beta)}J)\,,\label{J-curv}
\end{eqnarray}
\end{subequations}
so that the self-dual Yang-Mills equations for this potential become \begin{equation}\d^{\dot\alpha}(J^{-1}\tilde\d_{\dot\alpha}J)=0\,.\label{eq:jmatrixeom}\end{equation}

\subsection{Twistor-space description}
To define twistor space, introduce homogeneous coordinates $Z^A=(\mu^{\dot\alpha},\lambda_\alpha)$ on $\CP^3$ subject to the equivalence relation $Z^A\sim t Z^A$ for $t\in\C^*$.  The twistor space $\PT$ of $\M$ is the open subset of $\CP^3$ given by
\begin{equation}
	\PT=\{[Z^A]\in\CP^3:\lambda_\alpha\neq 0\}\,,\label{eq:twistorspace}
\end{equation}
and can thus be described as the total space of the holomorphic bundle $\mathcal{O}(1)\oplus\mathcal{O}(1)\to\CP^1$ over the Riemann sphere. Its relationship with $\M$ is encoded in the incidence relations
\begin{equation}
	\mu^{\dot\alpha}=x^{\alpha\dot\alpha}\lambda_\alpha\,.\label{eq:incidencerelation}
\end{equation}
For fixed  $x\in \M$, the incidence relations describe a \emph{twistor line}, that is a linearly and holomorphically embedded Riemann sphere  $X\subset \PT$, while for constant $Z^A\in\PT$ they give an $\alpha$-plane in $\M$ \cite{Penrose:1967wn}, i.e. a totally null 2-planes with self-dual tangent bivectors.  The definition \eqref{eq:twistorspace} removes  the twistor line $I\subseteq\CP^3$ corresponding to spatial infinity $i^0$. 

Radiative linear  fields admit a natural description in twistor space \cite{Mason:1986}. This can be obtained by pulling back asymptotic data on $\scri^+_\C$ to twistor space  $\PT$ via the natural projection $p$ to $\scri^+_\C$ given by
\begin{equation}
p:(\mu^{\dot\alpha},\lambda_\alpha)\mapsto(u=\mu^{\dot\alpha}\bar\lambda_{\dot\alpha},\lambda_\alpha,\bar\lambda_{\dot\alpha})\, .    
\end{equation}
We can define line bundles $\mathcal{O}(n)\rightarrow\PT$ whose sections can be represented by functions of homogeneity-degree $n$ in the homogeneous coordinates. These line bundles are identified with both the pull-backs $p^*\mathcal{O}(n,0)$ by $p$ from $\scri^+_\C$ and the pull-backs of the line bundles $\mathcal{O}(n)\to\CP^1$ by the holomorphic projection $\PT\to\CP^1$. For fields of helicity $h\leq 0$, we can pull back the characteristic data at $\scri^+_\C$ to twistor space to give $(0,1)$-forms of holomorphic weight $2h-2$ that defines a Dolbeault cohomology  class  
\begin{equation}
    \omega_{2h}=p^*\left(\frac{\p\Phi_{2h}}{\p u} \D\bar\lambda\right)\, \in H^{0,1}(\PT,\mathcal{O}(2h-2)).
\end{equation}
Following \cite{Mason:1986}, the Kirchhoff-d'Adhémar integral formula \eqref{K-dA} can now be re-interpreted as a version of the twistor integral formula or Penrose transform \cite{Penrose:1969ae, Eastwood:1981jy} using Dolbeault cohomology
\begin{equation}
    \Phi_{\alpha_1\ldots\alpha_{2|h|}}(x)=\int_X \D\lambda\wedge \lambda_{\alpha_1}\ldots\lambda_{\alpha_{2|h|}}\left.\omega_{2h}\right|_X\,.
\end{equation}
Holomorphicity of $\omega_{2h}$ then implies that $\Phi_{\alpha_1\dots\alpha_{2|h|}}$ solves the ZRM equation, so that it correctly represent a helicity-$h$ field. For $h=1/2$, we take $\omega_1=p^*(\Phi_{1} \D\bar \lambda) \in H^{0,1}(\PT,\mathcal{O}(-1))$, but the integral formula now requires a derivative
\begin{equation}
    \Phi_{\dot \alpha}(x)=\int_X  \left.\D\lambda\wedge\frac{\p\omega_1}{\p \mu^{\dot\alpha}}\right|_{X}\,,
\end{equation}
and similarly for other positive-helicity linear fields.

In order to extend this construction to fully non-linear, non-abelian self-dual gauge fields, we follow a  strategy due to Sparling \cite{Sparling:1990}.
We first use the asymptotic gauge field at $\scri^+_\C$ to  construct the $\mathfrak{g}$-valued $(0,1)$-form 
\begin{equation}
\twista\coloneqq p^*(A_+\,\D\bar\lambda)=A_+(\mu^{\dot\alpha}\bar\lambda_{\dot\alpha},\lambda,\bar\lambda)\D\bar\lambda\,,
\end{equation} 
valued in $\Omega^{0,1}(\PT,\mathcal{O}(0)\otimes\mathfrak{g})$ on twistor space. Since $\twista$ points only along the $\D\bar\lambda$ direction and is holomorphic in $\mu^{\dot\alpha}$, the $\dbar$ operator $\bar D=\dbar+\twista$ satisfies $\bar D^2=0$, so it defines a holomorphic vector bundle $E$ on twistor space and gives a direct method to construct the Ward transform of the self-dual gauge field from the characteristic data at null infinity.  More in detail, recall that
self-dual gauge fields are described on twistor space by the Ward correspondence \cite{Ward:1977ta, Newman:1978ze}:
\begin{thm}
There exists a one-to-one correspondence between 
\begin{itemize}
    \item self-dual gauge fields on $\M$ with gauge group $G=\GL(r,\C)$,
\item holomorphic rank-$r$ vector bundles $E\to\PT$ such that $\left.E\right|_X$ is trivial for each $x\in\M$.
\end{itemize}
\end{thm} 

We can use the reconstruction part of the Ward construction to solve the characteristic data initial value problem.  We give some details here as they will be needed in the calculations that follow. If $X\subseteq\PT$ is any line in twistor space, $\left.E\right|_X$ is topologically trivial by assumption, and holomorphic with $\bar\partial$-operator $\left.\bar D\right|_X$.  
For small $\twista$ -- that is, in perturbation theory-- this implies that $\left.E\right|_X$ is \emph{holomorphically} trivial, so  that there exists a frame $\H\colon \left.E\right|_X\to\C^r$ satisfying the Sparling equation \cite{Sparling:1990}
\begin{equation}
	\left.\bar D\right|_X \H(x,\lambda,\bar\lambda)\coloneqq \left. (\bar \p + \twista )\right|_X\H=0\,.\label{eq:sparling}
\end{equation}
It is possible to understand this equation in both geometric and holographic terms as follows \cite{Newman:1980fr}: the image of $X$ under $p$ is the light-cone cut of $x$ and it can be shown that the Sparling equation is satisfied when  $\H$ is taken to be the parallel propagator from the point $x$ to $\scri_\C^+$ along the light-cone. As we will see below, this parallel propagator determines the bulk gauge field, giving a holographic interpretation to the Ward correspondence. Note also that the frame is defined up to a matrix-valued function $g(x)$ on $\M$, $\H(x,\lambda)\to\H(x,\lambda)g(x)$, with the resulting ambiguity identified with gauge transformations on $\M$. We can remove the ambiguity by requiring that $\H(x,\iota)$ is the identity matrix for some fixed   spinor $\iota_\alpha$. The incidence relations and the chain rule imply that
$\lambda^\alpha\p_{\alpha\dot\alpha}\left.\twista\right|_X=0$, as $\left.\twista\right|_X$ only depends on $x$ through  $\mu^{\dot\alpha}$, which is annihilated by $\lambda^\alpha\partial_{\alpha\dot\alpha}$ on the support of the incidence relations \eqref{eq:incidencerelation}. Thus differentiating \eqref{eq:sparling} along $\lambda^\alpha\partial_{\alpha\dot\alpha}$ we quickly find
\begin{equation}
    \dbarx(\H^{-1}\lambda^\alpha\partial_{\alpha\dot\alpha}\H)=0\,,
\end{equation}
that is,  $\H^{-1}\lambda^\alpha\partial_{\alpha\dot\alpha}\H$ is a holomorphic function on $X$ of weight $+1$ in $\lambda_\alpha$. Liouville's theorem ensures the existence of a $\mathfrak{g}$-valued function $A_{\alpha\dot\alpha}$ on $\M$ such that
\begin{equation}
	\H^{-1}(x,\lambda)\lambda^\alpha\partial_{\alpha\dot\alpha}\H(x,\lambda)=-\im\lambda^\alpha A_{\alpha\dot\alpha}(x)\,.\label{eq:spacetimeconnection}
\end{equation}
$A_{\alpha\dot\alpha}(x)$ is the desired self-dual gauge potential transforming in the normal way under the gauge transformation $g$. Defining the covariant derivative $\nabla=\d -\im A$,  equation \eqref{eq:spacetimeconnection} can be recast as  
\begin{equation}
    \lambda^\alpha\nabla_{\alpha\dot\alpha}\H^{-1}\coloneqq \lambda^\alpha(\p_{\alpha\dot\alpha} -iA_{\alpha\dot\alpha})\H^{-1}=0\,.\label{eq:framecovariantderivative}
\end{equation}  
This directly implies
\begin{equation}
    \lambda^\alpha\lambda^\beta F_{\alpha\dot\alpha\beta\dot\beta}\H^{-1}=[\lambda^\alpha\nabla_{\alpha\dot\alpha},\lambda^\beta\nabla_{\beta\dot\beta}]\H^{-1}=0\,.
\end{equation}
Since this equation holds for any value of $\lambda_\alpha$, we deduce that $A_{\alpha\dot\alpha}$ is indeed self-dual. 
As promised, this equation implies that $\H$ is parallel propagated along light-rays from $x$ to infinity in the direction $\lambda_{\alpha}\bar\lambda_{\dot\alpha}$. If we fix the gauge freedom with the choice $\H(x,\iota)=1$ for the frame, \eqref{eq:spacetimeconnection} can be used to express the scalar potentials as well, namely
\begin{subequations}
    \begin{eqnarray}J&=&\H(x,o)\,,\label{J-from-H}\\
    K&=&\im \,o^\alpha \left.\frac{\partial}{\partial\lambda^\alpha}\H \right|_{\lambda=\iota}\,.\label{K-from-H}
    \end{eqnarray}
\end{subequations}
Finally, in the following, it will be important to know the Green's function for $\left.\bar D\right|_X$ acting on sections of $\mathcal{O}(-1)$, which can be immediately found in terms of the holomorphic frame as
\begin{equation}
    \K_X(\lambda,\lambda')=\frac{1}{2\pi \im}\frac{\H(x,\lambda)\H^{-1}(x,\lambda')}{\langle \lambda\lambda'\rangle}\,.\label{eq:Kpropagator}
\end{equation}

\section{Self-dual radiative backgrounds in $\N=4$ SYM}\label{sec:sd_fields_N=4}
We now extend the discussion from the previous section to $\N=4$ super-Yang-Mills. In order to have a discussion adapted to twistor theory, we consider the chiral formulation of the theory \cite{Chalmers:1996rq, Chalmers:1997sg}, where the $\N=4$ supermultiplet can be described by fields 
\begin{equation}
\{ A_{\alpha\dot\alpha}, \tilde \psi_{a\dot\alpha}, \phi_{ab}, \psi^a_\alpha, B_{\alpha\beta}\}\,.
\end{equation}
In the full $\N=4$ theory, $B_{\alpha\beta}=B_{(\alpha\beta)}$ is an ASD 2-form proportional to the ASD part of the curvature $F_{\alpha\beta}$, while in the self-dual theory it is a linear field imposing the self-duality condition. In both cases, we can make supersymmetry manifest by working with a superspace description: the superspace that is best adapted to make contact with twistor theory is chiral 
Minkowski superspace.

\subsection{Chiral super-fields}
We enlarge Minkowski space $\M$ to chiral Minkowski super-space $\M^{4|8}$ with coordinates $x^{\alpha A}\coloneqq(x^{\alpha\dot\alpha},\theta^{\alpha a})$ and also define the corresponding coordinate derivatives $\p_{\alpha A}\coloneqq(\p_{\alpha\dot\alpha},\p_{\alpha a})$, where we used the $2|4$ index $A=(\dot \alpha,a)$. We can then introduce the super-connection\footnote{We will underline super-fields on $\M^{4|8}$ to make the distinction with space-time fields on $\M$ clear.} $\underline\nabla_{\alpha A}=\p_{\alpha A}-\im \underline A_{\alpha A}$, with $\underline A_{\alpha A}(x,\theta)$ taking values in the Lie algebra of the gauge group.

On non-chiral superspace, the $\N=4$ equations of motion for the non-chiral super-fields are equivalent to constraint equations for the super-connection \cite{Witten:1978xx,Harnad:1985bc}. On chiral superspace, an analogue statement can be derived by imposing the following constraints on the super-connection
\begin{equation}
    [\unabla_{a(\alpha}, \unabla_{\beta)A}\}=0\,,\label{eq:N=4_constraints}\,.
\end{equation}
However, if we wish to ensure that the super-fields are a solution to the full $\N=4$ SYM equations of motion, we must require the further  constraint on the ASD part of the bosonic supercurvature
\begin{equation}
\uF_{\alpha\beta}=\lambda\uB_{\alpha\beta}\,\label{eq:N=4_ASD_constraint}\,,
\end{equation}
where the fields can be scaled so that $\lambda$ is the 't Hooft coupling and $\uB_{\alpha\beta}$ is defined as a consequence of the first set of constraints \eqref{eq:N=4_constraints}, see Equation \eqref{eq:B_definition} below and appendix \ref{app:N=4_constraints} for an extensive discussion. The self-dual theory is recovered in the limit $\lambda\to0$: in this limit, it is well known that the constraint equations can be supplemented with $[ \underline \nabla_{A(\alpha}, \underline \nabla_{\beta)B}\}=0$ and give the $\N=4$ self-dual SYM equations of motion for the super-fields, both on non-chiral \cite{Witten:1978xx, Witten:1985nt, Harnad:1985bc} and chiral \cite{Devchand:1996gv, Adamo:2013cra} superspace. 

In terms of the super-connection, $\utF_{\dot\alpha\dot\beta}$ and $\uF_{\alpha\beta}$ are defined as usual. $\uphi_{ab}$ and $\utpsi_{\dot\alpha a}$ are also defined as superspace curvatures by 
\begin{subequations}
    \begin{eqnarray}
    -\varepsilon_{\alpha\beta}\uphi_{ab}&\coloneqq&\{\unabla_{\alpha a},\unabla_{\beta b}\}\,,\label{eq:phi_supercurv}\\
        \varepsilon_{\alpha\beta}\utpsi_{a\dot\alpha}&\coloneqq&[\unabla_{\alpha a},\unabla_{\beta\dot\alpha}]\,,\label{eq:psi_supercurv}
    \end{eqnarray}
\end{subequations}
The remaining super-fields $\upsi^a_\alpha$ and $\uB_{\alpha\beta}$ are given by consistency conditions following from the constraints \eqref{eq:N=4_constraints} and suitable Jacobi identities and are defined by
\begin{subequations}
\begin{eqnarray}
\epsilon_{abcd}\upsi_\alpha^d&\coloneqq&\unabla_{\alpha a}\uphi_{bc}\,,\label{eq:psi_definition} \\ \uB_{\alpha\beta}&\coloneqq&\frac{1}{4}\unabla_{a(\alpha}\upsi_{\beta)}^a\,,\label{eq:B_definition}
\end{eqnarray}
\end{subequations}
see appendix \ref{app:N=4_constraints} for more details.  In each case the super-field at $\theta^{a\alpha}=0$  will be the corresponding $\N=4$ field on $\M$, for example the scalar fields $\phi_{ab}$ are the lowest components terms of $\uphi_{ab}$.  The Jacobi identity for $\unabla_{a\alpha}$, $\unabla_{b\beta}$ and $\unabla_{\gamma\dot\gamma}$ gives
\begin{equation}   
\unabla_{\alpha\dot\alpha}\uphi_{ab}=-\unabla_{a\alpha}\utpsi_{b\dot\alpha}\, .  \label{Jacobi}
    \end{equation}

\subsection{Chiral super-fields at $\scri$}
At null infinity, we take the tangent to the generators of $\scri^+_\C$ to be $\iota^\alpha\iota ^{\dot\alpha}$ and impose the gauge condition
\begin{subequations}
    \begin{eqnarray}
    \underline{A}_{\alpha a}&=&\iota_\alpha\underline A_a\,,\\\underline A_{\alpha\dot\alpha}&=&\iota_\alpha\underline A_{\dot\alpha}+\tilde \iota_{\dot\alpha}\underline A_{\alpha}\,.
    \end{eqnarray}
\end{subequations}
 In this gauge we can now investigate the $\theta$ dependence of the various super-fields at null infinity. It's useful to separate the fermionic variables the variables
\begin{equation}
\chi^a\coloneqq\theta^{a\alpha}o_\alpha\,,\qquad \tilde\chi^a\coloneqq-\theta^{a\alpha}\iota_\alpha\,.
\end{equation}
At null infinity, we focus on the $\chi^a$ dependence of the super-fields and set $\tilde\chi^a=0$ accordingly. Peeling means that as one approaches $\scri$, fields align with $o_\alpha$ to leading order $B_{\alpha\beta}\sim B_2 o_\alpha o_\beta/r + O(1/r^2)$.  Equivalently, we are considering a supersymmetrization $\scri_{\C,\,\N=4}^+$ of $\scri^+_\C$ coordinatized by homogeneous coordinates $(u,\lambda_\alpha,\bar\lambda_{\dot\alpha},\chi^a)$ defined up to the equivalence relation
\begin{equation}
(u,\lambda_\alpha, \bar\lambda_{\dot\alpha},\chi^a)\sim     (b\bar b \, u,b\lambda_\alpha, \bar b \bar \lambda_{\dot\alpha},b\chi^a)\,,
\end{equation}
for any $b\in\C^*$.
To obtain the $\chi^a$ dependence, we contract the spinor indices in \eqref{eq:psi_definition}, \eqref{eq:B_definition}, and \eqref{Jacobi} with $\iota^\alpha$ and, using  $\p/\p\chi^a=\iota^\alpha\p_{a\alpha}$, integrate with respect to $\chi^a$ to get
\begin{subequations}
    \begin{eqnarray}
        \langle \iota\upsi^a\rangle&=&\langle \iota\psi^a\rangle+\chi^a B_2+\mathcal{O}(\chi^2)\,,\\
        \uphi_{ab}&=&\phi_{ab}+\epsilon_{abcd}\chi^c\langle \iota\psi^d\rangle+\frac{1}{2}\epsilon_{abcd}\chi^c\chi^dB_2+\mathcal{O}(\chi^3)\,,\\{[\tilde \iota\utpsi_a]}&=&[\tilde \iota\tilde\psi_a]+\chi^b\p_u\phi_{ab}+\frac{1}{2}\epsilon_{abcd}\chi^b\chi^c\langle \iota\psi^d\rangle+\chi_a^3B_2+\mathcal{O}(\chi^4)
    \end{eqnarray}
\end{subequations}
where 
\begin{subequations}
    \begin{eqnarray}
    B_2&\coloneqq&\iota^\alpha \iota^\beta B_{\alpha\beta}\, , \\ \chi_a^3&\coloneqq&\frac{1}{3!}\epsilon_{abcd}\chi^b\chi^c\chi^d\,,
    \end{eqnarray}
\end{subequations}
and where we have taken the  integration constants to be the space-time field associated with the corresponding super-field.

The same computation applied to \eqref{eq:phi_supercurv} and \eqref{eq:psi_supercurv} leads to the expansion of the super-connection
\begin{subequations}
    \begin{eqnarray}
        \underline{A}_a&=&\p_u^{-1}[ \tilde \iota\,\tilde \psi_a]+\phi_{ab}\chi^b+\langle\iota\psi^d\rangle \frac12\epsilon_{abcd}\chi^b\chi^c+B_2\chi_a^3+O(\chi^4)\,,\\
        \tilde \iota^{\dot\alpha}\underline{A}_{\dot\alpha}&=&\tilde \iota^{\dot\alpha} A_{\dot\alpha}+[\tilde \iota\tilde\psi_a]\chi^a+ \p_u\phi_{ab}\frac12 \chi^a\chi^b+\langle \iota\,\p_u\psi^a\rangle\chi_a^3+ \p_u B_2 \chi^1\chi^2\chi^3\chi^4 \,.
    \end{eqnarray}\label{A-exp}
\end{subequations}
On restriction to $\scri^+_\C$, the super-connection thus determines the 1-form $\underline{\mathcal{A}}\,\D\bar\lambda$, where we identify $\underline{\mathcal{A}}$ with $\left.\underline A_{\dot\alpha}\right|_{\scri_{\C}}$. This means that we take
\begin{equation}
    \underline{\mathcal{A}}(u,\lambda,\bar\lambda,\theta)= A_+ + \Phi_{1,a}\chi^a + \p_u \Phi_{0,ab}\chi^a\chi^b+ \p_u\Phi_{-1}^a\chi^3_a +\p_u \Phi_{-2} \chi^1\chi^2\chi^3\chi^4\,,\label{eq:N=4_radiative_data}
\end{equation}
where $\chi^a\coloneqq\theta^{a\alpha}\lambda_\alpha$ and the coefficients are the characteristic data at $\scri^+_\C$ for the $\N=4$ super Yang-Mills multiplet
\begin{equation}
    \{\Phi_2=\p_u A_+,\Phi_{1,a}, \Phi_{0,ab}=\Phi_{0,[ab]}, \Phi_{-1}^a
    , \Phi_{-2}\}\, .
\end{equation}
Note that the definition of the radiative data doesn't require any self-duality condition to be valid, in complete analogy with \eqref{eq:radiative-data-pure-yang-mills}.

On the other hand, if we consider $\N=4$ \emph{self-dual} super Yang-Mills, the equations of motion are equivalent to the graded integrability conditions \cite{Witten:1978xx, Witten:1985nt, Harnad:1985bc, Devchand:1996gv}
\begin{equation}
    [\underline\nabla_{A (\alpha}, \underline\nabla_{\beta) B}\}=0\, , \label{SSDYM}
\end{equation}
and they imply that there exists a gauge for which $\underline A_\alpha=0$ as for the bosonic case. 
The integrability condition \eqref{SSDYM} implies the existence of supersymmetrized versions of the scalar potentials $\underline J(x,\theta)$ and $\underline K(x,\theta)$, so that 
\begin{equation}
    \underline A_{\alpha A}=\iota_\alpha \iota^\beta \p_{\beta A}\underline K=-\im\, \iota_\alpha o^\beta \underline J^{-1} \p_{\beta A} \underline J\, .\label{super-KJ}
\end{equation}
As before, the space-time fields $\tilde F_{\dot\alpha\dot\beta}$, $\tilde\psi_{\dot\alpha a}$, and $\phi_{ab}$ can also be understood as the lowest components of the self-dual super-curvature, defined by $\varepsilon_{\alpha\beta}\underline{\mathcal{F}}_{AB}\coloneqq[\underline\nabla_{\alpha A},\underline\nabla_{\beta B}\}$. We now have, with the definitions $\d_A\coloneqq \iota^\beta\p_{\beta A}$ and $ \tilde \d_A\coloneqq o^\beta\p_{\beta A}$
\begin{equation}
        \underline{\mathcal{F}}_{AB}=\d_A\d_B\underline K=-\im\,\d_{(A}(\underline J^{-1}\tilde\d_{B\}}\underline J)\,,
\end{equation}
in addition to \eqref{K-curv} and \eqref{J-curv}. Similarly, the $\N=4$ self-duality equations are the obvious super-symmetrizations of \eqref{eq:kmatrixeom} and \eqref{eq:jmatrixeom}
\begin{subequations}
    \begin{eqnarray}
2\tilde\d_{[A}\d_{B\}}\underline K+\im[\d_A\underline K,\d_B\underline K\}&=&0\,,\\\d_{[A}(\underline J^{-1}\tilde\d_{B\}}\underline J)&=&0\,.
    \end{eqnarray}
\end{subequations}
In this gauge, the expansion of the $\underline K$ matrix at $\tilde\chi=0$ can be obtained by integrating the fermionic part of \eqref{super-KJ} using \eqref{A-exp} to obtain
\begin{equation}
    \underline{K}=K-\p_u^{-1}[\tilde\iota\tilde\psi_a]\chi^a+\frac{1}{2}\phi_{ab}\chi^a\chi^b-\langle\iota\psi^a\rangle \chi_a^3+\iota^\alpha\iota^\beta B_{\alpha\beta} \chi^1\chi^2\chi^3\chi^4\,.
\end{equation}

\subsection{Super-twistor space description}
As in the pure Yang-Mills case, solutions to the $\N=4$ self-dual SYM  equations are compactly obtained from a supersymmetric version of the Ward correspondence.  Introducing homogeneous coordinates $Z^A=(\mu^{\dot\alpha},\lambda_\alpha,\chi^a)$, $a=1,\ldots,4$, on $\CP^{3|4}$ subject to $Z^A\sim t Z^A$ for $t\in\C^*$, the super-twistor space of $\M^{4|8}$ is defined to be  \cite{Ferber:1977qx}
\begin{equation}
\PT=\{[Z]\in\CP^{3|4}:\lambda_\alpha\neq 0\}\,,
\end{equation}
and we still denote it as $\PT$. The incidence relations are 
\begin{equation}
\mu^A\coloneqq(\mu^{\dot\alpha},\chi^a)=(x^{\alpha\dot\alpha}\lambda_\alpha,\theta^{\alpha a}\lambda_\alpha)\,,
\end{equation}
and we denote the line in super-twistor space again by $X$, even though it now depends on $(x,\theta)$. The Ward correspondence becomes \cite{Volovich:1983aa,Volovich:1983ii}
\begin{thm}
There exists a one-to-one correspondence between 
\begin{itemize}
    \item solutions to the $\N=4$ self-dual Yang-Mills equations on $\M^{4|8}$ with gauge group $G=\GL(r,\C)$,
\item holomorphic rank-$r$ vector bundles $E$ over super-twistor space such that $\left.E\right|_{X}$ is trivial for each $(x,\theta)\in\M^{4|8}$.
\end{itemize}
\end{thm}

In the Dolbeault framework, these bundles are equipped with an integrable super-connection $\A$ and Dolbeault operator $\bar D=\dbar+\A$, which can be obtained using the characteristic data built out of the null data. 
Together with the gluonic background twistor connection $\twista$ of the previous section, we can construct the connection on supertwistor space 
\begin{equation}
\A\coloneqq\twista+\twista_a\chi^a+\frac{1}{2}\twista_{ab}\chi^{a}\chi^b
+\twista^a\chi^3_a+\tilde \twista\chi^1\chi^2\chi^3\chi^4 \,,
\end{equation}
valued in $\Omega^{0,1}(\PT,\mathcal{O}(0)\otimes\mathfrak{g})$.  Here $\{\twista,\twista_a,\twista_{ab},
\twista^a,\tilde \twista\}$ are $(0,1)$-forms of  respective homogeneity $0,-1,-2,-3,-4$ in the bosonic twistor variables and can be obtained as pullbacks of the radiative data for the multiplet respectively $p^*\{A_+,\Phi_{1, a},\p_u \Phi_{0, ab}, \p_u\Phi_{-1}^a,\p_u\Phi_{-2}\}$ on $\scri^+_\C$.

The reconstruction part of the Ward correspondence is unaltered from the pure Yang-Mills case, the only difference being the promotion of every field to a super-field. In this way, at least for small data,  there exists a holomorphic frame $\underline\H(x,\theta,\lambda)$ satisfying
\begin{equation}
    \left.\bar D\right|_{X}\underline\H(x,\theta,\lambda)=0\,,\label{D-bar-H}
\end{equation}
in terms of which we can construct the super-connection $\underline A_{\alpha A}$ on $\M^{4|8}$ as
\begin{equation}
    \underline\H^{-1}(x,\theta,\lambda)\lambda^\alpha\p_{\alpha A}\underline\H(x,\theta,\lambda)=-\im\lambda^\alpha \underline A_{\alpha A}(x,\theta)\,.
\end{equation}

\section{Integral formulae for the curvature and background coupled fields}\label{sec:integral}
In order to lift space-time formulae to twistor space, in this Section we  obtain explicit expressions for the space-time gauge field in terms of the twistor connection $\twista$ and the frame $\H$; although these are in principle already determined by the previous Section, we will need more explicit formulae. Similarly, it has been known for some time how to  obtain formulae for background-coupled linear fields and super-fields, but here we give more detailed formulae for momentum eigenstates.

\subsection{Connection and curvature formulae}
We first introduce Green's functions on the Riemann sphere for inverting the $\dbar$ operators on different line bundles. Let $\mathcal{O}(n)\to\CP^1$ be the line bundle of homogenous functions of degree $n$ in $\lambda_\alpha$.  Provided $n\geq-1$,  we can invert 
\begin{equation}
    \dbar g_n =f_n\,,
\end{equation}
for  any $f_n\in\Omega^{0,1}(\CP^1,\mathcal{O}(n))$ to find a solution $g_n\in\Omega^{0,0}(\CP^1,\mathcal{O}(k))$, namely
\begin{equation}
    g_n(\lambda)=\int\frac{\D\lambda'}{2\pi \im}\; f_n(\lambda')\frac{1}{\langle\lambda\lambda'\rangle}\left(\frac{\langle \iota\lambda\rangle}{\langle  \iota\lambda'\rangle}\right)^{n+1}\,.\label{eq:greensfunction}
\end{equation}
The integral is over $\CP^1$, on which $\lambda_\alpha,\lambda'_\alpha$ are homogeneous coordinates, and the reference spinor $\iota_{\alpha}$ is used to fix the  freedom in adding polynomials of degree $n$ in $\lambda$ to $g_n$ by making it vanish to $n$-th order at $\iota_\alpha$; note that for $n=-1$ the solution is unique, while for $n\geq0$ the ambiguity in $g_n$ is a  consequence of $H^{0}(\CP^1,\mathcal{O}(n))\cong\C^{n-1}$. 

We can now find an integral formula for $A_{\alpha\dot\alpha}$ by differentiating the Sparling equation \eqref{eq:sparling} and eliminating the twistor connection via the definition of the holomorphic frame. In this way, we find
\begin{equation}
 \dbarx(\p_{\alpha\dot\alpha}\H^{-1}\,\H)=\lambda_\alpha \H^{-1}\frac{\p\twista}{\p\mu^{\dot\alpha}}\H\,,
 \end{equation}
 and using the Green's functions \eqref{eq:greensfunction}
 \begin{equation}
   \partial_{\alpha\dot\alpha}\H^{-1}(x,\lambda)\,\H(x,\lambda)=\frac{1}{2\pi \im}\int_X\frac{\D\lambda'}{\langle\lambda\lambda'\rangle}\frac{\langle \iota\lambda\rangle}{\langle \iota\lambda'\rangle}\lambda'_\alpha \H^{-1}(x,\lambda')\left.\frac{\partial\twista}{\partial\mu^{\dot\alpha}}\right|_X\H(x,\lambda')\,.\label{eq:dh-1h}
\end{equation}
The possible ambiguity in adding a constant to the right-hand side at homogeneity degree zero is fixed by the vanishing of both sides of the equation at $\lambda_\alpha=\iota_\alpha$ for the gauge  $\H(x,\iota)=I$ where $A_{\alpha\dot\alpha}=\iota_\alpha A_{\dot\alpha}$.   
Contracting this equation with $\lambda^\alpha$ using \eqref{eq:framecovariantderivative}  yields 
\begin{equation}
    A_{\alpha\dot\alpha}(x)=\frac{\iota_\alpha}{2\pi}\int_X\frac{\D\lambda'}{\langle \iota \lambda'\rangle}\H^{-1}(x,\lambda')\left.\frac{\partial\twista}{\partial\mu^{\dot\alpha}}\right|_X\H(x,\lambda')\,.\label{eq:gaugefieldintegral}
\end{equation}
Comparing this last equation with \eqref{eq:kmatrix}, we can identify the integral in \eqref{eq:gaugefieldintegral} with $\d_{\dot\alpha}K$. The associated field strength can be straightforwardly checked to be self-dual, with SD component
\begin{equation}
    \tilde F_{\dot\alpha\dot\beta}=\int_X\frac{\D\lambda_1}{2\pi \im}\H_1^{-1}\partial_{{\dot\alpha}}\partial_{\dot\beta}\twista_1\H_1
   -\int_{X^2}\frac{\D\lambda_1\D\lambda_2}{(2\pi \im)^2\langle\lambda_1\lambda_2\rangle}[\H_1^{-1}\partial_{{\dot\alpha}}\twista_1\H_1,\H_2^{-1}\partial_{{\dot\beta}}\twista_2\H_2]\,,
\label{eq:fieldstrength}
\end{equation}
where $\H_i=\H(x,\lambda_i)$, $\twista_i=\left.\twista\right|_X(x,\lambda_i)$, and we denoted $\mu^{\dot\alpha}$ derivatives as $\partial_{{\dot\alpha}}\coloneqq\partial/\partial\mu^{\dot\alpha}$. The expression for $\tilde F_{\dot\alpha\dot\beta}$ has the advantage of being now both Lorentz and gauge invariant; the linear  Penrose transform would lead to the first term in \eqref{eq:fieldstrength} only, but for the fully non-linear field we need also the second, double integral over $X$. 

For $\N=4$ SYM, we similarly have
\begin{equation}
    \underline A_{\alpha A}(x,\theta)=\frac{\iota_\alpha}{2\pi}\int_{X}\frac{\D\lambda'}{\langle \iota\lambda'\rangle}\underline\H^{-1}(x,\theta,\lambda)\left.\p_A\A\right|_{X}\underline\H(x,\theta,\lambda)\,.
\end{equation}
Here $\p_A\coloneqq(\p/\p\mu^{\dot\alpha},\p/\p\chi^a)$. The supersymmetrized $J$- and $K$-matrices and the propagator on the line $X$ can be defined in this context as well by replacing the holomorphic frame with its supersymmetrized version in \eqref{J-from-H}, \eqref{K-from-H}, \eqref{eq:Kpropagator}. The covariant derivative $\underline\nabla_{\alpha A}$ is self-dual by construction with $[\nabla_{aA},\nabla_{\beta B}\}=\varepsilon_{\alpha\beta} \underline F_{AB}$; the gluon self-dual curvature, the positive-helicity fermions and the scalar fields arise as the lowest components of this self-dual super-curvature. Moreover, the supersymmetrized Ward correspondence now gives gauge and Lorentz invariant expressions for both the fermions and the scalars, in complete analogy with the pure Yang-Mills case. Explicitly, the space-time fermion $\tilde\psi_{a\dot\alpha}$ is the lowest component of the superfield
\begin{equation}
\underline{\tilde\psi}_{a\dot\alpha}=\int_{X}\frac{\D\lambda_1}{2\pi \im}\underline\H_1^{-1}\partial_{{a}}\partial_{\dot\alpha}\A_1\underline\H_1 
   -\int_{X^2}\frac{\D\lambda_1\D\lambda_2}{(2\pi \im)^2\langle\lambda_1\lambda_2\rangle}[\underline\H_1^{-1}\partial_{a}\A_1\underline\H_1,\underline\H_2^{-1}\partial_{\dot\alpha}\A_2\underline\H_2]\,,
\label{eq:fermionsuper-field}
\end{equation}
whilst the scalar $\phi_{ab}$ is the lowest component of
\begin{equation}
\underline\phi_{ab}=\int_{X}\frac{\D\lambda_1}{2\pi \im}\underline\H_1^{-1}\partial_{{a}}\partial_{b}\A_1\underline\H_1 
   -\int_{X^2}\frac{\D\lambda_1\D\lambda_2}{(2\pi \im)^2\langle\lambda_1\lambda_2\rangle}\{\underline\H_1^{-1}\partial_{a}\A_1\underline\H_1,\underline\H_2^{-1}\partial_{b}\A_2\underline\H_2\}\,.
\label{eq:scalarsuper-field}
\end{equation}
This is also the prescription given in \cite{Koster:2016ebi, Koster:2016loo, Koster:2016fna} for the construction of vertices for composite operators in $\N=4$ SYM.


\subsection{Perturbations, linearized modes and momentum eigenstates}
Massless fields of helicity $n/2$ on a self-dual background are well-known to be given as first cohomology classes on twistor space with values in the appropriate representation of the Ward bundle $E$ twisted by $\mathcal O(n-2)$.  In our radiative framework, for $n\leq 0$, these can be represented by their $\scri$ data $f_{n-2}\coloneqq\p_u\Phi_n \D\bar\lambda$.  These are $\dbar$ closed around the non-trivial self-dual background too, because both $\twista$ and and $ f_{n-2}$ are pointing only along the $\D\bar\lambda$ direction and $\Phi_n$ is holomorphic in $\mu^{\dot\alpha}$.  For $n\leq 0$, we can obtain a space-time linear field via the standard integral representation of the Penrose transform; the coupling with the background arises because the bundle $E$ must first be trivialized before performing the twistor integral.  Taking $f_{n-2}$ to be in the adjoint, we obtain
\begin{equation}
    \Phi_{\alpha_1\ldots\alpha_{|n|}}=\frac{1}{2\pi\im }\int_X \lambda_{\alpha_1} \H^{-1}(x,\lambda) \left.f_{n-2}\right|_X\H(x,\lambda)\wedge \D\lambda\, .
\end{equation}
For concrete calculations, we will take our $f_{n-2}$ to be momentum eigenstates
with colour $T_{j}\in \mathfrak{g}$ and null momentum $k^{\alpha\dot\alpha}_j=\kappa_j^\alpha\tilde\kappa_j^{\dot\alpha}$.
\begin{equation}
    f^j_{n-2}(Z)=T_{j}\int_{\C^*}\frac{\d s}{s^{n-1}}\bar\delta^{2}(\kappa_j-s\lambda)e^{\im s[\mu\tilde\kappa_j]}\,,\label{eq:negativemomentumeigenstates}
\end{equation}
where the holomorphic $\delta$-function is defined by
\begin{equation}
    \bar\delta^2(\kappa_j-s\lambda)\coloneqq\frac{1}{(2\pi \im)^2}\bigwedge_{\alpha=1,2}\dbar\left(\frac{1}{\kappa_{j\,\alpha}-s\lambda_\alpha}\right)\,.
\end{equation}

\paragraph{Negative-helicity gluons.}
The ASD linearized field strength itself is the case $n=-2$ where the integral formula is immediately performed against the delta functions to give the ASD field strength 
\begin{equation}
    f^j_{\,\alpha\beta}(x)=\kappa_{j\alpha}\kappa_{j\beta}\H^{-1}_j
    T_{j}
    \H_j\, e^{\im k_j\cdot x}\,, \qquad\qquad \H_j\coloneqq\H(x,\kappa_j)\, .
    \label{eq:negativemomentumeigenstates_spacetime}    
\end{equation}
It is now easily checked that the spin-1 equation \eqref{ASD-pert} below follows using \eqref{eq:framecovariantderivative} under the integral sign. We will see that this only gives the perturbation  $f_{j\,\alpha\beta}$ of the curvature, whilst the construction of the corresponding  gauge field perturbation $a^j_{\alpha\dot\alpha}$ is not so straightforward as we now describe.

\paragraph{Backgound perturbations}
The linearized equations of motion for a general perturbation $a_{\alpha\dot\alpha}$ of a  self-dual background gauge field read
\begin{subequations}
\begin{eqnarray}
 \nabla_{\dot \alpha(\alpha}a_{\beta)}^{\dot\alpha}&=&f_{\alpha\beta}\,,\label{pot-pert}\\
    \nabla_{\alpha\dot\alpha}f^{\alpha\beta}&=&0\,.\label{ASD-pert}
\end{eqnarray}
\end{subequations}
In a general background, it is no longer consistent to decompose a linear field on the background into those whose perturbation of the curvature is self-dual and anti-self-dual  as \eqref{ASD-pert}   would have an extra term from the background ASD curvature \cite{Adamo:2021rfq}.  However, on a self-dual background,  it is still consistent to require that $f_{\alpha\beta}=0$ in which case  the perturbation preserves the self-duality condition. Such solutions are still naturally identified with positive-helicity gluons. Conversely, non-trivial solutions to \eqref{ASD-pert} are interpreted as negative-helicity gluons around the SD background, but the corresponding potential that solves \eqref{pot-pert} will generally lead to a non-trivial self-dual curvature perturbation too: even if it is imposed it to be zero asymptotically, it will develop a non-zero value as the field is evolved through the space-time. This follows because the MHV tree amplitude can be understood as being generated by the self-dual field at $\scri^+$ associated with a potential crossing space-time on an SD background whose data at $\scri^-$ is purely ASD, see  \cite{Mason:2008jy, Adamo:2021rfq} for details.

\paragraph{Positive-helicity gluons.}
In the following, we focus on  MHV form factors, which contain arbitrarily many positive-helicity external states. The Penrose transform relates these self-dual gluon perturbations to the cohomology classes of weight 0 whose representative $a_j$ with colour $T_i
\in \mathfrak{g}$ and null momentum $k^{\alpha\dot\alpha}_j=\kappa_j^\alpha\tilde\kappa_j^{\dot\alpha}$ is as above
\begin{equation}
    a_j(Z)=T_{j}
    \int_{\C^*}\frac{\d s}{s}\bar\delta^{2}(\kappa_j-s\lambda)e^{\im s[\mu\tilde\kappa_j]}\,.\label{eq:momentumeigenstates}
\end{equation}
These will be used around a non-trivial background as well as in the radiative framework, the  construction provides a perturbation that is asymptotic to a positive-helicity plane wave at $\scri_\C$. The corresponding space-time perturbation can be reconstructed by perturbing  the formulae above.  Considering just one of the perturbations, we perturb $\twista\to \twista+\epsilon_j a_j $ so that   $\H\to\H+\epsilon_j \,\delta_j\H$. The perturbation of \eqref{eq:sparling} to first order in $\epsilon_j$ yields
\begin{equation}
    \dbarx(\H^{-1}\delta_j\H)=\H^{-1}a_j\H\,,
\end{equation}
where we used the definition of $\H$ to eliminate the background twistor connection. Using the Green's function \eqref{eq:greensfunction} and integrating against the delta function then gives
\begin{equation}
    \H^{-1} \delta_j\H=\frac{\langle \iota\lambda\rangle}{\langle \iota j\rangle\langle\lambda j\rangle}\H^{-1}_jT_{j}\H_je^{\im k_j\cdot x}\,.\label{eq:sd_perturbation_frame}    
\end{equation}
where insertions of $\kappa_j$ into angle-brackets are denoted just by $j$. The variations  of the $J$- and $K$-matrices can be obtained via \eqref{K-from-H} and \eqref{J-from-H} to give 
\begin{subequations}
\begin{eqnarray}
    \delta_jK &=& -\frac{\im}{\langle \iota j\rangle^2} \H^{-1}_jT_{j}
    \H_je^{\im k_j\cdot x}\,, \\
   J^{-1}\delta_j J&=& -\frac{1}{\langle \iota j\rangle\langle o j\rangle}\H^{-1}_jT_{j}
    \H_je^{\im k_j\cdot x}  \, .\end{eqnarray}
\end{subequations}
These then yield the space-time perturbation
\begin{align}
a_{j\,\alpha\dot\alpha}(x)=\frac{\iota_\alpha}{\langle \iota j\rangle}\H^{-1}_j\left(\tilde{\kappa}_{j\,\dot\alpha}T_{j}
+[g_{j\dot\alpha}(x),T_{j}
]\right)\H_j e^{\im k_j\cdot x}\,,\label{eq:spacetimeperturbation}
\end{align}
where we have used \eqref{eq:framecovariantderivative} to define $    g_{\dot\alpha}(x,\lambda)$ by
\begin{equation}
\nabla_{\alpha\dot\alpha }\H^{-1}= \lambda_\alpha    \H^{-1}  g_{\dot\alpha}\,,
\end{equation} 
and the subscript $j$ denotes  evaluation  at $\lambda=\kappa_j$. Similar formulae for the variation of the curvature can be obtained. Here  the usual gauge dependent undotted spinor present in the vector polarization of a spin-1 momentum eigenstate is taken to be $\iota_\alpha$. Note in particular that the reconstruction of the space-time gauge field, rather than of its curvature, relies heavily on the existence of (perturbed) $J$- and $K$-matrices, in other words it's possible only for positive-helicity perturbations that preserve the self-duality condition. For negative-helicity fields, we can at best construct the linearized field strength \eqref{eq:negativemomentumeigenstates_spacetime}.

Perturbations corresponding to positive-helicity gluons also modify the propagator \eqref{eq:Kpropagator}. For a general perturbation $\delta a$ --that is, not necessarily a momentum eigenstate-- the variation $\delta\K_X$ follows again from \eqref{eq:sd_perturbation_frame}
\begin{align}
    \delta\K_X(\lambda,\lambda')&=-\int_X\D\lambda''\,\K_X(\lambda,\lambda'')\left.\delta a\right|_X\K_X(\lambda'',\lambda')\,.
    \label{eq:pertubedKpropagator}
\end{align}
This variation can be iterated $n$ times to obtain the colour-ordered $n$th perturbation of the propagator on $X$ as
\begin{equation}
    \K_X(\lambda,\lambda')=\sum_{n=0}^\infty\left(\frac{-1}{2\pi\im}\right)^n\H(x,\lambda)\int_{X^n}\D\lambda_1\ldots\D\lambda_n\frac{\H_1^{-1}\delta a_{1}\H_1\ldots\H_n^{-1}\delta a_{n}\H_n}{\langle\lambda\lambda_1\rangle\langle\lambda_1\lambda_2\rangle\ldots\langle\lambda_n\lambda'\rangle}\H^{-1}(x,\lambda')\,,\label{eq:Kperturbed}
\end{equation}
where $\delta a_{j}\coloneqq\left.\delta a\right|_X(x,\lambda_j)$.
The frames in \eqref{eq:Kperturbed} are the frames for the background $\twista$, and as before each term is evaluated on the line $X$. If the perturbations are taken to be momentum eigenstates, the integrations can be directly performed against the delta functions.

\paragraph{$\N=4$ super-momentum eigenstates.} The Penrose transform provides momentum eigenstates for different helicities as well. In the following, we will consider super form factors in $\N=4$ where we super-symmetrize the external states (but not the local operator), so we arrange the possible external states on space-time in terms of Nair's super-field \cite{Nair:1988bq}, i.e. we consider the external state
\begin{equation}
    \Phi_j=g_j^++\psi^+_{j\,a}\eta^a_j+\frac{1}{2}\phi_{j\,ab}\eta^a_j\eta^b_j+\frac{1}{3!}\varepsilon_{abcd}\psi^{-\,a}_j\eta_j^b\eta_j^c\eta_j^d+g^-_j\eta_j^1\eta_j^2\eta_j^3\eta_j^4\,,
\end{equation}
with super-momentum $k_{j\,\alpha A}\coloneqq\kappa_{j\,\alpha}\tilde\kappa_{j\,A}=(\kappa_{j\,\alpha}\tilde\kappa_{j\,\dot\alpha},\kappa_{j\,\alpha}\eta_{j\,a})$. The associated twistor representative around the flat background is
\begin{equation}
    \underline{a}_i(Z)=T_i\int_{\C^*}\frac{\d s}{s}\bar\delta^2(\kappa_i-s\lambda)e^{\im s[\mu\tilde\kappa_i]+\im s\{\chi\eta_i\}}\,,\label{eq:n=4_on_shell_state}
\end{equation}
and since we are considering horizontal background fields on twistor space, such a representative can be used around these backgrounds as well. The corresponding linear perturbation on chiral superspace is 
\begin{equation}
    \underline{a}_{j\,\alpha A}(x,\theta)=\frac{\iota_\alpha}{\langle \iota j\rangle}\underline\H_j^{-1}(\tilde\kappa_{j\,A}+[\underline g_{j\,A}(x,\theta),T_j])\underline\H_je^{\im k_j\cdot x+\im \kappa_{j\,\beta}\eta_{j\,b}\theta^{b\beta}}\,,
\end{equation}
where $\underline g_A$ is the natural supersymmetrization of $g_{\dot\alpha}$. Notice in particular that super-momentum eigenstates will perturb the propagator on the line $X$ as well, the variation being given by the supersymmetrization of \eqref{eq:Kperturbed}.

\section{MHV (super) form factors}\label{sec:form_factors}
Given a local operator $\mathscr{O}(x)$, its form factor $\mathscr{F}_{\mathscr{O}}=\mathscr{F}_{\mathscr{O}}(1^{h_1},\ldots,n^{h_n};q)$ in presence of $n$ external gluons is defined as the Fourier transform of the matrix element of $\mathscr{O}(x)$ between the vacuum and the $n$-gluon multiparticle state
\begin{equation}
    \mathscr{F}_{\mathscr{O}}(1,\ldots, n;q)\coloneqq\int_\M\d^4x\,e^{-\im q\cdot x}\langle 1^{h_1},\ldots,n^{h_n}|\mathcal{O}(x)|0\rangle\,,
\end{equation}
where we implicitly assumed to take our gluons to be outgoing plane waves. Since we focus on MHV form factors, the helicities are almost all positive, the number of negative-helicity gluons being equal to the number of $B$ fields appearing in $\mathscr{O}$. 
 
\subsection{MHV form factors and Cartan backgrounds} 
 If $\mathscr{O}$ is a composite operator depending only on $\tilde F_{\dot\alpha\dot\beta}$ but not on $B_{\alpha\beta}$, it's straightforward to show that its tree-level MHV form factor can be readily computed in self-dual Yang-Mills \cite{Chalmers:1996rq} by simply putting it on-shell \cite{Boels:2013bi}.
 More generally, in the MHV sector, we can treat form factors involving  $B_{\alpha\beta}$ as well by treating these ASD terms as perturbations away from the self-dual sector of $S_{\text{SDYM}}$ below.  
 The generating functional for such a form factor is the path integral with action 
\begin{equation}S_\text{SDYM}+\int_\M\d^4x\,\mathscr{J}\mathscr{O}\,,\end{equation} $S_\text{SDYM}$ being the action for self-dual Yang-Mills \cite{Chalmers:1996rq}
\begin{equation}
    S_\text{SDYM}=\int_\M\d^4x\,\tr B_{\alpha\beta} F^{\alpha\beta}\,,
\end{equation}
and $\mathscr{J}$ being a source for $\mathscr{O}$. At tree-level, the generating functional reduces to (the exponential of) the on-shell action in the presence of the source, but even for non-trivial $\mathscr{J}$ the existence of the $J$- and $K$-matrices is not affected, as long as $\mathscr{O}$ is a polynomial in $\tilde{F}_{\dot\alpha\dot\beta}$ and its derivatives.
The argument for  the generating functional still holds for operators involving the $B$ field, because in the MHV sector, the number of negative-helicity gluons is then precisely the number of $B_{\alpha\beta}$'s appearing in the operator under study, so for non-exceptional kinematical configurations we can treat each $B_{\alpha\beta}$ as a linear perturbation away from self-duality. Similar considerations hold for form factors in the self-dual sector of $\N=4$ SYM. From the perspective of twisted holography \cite{Costello:2022wso}, this procedure can be interpreted as an explicit realization \cite{Bu:2022dis} of the correspondence between local operators in four dimensions and conformal blocks of a two-dimensional chiral algebra living on the celestial sphere. 

If the form factor is computed in the MHV sector, the set of external positive-helicity gluons defines a self-dual background obtained as the coherent state whose data is a sum of plane waves at $\scri_\C^+$, thus the on-shell expression of $\mathscr{O}$ around a general self-dual radiative background is the generating functional for the MHV form factor of $\mathscr{O}$.  In practice, $\mathscr{F}_{\mathscr{O}}(1^+,\ldots,n^+)$ can be computed by considering the bulk gauge field that reduces to $\epsilon_1a_1+\ldots \epsilon_na_n$ at $\scri_\C$, where $\{\epsilon_j\}$ are formal parameters, and extracting the term in $\mathscr{O}$ proportional to $\epsilon_1\ldots\epsilon_n$ -- see also \cite{Rosly:1996vr, Selivanov:1998hn} for a related approach.
The twistor theory developed in the previous sections reduces  the task of finding such coherent states from the non-linear equations of motion on $\M$ with boundary condition at $\scri^+_\C$ to a linear problem on $\PT$, leading to the construction of all-multiplicity formulae for the form factor from integrability rather than perturbation theory. The same strategy can be used for form factors around \emph{any} self-dual radiative background, as long as we include the twistor data for the radiative background together with the momentum eigenstates representatives, and treat the former non-perturbatively.

For the most part,  we restrict ourselves to backgrounds valued in a Cartan subalgebra $\mathfrak{h}\subseteq\mathfrak{g}$. Although our methods naturally yield concrete formulae for more general backgrounds, this restriction leads to simpler formulae in which the background is encoded into abelian factors obtainable by quadratures, i.e., direct integral formulae. These formulae   still  decompose an observable into colour-ordered components, with the additional  information of a set of charges $e^i$ of the positive helicity gluons relative to the background.  These are determined by the relation $[t^i,T_j]=e^iT_j$, where $\{t^i\}$ are a basis of $\mathfrak{h}$ and $T_j$ the colour of the $j$-th gluon. If we compute observables around non-Cartan backgrounds, we need to introduce non-abelian $\H$ factors associated with the background which, although determined as above, are no longer expressible as quadratures and further interleave the colour-ordered expression.


When the background is valued in the Cartan subalgebra, we can express the holomorphic frame $\H=e^{-g}$ explicitly in terms of the Cartan-valued function $g$ \cite{Ward:1977ta, Adamo:2020yzi} 
\begin{equation}
    g(x,\lambda)=\frac{1}{2\pi \im}\int_X\frac{\D\lambda'}{\langle\lambda\lambda'\rangle}\frac{\langle \iota\lambda\rangle}{\langle \iota\lambda'\rangle}\left.\twista\right|_X\,,
\end{equation}
and further use it to provide integral formulae for the background coupled fields. For example, the $J$- and $K$-matrices are given by
\begin{subequations}
\begin{eqnarray}
\log J&=&-\frac{1}{2\pi\im}\int_X\frac{\D\lambda'}{\langle \iota\lambda'\rangle\langle o\lambda'\rangle}\left.\twista\right|_X\,,\\K&=&\frac{1}{2\pi}\int_X\frac{\D\lambda'}{\langle \iota\lambda'\rangle^2}\left.\twista\right|_X\,.
\end{eqnarray}
\end{subequations}
Similarly, one can further simplify the field strength for linear perturbations around a Cartan-valued background. Negative-helicity gluons have a linearized ASD field strength
\begin{equation}
    b_{j\,\alpha\beta}(x)=\kappa_{j\alpha}\kappa_{j\beta}T_je^{\im k_j\cdot x+e_jg(x,\kappa_j)}\,,
\end{equation}
whilst positive-helicity gluons have linearized potential
\begin{equation}
        a_{j\,\alpha\dot\alpha}(x)=\frac{\iota_\alpha}{\langle \iota j\rangle}(\tilde\kappa_{j\dot\alpha}+e_jg_{\dot\alpha}(x,\kappa_j))T_je^{\im k_j\cdot x+e_jg(x,\kappa_j)}\,.
\end{equation}
In particular, there is a factorization between the colour and kinematical degrees of freedom (as opposed to the more general structure in \eqref{eq:spacetimeperturbation}), the background dresses the dotted component of the momentum as
\begin{equation}
    \kappa_{j\,\alpha}\tilde{K}_{j\,\dot\alpha}(x)=\kappa_{j\,\alpha}(\tilde\kappa_{j\,\dot\alpha}+e_jg_{\dot\alpha}(x,\kappa_j))\,,\label{eq:dressed_momentum}
\end{equation}
while leaving invariant the undotted component. This is expected as a consequence of self-duality. The corresponding linearized SD field strength for a positive-helicity gluon is
\begin{equation}
    \tilde f_{j\,\dot\alpha\dot\beta}(x)=\tilde{K}_{j\,\dot\alpha}(x)\tilde{K}_{k\,\dot\beta}(x)T_je^{\im k_j\cdot x+e_jg(x,\kappa_j))}-\frac{\im}{\langle \iota j\rangle}\iota^\alpha\p_{\alpha\dot\alpha}g_{\dot\beta}(x) T_je^{\im k_j\cdot x+e_jg(x,\kappa_j)}\,.
\end{equation}

\subsection{Form factors for powers of $B$ and $\tilde F$}

We first consider the form factor for $\tr B^2$.  At $q=0$, this is  the interaction term in the Chalmers-Siegel action that extends the SDYM action to the full YM action.  Its generating functional is
\begin{equation}
    \int_{\M}\d^4x\,e^{\im(k_i+k_j- q)\cdot x+e_ig(x,\kappa_i)+e_jg(x,\kappa_j)}\langle ij\rangle^4\tr( \K_X(\kappa_i,\kappa_j)T_i\K_X(\kappa_j,\kappa_i)T_j)\,,
\end{equation}
where we assumed that the negative-helicity gluons have momenta $k_i,k_j$ and colours $T_i,T_j$. Since the addition of $\tr B^2$ to $S_{\text{SDYM}}$ yields an action perturbatively equivalent to the ordinary full Yang-Mills action, the $q\to 0$ limit of the generating functional is interpreted as the amplitude for the helicity flip of a single negative-helicity gluon traversing an SD background \cite{Mason:2009afn}. If the background connection on twistor space is further taken to be in the form $\twista+\epsilon_1a_1+\ldots+\epsilon_{n-2}a_{n-2}$, where $a_i$ are twistor representatives for positive-helicity momentum eigenstates, the 2-point amplitude expands into the $n$-point MHV amplitude around the background defined by $\twista$ \cite{Adamo:2020yzi}, in particular the Parke-Taylor denominator arises from the perturbation \eqref{eq:Kperturbed} of the propagator on the line $X$, the integrals over $X$ being saturated by the holomorphic $\delta$ functions of the momentum eigenstates. More generally, the same expansion allows us to obtain the $\tr B^2$ form factor away from $q=0$
\begin{equation}
    \mathscr{F}_{\tr B^2}(1^+,\ldots,i^-,\ldots,j^-,\ldots n;q)=\frac{\langle ij\rangle^4}{\langle 12\rangle\ldots\langle n1\rangle}\int_{\M}\d^4x\,e^{\im (Q-q)\cdot x+\sum_je_jg(x,\kappa_j)}\,.
\end{equation}
The generalization to the form factor of an arbitrary power $\tr B^k\coloneqq\tr \tensor{B}{_{\alpha_1}^{\alpha_2}}\ldots\tensor{B}{_{\alpha_k}^{\alpha_1}}$ is straightforward
\begin{equation}
    \mathscr{F}_{\tr B^k}(1^+,\ldots,i_1^-\ldots,i_k^-,\ldots,n^+;q)=\frac{(\langle i_1i_2\rangle\ldots\langle i_ki_1\rangle)^2}{\langle 12\rangle\ldots\langle n1\rangle}\int_{\M}\d^4x\,e^{\im (Q-q)\cdot x+\sum_je_jg(x,\kappa_j)}\,.\label{eq:trb^kformfactor}
\end{equation}

For generic operators containing $\tilde F_{\dot\alpha\dot\beta}$ as well, the resulting formulae can still be considerably involved, so in the present section we first consider the form factor for $\tr\tilde{F}_{\dot\alpha\dot\beta}\tilde{F}^{\dot\alpha\dot\beta}$. Around the flat background, the tree-level colour-ordered MHV form factor is \cite{Dixon:2004za}
\begin{equation}
    \mathscr{F}_{\tr\tilde F^2}(1^+,\ldots,n^+;q)=\frac{(q^2)^2}{\langle 12\rangle\ldots\langle n1\rangle}\delta^4(Q-q)\,,\label{eq:flatformfactor}
\end{equation}
where $Q=k_1+\ldots+k_n$ is the sum of the external gluon momenta. If one interprets the form factor as the amplitude for a massive complex scalar chirally coupled to the SD field strength, this beautiful formula can be proved by Berends-Giele recursion \cite{Dixon:2004za}; alternatively, the parity-conjugate form factor $\mathscr{F}_{\tr B^2}(1^-,\ldots,n^-;q)$ can be computed as the "maximally-non-MHV" form factor using the MHV formalism \cite{Brandhuber:2011tv}. It's remarkable that such a compact formula exists at all for a maximally googly form factor. We now show that the simplicity of this form factor is a consequence of the existence of the $K$-matrix. Expressing the SD field strength in terms of the $K$-matrix and integrating by parts twice, it is straightforward to rewrite the Fourier transform of $\tr\tilde{F}^2$ as
\begin{equation}
    \int_\M\d^4x\,e^{-\im q\cdot x}\iota_\alpha q^{\alpha\dot\alpha}\iota_\beta q^{\beta\dot\beta}\tr\d_{\dot\alpha}K\,\d_{\dot\beta}K\,,
\end{equation}
and the lifting of this expression to twistor space reads
\begin{equation}
    \int_\M\d^4x\,e^{-\im q\cdot x}\iota_\alpha q^{\alpha\dot\alpha}\iota_\beta q^{\beta\dot\beta}\\\quad\int_{X^2}\frac{\D\lambda_1\D\lambda_2\,\langle\lambda_1\lambda_2\rangle^2}{\langle \iota\lambda_1\rangle\langle \iota\lambda_2\rangle}\tr\left(\left.\frac{\p\twista_1}{\p\mu_1^{\dot\alpha}}\right|_X\K_{12}\left.\frac{\p\twista_2}{\p\mu^{\dot\beta}_2}\right|_X\K_{21}\right)\,,\label{eq:generatingfunctional}
\end{equation}
where we defined $\K_{ij}\coloneqq\K_X(\lambda_i,\lambda_j)$ for the propagator. Since $\tr\tilde{F}^2$ coincides on-shell with the topological term $\tr F\wedge F$, one {na\"{i}vely} expects to write $\tr\tilde F^2$ as the divergence of the Chern-Simons current, and indeed one can check that the latter simply takes the form
\begin{equation}
    \mathcal{J}^{\alpha\dot\alpha}=\iota^\alpha\tr(\d_{\dot\beta}K\,\d^{\dot\alpha}\d^{\dot\beta}K)\,.
\end{equation}
From this relation, it's clear that expressing the gauge field via the $K$-matrix allows to extract a further total derivative from $\mathcal{J}^{\alpha\dot\alpha}$. Notice also that \eqref{eq:generatingfunctional} seemingly depends on the gauge spinor $\iota_\alpha$, but it is actually gauge invariant because it coincides with the generating functional computed with \eqref{eq:fieldstrength}; this latter expression, although gauge invariant, is more involved and leads to less compact formulae for the form factors.

We now evaluate the generating functional \eqref{eq:generatingfunctional} on a perturbed background, $\twista\to\twista +a$. The perturbation $a$ is assumed to be the sum of $n$ momentum eigenstates \eqref{eq:momentumeigenstates}, and the term in \eqref{eq:generatingfunctional} containing precisely $n$ distinct momentum eigenstates is the form factor. Expanding the twistor propagators with the aid of \eqref{eq:Kperturbed} and decomposing the form factor into colour-ordered terms, we obtain the expression
\begin{equation}
	\begin{aligned}\mathscr{F}=&\frac{1}{\langle 12\rangle\ldots\langle n1\rangle}\int_\M\d^4x\,e^{\im(Q-q)\cdot x+\sum_je_jg(x,\kappa_j)}\iota_\alpha q^{\alpha\dot\alpha}\iota_\beta q^{\beta\dot\beta}\sum_{i,j}\Biggl\{\frac{\langle ij\rangle^2}{\langle \iota i\rangle\langle \iota j\rangle}\tilde\kappa_{i\dot\alpha}\tilde\kappa_{j\dot\beta}\\&-2\int_X\frac{\D\lambda}{2\pi \im}\frac{\langle i-1,i\rangle\langle \lambda j\rangle^2\tilde\kappa_{j\dot\beta}}{\langle i-1,\lambda\rangle\langle\lambda i\rangle\langle \iota\lambda\rangle\langle \iota j\rangle}\left.\frac{\partial\twista}{\partial\mu^{\dot\alpha}}\right|_X\\&+\int_{X^2}\frac{\D\lambda\D\lambda'}{(2\pi \im)^2}\frac{\langle i-1,i\rangle\langle j-1,j\rangle\langle\lambda\lambda'\rangle^2}{\langle i-1,\lambda\rangle\langle\lambda i\rangle\langle j-1,\lambda\rangle\langle \lambda j\rangle\langle \iota\lambda\rangle\langle \iota\lambda'\rangle}\left.\frac{\partial\twista}{\partial\mu^{\dot\alpha}}\right|_X\left.\frac{\partial\twista}{\partial{\mu'}^{\dot\beta}}\right|_X\Biggr\}\,.\end{aligned}\label{eq:pre-formfactor}
\end{equation}
In principle, the background twistor connection can give derivative contributions to the form factor, namely the second and third line in \eqref{eq:pre-formfactor}. However, one can use the integral
\begin{equation}
\begin{aligned}
\int_X\frac{\D\lambda}{2\pi \im}\frac{\lambda_\alpha\lambda_\beta}{\langle \iota\lambda\rangle\langle \ell-1,\lambda\rangle\langle\lambda \ell\rangle}\left.\frac{\partial\twista}{\partial\mu^{\dot\alpha}}\right|_X=-\im&\left(\frac{\kappa_{\ell-1\,\alpha} \kappa_{\ell-1\,\beta}}{\langle \ell-1,\iota\rangle\langle \ell-1,\ell\rangle}g_{\dot\alpha}(x,\kappa_{\ell-1})\right.\\&+\frac{\kappa_{\ell\,\alpha}\kappa_{\ell\,\beta}}{\langle \ell \iota\rangle\langle \ell,\ell-1\rangle}g_{\dot\alpha}(x,\kappa_\ell)\\&\left.+\frac{\iota_\alpha \iota_\beta}{\langle \iota,\ell-1\rangle\langle \iota\ell\rangle}g_{\dot\alpha}(x,\iota)\right)\,,
\end{aligned}
\end{equation}
to realize that the terms containing derivatives of the background actually give a vanishing contribution to the form factor. The final expression for the colour-ordered form factor around a Cartan-valued self-dual radiative background is
\begin{equation}
    \mathscr{F}_{\tr\tilde F^2}(1^+,\ldots,n^+;q)=\frac{(q\cdot Q)^2}{\langle 12\rangle\ldots\langle n1\rangle}\int_\M\d^4x\,e^{\im(Q-q)\cdot x+\sum_j e_jg(x,\kappa_j)}\,,\label{eq:curvedformfactor}
\end{equation}
see Appendix \ref{app:computation} for more details on the derivation. As previously anticipated, the dependence on the gauge spinor $\iota_\alpha$ dropped out and the result is fully gauge-invariant. Around a non-trivial background, we expect translations to be broken, and therefore the "momentum-conserving" $\delta$-function in \eqref{eq:flatformfactor} is replaced by the residual integral over space-time in \eqref{eq:curvedformfactor}. This integral cannot be performed analytically for a generic background, but for specific, highly-symmetric examples, it is possible to further simplify it. For example, three of the four integrals can be evaluated around a self-dual plane wave background \cite{Adamo:2017nia, Adamo:2018mpq, Adamo:2019zmk}, and we recover three momentum-conserving $\delta$-functions on the directions along which the gauge field is constant. Furthermore, we note that for the form factor around the trivial background, $q^2=q\cdot Q$ on the support of the momentum conserving $\delta$ function,
so that the form factor around a non-trivial background can be obtained simply by replacing the $\delta$ function with the space-time integral in \eqref{eq:curvedformfactor} while leaving the prefactor containing the spinor brackets unaffected.

A similar analysis can be set up for form factors of other polynomials in $\tilde{F}_{\dot\alpha\dot\beta}$ and its derivatives. Let us now consider the cubic operator 
\begin{equation}
    \tr\tilde F^3\coloneqq\tr\tensor{\tilde F}{_{\dot\alpha}^{\dot\beta}}\tensor{\tilde{F}}{_{\dot\beta}^{\dot\gamma}}\tensor{\tilde{F}}{_{\dot\gamma}^{\dot\alpha}}\,.
\end{equation}
Note that this is the unique cubic operator not involving derivatives of the SD field strength, up to a sign. The lifting of this operator to twistor space cannot be simplified using the $K$-matrix anymore but can be straightforwardly obtained using \eqref{eq:fieldstrength} and again the generating functional is the Fourier transform of such lifting. Since the expression of this lifting is rather involved, we prefer to display it in Equation \eqref{eq:trf^3generatingfunctional} in Appendix \ref{app:computation}. The perturbative expansion proceeds as before: at arbitrary multiplicity, the tree-level, colour-ordered MHV form factor around a self-dual, Cartan-valued, radiative background is
\begin{equation}\begin{aligned}
    \mathscr{F}_{\tr\tilde F^3}=&\frac{1}{\langle 12\rangle\ldots\langle n1\rangle}\int_{\M}\d^4x\,e^{\im(Q-q)\cdot x+\sum_j e_jg(x,\kappa_j)}\times\\&\times\Biggl(\sum_{i,j,k}\langle ij\rangle\langle jk\rangle ki\rangle[ij][jk][ki]\\&+3\sum_{i,j,k,\ell}\langle jk\rangle\langle k\ell\rangle\langle \ell i\rangle[k\ell]([\ell i][jk]+[ik][\ell j])\\&+3\sum_{i,j,k,\ell,m}\langle jk\rangle\langle \ell m\rangle\langle mi\rangle\Bigl([mi]([jk][\ell m]+[j\ell][k m])+(i\leftrightarrow j)\Bigr)\\&+\sum_{i,j,k,\ell,m,n}\langle jk\rangle\langle\ell m\rangle\langle ni\rangle\Bigl([jk]([ni][\ell m]+[mi][\ell n])+(k\leftrightarrow \ell)\\&+(i\leftrightarrow j)+(i\leftrightarrow j,k\leftrightarrow \ell)\Bigr)\Biggr)\,.\end{aligned}\label{eq:trf^3formfactor}\end{equation}
In particular, the colour-ordered formulae obtained with our generating functional match the first few expressions around the trivial background, namely the minimal form factor
\begin{equation}
    \mathscr{F}_{\tr\tilde{F}^3}(1^+,2^+,3^+;q)=[12][23][31]\,\delta^{4}(Q-q)\,,
\end{equation}
and the 4-point form factor \cite{Brandhuber:2017bkg, Koster:2016loo}
\begin{equation}
    \mathscr{F}_{\tr\tilde F^3}(1^+,2^+,3^+,4^+;q)=\frac{[12][23][34][41]}{\langle 12\rangle[21]}\left(1+\frac{[31][4|q|3\rangle}{\langle 23\rangle[32][41]}\right)\delta^{4}(Q-q)+\text{cyclic}\,.
\end{equation}
\subsection{Generic MHV form factors}\label{sec:genericformfactors}
The most remarkable feature of the expressions \eqref{eq:trb^kformfactor}, \eqref{eq:curvedformfactor}, and \eqref{eq:trf^3formfactor} is that the form factor around a non-trivial background is obtained by a simple dressing of the form factor around the trivial background, namely one only has to replace the $\delta$ function with the space-time integral in the last line of \eqref{eq:trf^3formfactor}, while keeping intact the kinematical prefactor. This property is generic of any form factor of a composite operator, as the lifting of any polynomial in $\tilde F_{\dot\alpha\dot\beta}$ and its derivatives\footnote{The possible presence of $B_{\alpha\beta}$ fields does not bring in any additional issue, as we are working in the MHV sector.} will be a sum of products of integrals over $X$ of terms of the form $\lambda^{\alpha_1}_i\ldots\lambda^{\alpha_s}_i\partial_{\mu_i^{\dot\alpha_1}}\ldots\partial_{\mu_i^{\dot\alpha_s}}\twista$, intertwined by propagators $\K_X(\lambda_i,\lambda_j)$ between adjacent positions on the sphere $X$ and with some contraction of the spinor indices. Once we expand such an expression in terms of momentum eigenstates, the background connection can contribute to the form factor in two distinct ways: it can either be present only in the holomorphic frames $\H(x,\lambda_i)$, or it can potentially give a contribution when present in a $\mu^{\dot\alpha}$ derivative. The first type of contribution gives a factor of $\exp(e_jg(x,\kappa_j))$ when the background acts by conjugation on the $j$th external gluon. Conversely, the second type of contribution is proportional to (possibly a spacetime derivative of)
\begin{equation}
	\frac{1}{2\pi \im}\int_X\frac{\D\lambda}{\langle j-1,\lambda\rangle\langle \lambda j\rangle}\lambda_\alpha\left.\frac{\partial\twista}{\partial\mu^{\dot\alpha}}\right|_X\,,\label{eq:integral_amudotalpha}
\end{equation}
when we consider a colour-ordered form factor. Such a term must be inserted in any possible position inside the perturbative expansion, and for a colour-ordered form factor, this means that we must sum over $j$. Moreover, any of these contributions come together with a partial Parke-Taylor denominator $1/(\langle 12\rangle\ldots\widehat{\langle j-1,j\rangle}\ldots\langle n1\rangle)$ from which the factor $1/\langle j-1, j\rangle$ is removed. The integral \eqref{eq:integral_amudotalpha} can be evaluated for generic external momenta by considering $\kappa_{j-1,\alpha}$ and $\kappa_{j\,\alpha}$ as a basis of undotted spinors, and it's equal to
\begin{equation}
	\frac{1}{2\pi \im}\int_X\frac{\D\lambda}{\langle j-1,\lambda\rangle\langle \lambda j\rangle}\lambda_\alpha\left.\frac{\partial\twista}{\partial\mu^{\dot\alpha}}\right|_X=\im\frac{\kappa_{j-1,\alpha}g_{\dot\alpha}(x,\kappa_{j-1})-\kappa_{j\,\alpha}g_{\dot\alpha}(x,\kappa_j)}{\langle j-1,j\rangle}\,.
\end{equation}
The denominator $\langle j-1,j\rangle$ is precisely the missing factor needed to reconstruct the Parke-Taylor denominator, so that the residual sum over $j$ is telescopic
\begin{equation}
	\im\sum_{j=1}^n(\kappa_{j-1,\alpha}g_{\dot\alpha}(x,\kappa_{j-1})-\kappa_{j\,\alpha}g_{\dot\alpha}(x,\kappa_j))=0\,.
\end{equation}
Then the only non-vanishing contribution to the form factor around a non-trivial background comes from terms where the background is present only in the holomorphic frames, and this means that the desired form factor coincides with the one around the trivial background, once we replace the $\delta$ function with the integral
\begin{equation}
	\int_\M\d^4x\,e^{\im(Q-q)\cdot x+\sum_je_jg(x,\kappa_j)}\,.\label{eq:space-time_integral}
\end{equation}

\subsection{MHV super form factors in $\N=4$ SYM}
The results from the previous section can be extended to super form factors in $\N=4$ super-Yang-Mills as well, starting with the space-time expression of a composite operator, lifting it to twistor space and expanding around the desired background using the on-shell states \eqref{eq:n=4_on_shell_state}. As in the pure Yang-Mills case, any MHV super form factor around a Cartan-valued gluonic background is obtained by replacing the momentum conserving $\delta$ function with the by-now usual background-dependent integral. The simplest case is the super form factor for $\frac{1}{2}\tr\phi_{ab}\phi_{ab}$, which can be computed starting from the supersymmetrized $K$-matrix, that is using the lifting formula
\begin{equation}
    \frac{1}{2}\tr\phi_{ab}\phi_{ab}=\frac{1}{2}\int\d^8\theta\,\iota_\alpha \iota_\beta\prod_{\substack{\gamma\neq \alpha,\beta\\c\neq a,b}}\theta^{\gamma c}\int_{X^2}\frac{\D\lambda_1\,\D\lambda_2\,\langle\lambda_1\lambda_2\rangle^2}{\langle \iota\lambda_1\rangle \langle \iota\lambda_2\rangle}\tr \left(\left.\frac{\p\A_1}{\p\chi_1^a}\right|_X\underline\K_{12}\left.\frac{\p\A_2}{\chi^b_2}\right|_X\underline\K_{21}\right)\,,\label{eq:super_form_factor_generating_functional}
\end{equation}
the fermionic integrals being necessary to extract the lowest component of $\frac{1}{2}\tr\underline\phi_{ab}\underline\phi_{ab}$. The corresponding MHV super form factor around a non-trivial background is given by
\begin{equation}
    \mathscr{SF}_{\frac{1}{2}\tr\phi^2}(1,\ldots,n;q)=\frac{1}{4}\frac{(\mathcal{Q}^2)^2}{\langle 12\rangle\ldots\langle n1\rangle}\int_{\M}\d^4x\,e^{\im (Q-q)\cdot x+\sum_je_jg(x,\kappa_j)}\,,\label{eq:phi_super_form_factor}
\end{equation}
where $\mathcal{Q}_{\alpha a}=\kappa_{1\,\alpha}\eta_{1\,a}+\ldots+\kappa_{n\,\alpha} \eta_{n\,a}$, as we show in Appendix \ref{app:computation}. In the flat-background limit, we recover the standard expression for the super form factor \cite{Brandhuber:2011tv}. For generic operators around purely gluonic backgrounds, the same argument of Section \ref{sec:genericformfactors} holds and the corresponding form factor can obtained by dressing the form factor around the trivial background with \eqref{eq:space-time_integral}. Conversely, for backgrounds including fermions and scalars, the grading in the fermionic coordinates prevents having results as simple as in the purely gluonic case,\footnote{Presumably, one could obtain interesting results by considering super form factors of supersymmetrized operators, i.e. Fourier transforms over the entire chiral superspace of matrix elements of a local, composite operator of superfields between the vacuum and an on-shell superstate.} but we can still obtain reasonably compact formulae. For example, one can consider fermionic space-time backgrounds: introducing the function
\begin{equation}
    g_a^{(f)}(x,\lambda)=\frac{1}{2\pi\im}\int_{X}\frac{\D\lambda'}{\langle \lambda\lambda'\rangle}\left.\twista_a\right|_X\,,\label{eq:g_function_fermionic}
\end{equation}
the super form factor in the presence of a mixed gluonic and fermionic background is
\begin{equation}
    \mathscr{SF}_{\frac{1}{2}\tr\phi^2}(1,\ldots,n;q)=\frac{1}{4\langle 12\rangle\ldots\langle n1\rangle}\int_{\M}\d^4x\, (\mathcal{Q}^\alpha_a\tilde{\mathcal{Q}}_{\alpha a})^2e^{\im (Q-q)\cdot x+\sum_je_jg(x,\kappa_j)}\label{eq:super_form_factor_around_fermions}\,, 
\end{equation}
where $\tilde{\mathcal{Q}}_{\alpha a}=\kappa_{1\,\alpha}\tilde\eta_{1\,a}+\ldots+\kappa_{n\,\alpha}\tilde\eta_{n\,a}$ is the dressed super-momentum defined in terms of the fermionic background by
\begin{equation}
\kappa_{i\,\alpha}\tilde\eta_{i\,a}(x)=\kappa_{i\,\alpha}(\eta_{i\,a}+e_ig^{(f)}_a(x,\kappa_i))\,,
\end{equation}
in complete analogy with \eqref{eq:dressed_momentum}.

\section{Discussion}\label{discussion}
We have extended the  twistor framework for constructing  form factors by exploiting the representation of the Ward correspondence at null infinity \cite{Sparling:1990, Newman:1978ze} and using asymptotic $\scri$-data for background fields to give formulae on such non-trivial self-dual radiative backgrounds. We have further extended the $\scri$ framework to incorporate supersymmetry. This presentation links directly into the celestial and twisted holography programmes. In particular, we gave a novel proof of the MHV form factor for $\tr\Tilde{F}^2$  extending it to the formula \eqref{eq:curvedformfactor} for the same form factor, but now evaluated on a general Cartan-valued, self-dual, radiative background. The framework is set up so that, in principle, we can  compute the tree-level MHV form factor for an arbitrary composite operator of the field strength and the MHV super form factor for an arbitrary composite operator in $\N=4$ SYM, but for general operators we don't expect to have such simple expressions, as exemplified by the form factor of $\tr\tilde{F}^3$. It's nevertheless remarkable that around any self-dual background, tree-level MHV form factors can be obtained by a simple dressing via a single, residual space-time integral encoding the details of the background, without any modification of the rational prefactor depending on the spinor brackets. This is to be contrasted with the expected complexity of the result, which should na\"ively consist of $n-2$ space-time integrals for an observables at $n$ points.

\subsection{N$^k$MHV observables}\label{n-kmhv}
The main restriction in this work was the restriction to tree-level MHV form factors. To go to higher MHV degrees or loop form factors on non-trivial backgrounds, we must incorporate twistor space propagators on such non-trivial backgrounds -- recall that single insertion raises the MHV degree by one at tree-level and each integration reduces the MHV degree by two. The main issue is the absence of a known, compact expression for the propagator around a radiative background: at the abstract level,  twistor space propagators have been studied for many years, see for example \cite{Atiyah:1981ey}.  More recently, in the context of twistor actions expressions for the twistor space propagators  have been found in \cite{Mason:2010yk, Adamo:2011cb, Adamo:2011pv} in a gauge that induces the axial gauge used in the MHV formalism \cite{Cachazo:2004kj, Boels:2007qn} when represented in momentum space.  This propagator was used for form factors and correlators in \cite{Chicherin:2014uca, Koster:2016ebi,Koster:2016fna,Koster:2016loo,Chicherin:2016fac,Chicherin:2016soh,Chicherin:2016fbj,Chicherin:2016ybl}.  A twistor space propagator $\Delta_\twista(Z,Z')$ on a background defined by a Lie algebra-valued $(0,1)$-form $\twista$  must satisfy  the defining relation
\begin{equation}
    (\dbar_0 +\twista)\Delta _\twista(Z,Z')=\bar\delta^3(Z,Z')
\end{equation}
This can be solved in Euclidean signature by starting from the MHV propagator $\Delta_0(Z, Z')=\bar\delta^{2|4}(Z,Z',Z_*)$ around the trivial background as in \cite{Mason:2010yk, Adamo:2011cb, Adamo:2011pv}, where $Z_*$ is the chosen reference twistor,  and conjugating with the holomorphic frame $\H(Z)$ for the line joining $(Z,Z')$.  This will give the expression 
\begin{equation}
 \Delta _\twista(Z,Z')=\H(Z,Z_*)\bar\delta^{2|4}(Z,Z',Z_*)\H(Z',Z_*)^{-1}
\end{equation}
where $\H(z,Z_*)$ satisfies $(\bar\p_0+\twista)\H(Z,Z_*)=0$ on the line joining $Z$ to $Z_*$.  This then satisfies the defining relations above as a consequence of \eqref{D-bar-H} on the support of the delta functions.   The fact that $\Delta_\twista(Z, Z')$ has four delta functions in its definition should then lead to easy integrations at higher MHV degrees, and even to find formulae for loop integrands \cite{Lipstein:2012vs}, although carrying out the loop integrals is possible but subtle, see \cite{Lipstein:2013xra}. With  these,  one should be able to compute both amplitudes and form factors at higher MHV degrees on a background  in pretty much the same way as in \cite{Adamo:2011cb}.

\subsection{The one-loop all plus amplitude}\label{one-loop}
This framework can be used to give new insights into loop amplitudes.  The simplest case is the all-plus one-loop amplitude in pure Yang-Mills and we can justify old generating function formulae, extending them to a nontrivial background, and to understand the more recent dual-conformal invariant formulae of \cite{Henn:2019mvc,Chicherin:2022bov}.  

The vanishing of the tree-level amplitude ensures that the one-loop amplitude around the trivial background is finite and a rational function of the spinor brackets. There are a number of versions of the all-multiplicity formula for this \cite{Bern:1993qk, Mahlon:1993si, Bern:1994ju}, for example, ignoring constant prefactors
\begin{equation}
    \mathcal{A}^{\text{1-loop}}=\sum_{i<j<s<t}\frac{\langle ij\rangle[js]\langle st\rangle[ti]}{\langle 12\rangle\ldots\langle n1\rangle}\,.\label{eq:one-loop-amplitude}
\end{equation}
The following generating functional, originally due to one of us, was briefly quoted without explanation
in \cite{Boels:2007gv} as
\begin{equation}
    \int_\M\d^4x\int_{X^2}\D\lambda_1 \D\lambda_2 \tr( \p_\alpha^{ \dot\alpha} \K_{21} \p^\alpha_{\dot\beta}\left.\twista_1\right|_X \p^{\dot\beta}_\beta\K_{12} \p^\beta_{\dot\alpha} \left.\twista_2\right|_X ) \,.\label{eq:oneloopgeneratingfunctional_2point}
\end{equation}
We can express the derivative of the propagator as the variation under a translation  using \eqref{eq:pertubedKpropagator} 
\begin{equation}
    \p_{\alpha\dot\alpha}\K_X(\lambda_i,\lambda_j)=-\int_X\D\lambda\K_X(\lambda_i,\lambda)\p_{\alpha\dot\alpha}\left.\twista\right|_X\K_X(\lambda,\lambda_j)\,. \label{U-derivative}
\end{equation}
With this, the generating functional becomes \begin{equation} \int_\M\d^4x\int_{X^4}\prod_{i=1}^4\D\lambda_i\,\langle\lambda_1\lambda_2\rangle\langle\lambda_3\lambda_4\rangle\tr(\partial_{{\dot\alpha}}\twista_1\,\K_{12}\,\partial^{{\dot\beta}}\twista_2\,\K_{23}\,\partial_{{\dot\beta}}\twista_3\,\K_{34}\,\partial^{{\dot\alpha}}\twista_4\,\K_{41})\,.\label{eq:oneloopgeneratingfunctional}
\end{equation}
This is more clearly a background-coupled 4-vertex generating  amplitudes with at least four external legs. As recently observed in \cite{Bu:2022dis}, it clearly gives \eqref{eq:one-loop-amplitude} at any multiplicity when expanded around the trivial background, since the angle brackets (respectively, $\mu^{\dot\alpha}$ derivatives) give the correct angle brackets (square brackets) in the numerator, whilst the expansion of the propagators as in \eqref{eq:Kperturbed} reproduces the Parke-Taylor denominator. The same expansion around a Cartan-valued background yields
\begin{equation}
     \mathcal{A}^{\text{1-loop}}_{\text{background}}=\sum_{i<j<s<t}\frac{\langle ij\rangle[js]\langle st\rangle[ti]}{\langle 12\rangle\ldots\langle n1\rangle}\int_\M\d^4x\,\exp\left(\sum_{j=1}^n(\im k_j\cdot x+e_jg(x,\kappa_j))\right)\,,\label{eq:1-loop-amplitude-background}
\end{equation}
in agreement with our results on form factors, as well as with the previously known results on gluon amplitudes \cite{Adamo:2020yzi}.

In a different direction, the two-point generating function \eqref{eq:oneloopgeneratingfunctional_2point}
neatly ties into region momenta formulae, and, in particular, the observation of \cite{Henn:2019mvc} that the 1-loop all plus rational amplitude can be derived from the 1-loop maximally supersymmetric MHV 1-loop integrand  in region momentum space where the region momentum for the loop is placed at infinity.  We can see this in a slightly different way here.  We first  introduce the  region momenta $y_i$'s by 
\begin{equation}
y_{ij}=y_i-y_j=\sum_{i\leq s<j} k_s\, .     
\end{equation}
 We can then expand $\p_{\alpha\dot\alpha} \K_{ij}$ in momentum eigenstates, using 
  \eqref{U-derivative} to obtain, for a given cyclic ordering, region momenta appearing at the  $(j-i)$-th order in the form 
\begin{equation}
    \p^{\alpha\dot\alpha}  \K_{ij}=\frac{y_{i\,j+1}^{\alpha\dot\alpha}}{\langle i i+1\rangle \ldots \langle j-1 j\rangle  }\, ,\qquad    \p^{\alpha\dot\alpha} \p^{\beta \dot\beta} \K_{ij}=\frac{y_{i\, j+1}^{\alpha\dot\alpha}y_{i\,j+1}^{\beta\dot\beta}}{\langle i i+1\rangle \ldots \langle j-1 j\rangle  }\, .
\end{equation}
Using the first,  we see that our generating function \eqref{eq:oneloopgeneratingfunctional_2point} expands to give
\begin{equation}
    \sum_{i<j} \frac{\langle i|y_{i\, j+
1}|j]\langle j|y_{j+1\, i}|i] }{\langle 12\rangle \langle 2 3\rangle \ldots \langle n 1\rangle}\, .
\end{equation}
and, using, for example, \S6 of \cite{Bullimore:2010pj}, this can be recognised to be the sum over $2$-mass easy and $1$-mass boxes \cite{Bern:1994zx} written in terms of region momenta, with the loop insertion point $x_0$ --represented by the line $(A, B)$ in \cite{Bullimore:2010pj}-- placed at infinity.
Of course, this formula could be seen directly by summing the $j$ and $t$ in \eqref{eq:one-loop-amplitude} to give region momenta.

An alternative version of the one-loop amplitude was observed in  \cite{Henn:2019mvc}
\begin{equation}
    \mathcal{A}^{\text{1-loop}}=\sum_{i<j}\frac{\langle 1| y_{1i} y_{ij}|1\rangle^2  }{\langle 12\rangle\ldots\langle n1\rangle 
 }\frac{\langle i-1 , i \rangle\langle j-1, j\rangle}{\langle1, i-1\rangle\langle1 i\rangle\langle1, j-1\rangle\langle1 j\rangle}\,.\label{eq:one-loop-amplitude_henn}
\end{equation}
As explained in \cite{Chicherin:2022bov}, this is the so-called \emph{Kermit} formula for the supersymmetric MHV loop integrand, with the loop integrand region momentum placed at infinity.  This is the form obtained from the all-loop BCFW recursion of \cite{ArkaniHamed:2010kv} and also the MHV-diagram version of \cite{Bullimore:2010pj}; in the former, leg 1 is the direction of the BCFW shift, whereas, in the latter, it can be taken to be an arbitrary reference twistor associated to the reference spinor of the MHV formalism.   In \cite{Bullimore:2010pj}, a detailed argument was given, based on the calculations in \cite{Brandhuber:2004yw} to relate this to the more familiar 2-mass-easy boxes of \cite{Bern:1994ju}. At the level of the loop integrand (here again with the loop insertion point placed at infinity) this follows from the algebraic relation
\begin{equation}
\begin{aligned}
\mathcal{A}^{\text{1-loop}}&=\sum_{i<j}\frac{1}{\langle 12\rangle\ldots\langle n1\rangle}\left(\frac{\langle 1|y_{1j}y_{ji}|i\rangle\langle 1|y_{1i}y_{ij}|j\rangle}{\langle 1i\rangle\langle 1j\rangle}-\frac{\langle 1|y_{1j}y_{j,i-1}|i-1\rangle\langle 1|y_{1i}y_{ij}|j\rangle}{\langle 1,i-1\rangle\langle 1j\rangle}\right.\\&\quad\left.-\frac{\langle 1|y_{1j}y_{ji}|i\rangle\langle 1|y_{1i}y_{i,j-1}|j-1\rangle}{\langle 1i\rangle\langle 1,j-1\rangle}+\frac{\langle 1|y_{1j}y_{j,i-1}|i-1\rangle\langle 1|y_{1i}y_{i,j-1}|j-1\rangle}{\langle 1,i-1\rangle\langle 1,j-1\rangle}\right)
\end{aligned}
\end{equation}
so it should be possible to write a two-point generating functional for this form of the amplitude, along the lines of \eqref{eq:oneloopgeneratingfunctional_2point}. However, this generating functional is unlikely to be local on space-time, as the summands for this version of the one-loop amplitude contain spurious poles (the poles cancel in the sum though, as they should \cite{Henn:2019mvc}).

An open question is how to construct these generating functionals as local expressions in the self-dual Yang-Mills potentials. The central puzzle is that \eqref{eq:oneloopgeneratingfunctional} contains four derivatives of the twistor connection $\twista$, so it corresponds to a local space-time expression containing two derivatives on the space-time gauge field. Gauge invariance and self-duality restrict the possible terms to the integral of $\tr\tilde F^2$, but this integral vanishes, as one can infer from the $q\to0$ limit of the form factor we presented in this work \footnote{Alternatively, the radiative backgrounds we are considering are topologically trivial, so the winding number vanishes.}. A more conservative approach towards a first-principle derivation of \eqref{eq:oneloopgeneratingfunctional} could be a computation of a one-loop determinant around the background gauge field, but at the moment only the \emph{variation} of the determinant seems to be easy to compute; it's however worth mentioning that a one-loop determinant is generically non-local, but its variation becomes local precisely around self-dual backgrounds \cite{Brown:1978yj,Corrigan:1979di}. It would also be nice to reproduce the chiral algebra computations of \cite{Costello:2022wso} directly from a space-time perspective, without referring to the 2d conformal blocks; indeed, the Green-Schwarz mechanism for the anomaly cancellation on twistor space \cite{Bardeen:1995gk,Costello:2021bah} suggests that the all-plus one-loop amplitude could be computed using tree diagrams only, as showed in \cite{Monteiro:2022nqt} for the parity-conserving terms of the amplitude. We hope to return to these issues in future work.

\paragraph{Acknowledgements:} We thank Atul Sharma and Arthur Lipstein for useful discussions and Atul Sharma for providing feedback on the draft. GB is supported by a joint Clarendon Fund and Merton College Mathematics Scholarship.
LJM is grateful to the IHES and ENS Paris for hospitality  while this work was progressing and to the STFC for support under grant ST/T000864/1.

\appendix
\section{$\N=4$ constraint equations on chiral superspace}\label{app:N=4_constraints}
It is well known that there exist an equivalence between the field equations for the super-fields and constraint equations for the super-connections. The precise statement depend on the superspace one is considering: the $d=10$, $\N=1$ SYM equations of motion are equivalent to the constraint equation \cite{Witten:1978xx, Harnad:1985bc}
\begin{equation}
    \{\unabla_\cA,\unabla_\cB\}=2\gamma^M_{\cA\cB}\unabla_\mu\,,\label{10d-constfraints}
\end{equation}
where $(x^M,\theta^\cA)$ are coordinates on $\C^{10|16}$. When compactified to 4 dimensions, this equivalence becomes an equivalence between the constraint equations and the $\N=4$ equations of motion on non-chiral superspace $\C^{4|16}$ with coordinates $(x^{\alpha\dot\alpha},\theta^{a\alpha},\bar\theta^{\dot\alpha}_a)$. If one further wants to reduce to chiral superspace, additional conditions are required \cite{Devchand:1996gv, Adamo:2013cra}: in this Appendix, we show the following result
\begin{thm}
There is an equivalence between
\begin{enumerate}
    \item the constraint equations on chiral superspace
\begin{subequations}
    \begin{eqnarray}
        [\underline\nabla_{a(\alpha},\underline\nabla_{\beta)B}\}&=&0\,,\label{eq:chiral_constraint}\\
        \underline F_{\alpha\beta}&=&\lambda \underline B_{\alpha\beta}\label{eq:F_B_lagrange_multiplier}\,,
    \end{eqnarray}
\end{subequations}
where $\underline F_{\alpha\beta}$ is the ASD component of the supercurvature $[\underline\nabla_{\alpha\dot\alpha},\underline\nabla_{\beta\dot\beta}]$, $\underline B_{\alpha\beta}$ will be determined below by the constraint equations \eqref{eq:chiral_constraint}, and $\lambda$ is the 't Hooft coupling,
\item super-fields $(\uA_{\alpha\dot\alpha},\utpsi_{a\dot\alpha},\uphi_{ab},\upsi^\alpha_a,\uB_{\alpha\beta})$ on chiral superspace satysfing
\begin{subequations}
    \begin{eqnarray}
        \unabla_{a\alpha}\utF_{\dot\alpha\dot\beta}&=&\unabla_{\alpha(\dot\alpha}\utpsi_{\dot\beta)a}\,,\label{eq:fer_der_tF}\\
        \unabla_{a\alpha}\utpsi_{\dot\alpha b}&=&2\unabla_{\alpha\dot\alpha}\uphi_{ab}\,,\label{eq:fer_der_tpsi}\\
        \unabla_{a\alpha}\uphi_{bc}&=&\epsilon_{abcd}\upsi^d_\alpha\,,\label{eq:fer_der_phi}\\
        \unabla_{a\alpha}\upsi^b_\beta&=&-\varepsilon_{\alpha\beta}[\uphi_{ac},\uphi^{bc}]+
    \delta^b_a\uB_{\alpha\beta}\,,\label{eq:fer_der_psi}\\
        \unabla_{a\alpha}\uB_{\alpha\beta}&=&-2\varepsilon_{\alpha(\beta}[\upsi^b_{\gamma)},\uphi_{ab}]\,,\label{eq:fer_der_B}\\
        \uF_{\alpha\beta}&=&\lambda\uB_{\alpha\beta}\label{eq:F_B_lagrangemultiplier2}\,,
    \end{eqnarray}
\end{subequations}
which in turn imply the $\N=4$ super-field equations of motion
\begin{subequations}
    \begin{eqnarray}
        \unabla_{\dot\alpha}^\beta\uB_{\alpha\beta}&=&-\{\utpsi_{a\dot\alpha},\upsi^a_\alpha\}-\frac{1}{2}[\uphi_{ab},\unabla_{\alpha\dot\alpha}\uphi^{ab}]\,,\label{eq:super_B_eom}\\
        \unabla_\alpha^{\dot\alpha}\utpsi_{a\dot\alpha}&=&2\lambda[\upsi^b_\alpha,\uphi_{ab}]\,,\label{eq:super_tpsi_eom}\\
        \usquare\uphi_{ab}&=&\{\utpsi_{\dot\alpha[b},\utpsi^{\dot\alpha}_{a]}\}+\lambda\epsilon_{abcd}\{\upsi^c_\alpha,\upsi^{d\alpha}\}+2\lambda[\uphi_{c[a},[\uphi^{cd},\uphi_{b]d}]]\,,\label{eq:super_phi_eom}\\
        \unabla_{\dot\alpha}^\alpha\upsi^a_\alpha&=&[\utpsi_{b\dot\alpha},\uphi^{ab}]\,,\label{eq:super_psi_eom}\\\uF_{\alpha\beta}&=&\lambda \uB_{\alpha\beta}\,,\label{eq:super_F_eom}
    \end{eqnarray}
\end{subequations}
\item component fields $(A_{\alpha\dot\alpha},\tilde\psi_{a\dot\alpha},\phi_{ab},\psi^\alpha_a,B_{\alpha\beta})$ on space-time satisfying the $\N=4$ equations of motion -- that is the $\theta^{\alpha a}=0$ truncation of the super-field equations of motion--, supplemented by the recursion relations \eqref{eq:rec_rel_tF}, \eqref{eq:rec_rel_tpsi}, \eqref{eq:rec_rel_phi}, \eqref{eq:rec_rel_psi}, \eqref{eq:rec_rel_B}, \eqref{eq:rec_rel_F}, \eqref{eq:rec_rel_Ab}, and \eqref{eq:rec_rel_Af}.
\end{enumerate}
\end{thm}
Our result is therefore a deformation away from self-duality of the results in \cite{Devchand:1996gv} for $\N=4$ and reduces to that construction in the $\lambda=0$ limit, and is more self-contained compared to the result of \cite{Adamo:2013cra}.

For the 1. $\Rightarrow$ 2. implication, we begin with the construction of the various super-fields appearing in the equations of motion, slightly varying the normalizations from the main body of the paper. The super-connection is taken to be $\unabla_{\alpha A}=\p_{\alpha A}+\uA_{\alpha A}$. We define the supercurvature components as implied by the constraint equations
\begin{subequations}
    \begin{eqnarray}
        [\underline\nabla_{\alpha\dot\alpha},\underline\nabla_{\beta\dot\beta}]&\coloneqq&\varepsilon_{\alpha\beta}\underline{\tilde F}_{\dot\alpha\dot\beta}+\varepsilon_{\dot\alpha\dot\beta}\underline{F}_{\alpha\beta}\,,\label{eq:bb_constraint}\\
        {[\underline\nabla_{\alpha a},\underline\nabla_{\beta\dot\alpha}]}&\coloneqq&\varepsilon_{\alpha\beta}\underline{\tilde\psi}_{\dot\alpha a}\,,\label{eq:bf_constraint}\\\{\underline\nabla_{\alpha a},\underline\nabla_{\beta b}\}&\coloneqq&2\varepsilon_{\alpha\beta}\underline\phi_{ab}\,,\label{eq:ff_constraint}
    \end{eqnarray}
\end{subequations}
as well as
\begin{equation}
    \underline\phi^{ab}\coloneqq\frac{1}{2}\epsilon^{abcd}\underline\phi_{cd}\,.
\end{equation}
The remaining super-fields are defined by
\begin{subequations}
    \begin{eqnarray}\underline\psi^a_\alpha&\coloneqq&-\frac{1}{3!}\epsilon^{abcd}\underline\nabla_{\alpha b}\underline\phi_{cd}\,,\\ 
    \underline B_{\alpha\beta}&\coloneqq& \frac{1}{4}\unabla_{a(\alpha}\psi^a_{\beta)}\,.
    \end{eqnarray}
\end{subequations}
The definitions for $\upsi^a_\alpha$ $\uB_{\alpha\beta}$ can be interpreted as arising from Jacobi identities for the super-connection. For example, the identity for $\underline\nabla_{\alpha a}$, $\underline\nabla_{\beta b}$, and $\underline\nabla_{\gamma c}$ implies 
\begin{equation}
    \underline\nabla_{\alpha a}\underline\phi_{bc}=\underline\nabla_{\alpha[a}\phi_{bc]}\,.
\end{equation}
Similarly, the constraint equation \eqref{eq:ff_constraint} implies that the super-field
\begin{equation}
    \underline\nabla_{a(\alpha}\underline\nabla_{\beta)b}\phi_{cd}\,,
\end{equation}
is totally skew in the $\SU(4)$ indices. We complete our definitions by introducing the 't Hooft coupling $\lambda$ and by requiring
\begin{equation}
    \underline F_{\alpha\beta}=\lambda \underline B_{\alpha\beta}\,,
\end{equation}
for the ASD part of the bosonic supercurvature.

With these definitions, we can derive the $\N=4$ equations of motion as follows: we first consider the equations of motion for $\upsi^a_\alpha$ and $\uB_{\alpha\beta}$ and notice that the Jacobi identity for $\unabla_{\alpha \dot\alpha}$, $\unabla_{\beta b}$, and $\unabla_{\gamma c}$ can be written as
\begin{equation}
2\underline\nabla_{\alpha\dot\alpha}\uphi_{ab}=\unabla_{\alpha a}\utpsi_{\dot\alpha b}\,,\label{eq:phi_tpsi_tpsi_jacobi}
\end{equation}
while the skew part of $\nabla_{\alpha a}\utpsi_{b\dot\alpha}$ in $a$, $b$ vanishes. Using \eqref{eq:bf_constraint}, \eqref{eq:ff_constraint} and  \eqref{eq:phi_tpsi_tpsi_jacobi}, we reproduce \eqref{eq:super_psi_eom}
\begin{equation}
    \unabla_{\dot\alpha}^\alpha\upsi_\alpha^a=[\utpsi_{\dot\alpha b},\phi^{ab}]\,.
\end{equation}
In the same way, we can use \eqref{eq:ff_constraint} to obtain for the fermionic derivative acting of $\upsi^a_\alpha$
\begin{equation}
    \unabla_{a\alpha}\upsi^b_\beta=-\varepsilon_{\alpha\beta}[\uphi_{ac},\uphi^{bc}]+\delta^b_a\uB_{\alpha\beta}\,.\label{eq:fermionic_derivative_psi}
\end{equation}
From \eqref{eq:bf_constraint}, \eqref{eq:phi_tpsi_tpsi_jacobi}, and \eqref{eq:fermionic_derivative_psi} one can straightforwardly show that
\begin{equation}
    \unabla^\beta_{\dot\alpha}\uB_{\alpha\beta}=-\{\utpsi_{a\dot\alpha},\upsi^a_\alpha\}-\frac{1}{2}[\uphi_{ab},\unabla_{\alpha\dot\alpha}\uphi^{ab}]\,,
\end{equation}
thus reproducing \eqref{eq:super_B_eom}. The fermionic derivative of $\uB_{\alpha\beta}$ be obtained via \eqref{eq:ff_constraint} and reads
\begin{equation}
    \unabla_{a\alpha}\uB_{\beta\gamma}=-2\varepsilon_{\alpha(\beta}[\upsi^b_{\gamma)},\uphi_{ab}]\,,\label{eq:fermionic_derivative_B}
\end{equation}
We can finally derive the equations of motion for the positive-helicity spinors and the scalars. Starting with the spinors, the Jacobi identity for $\underline\nabla_{\alpha a}$, $\underline\nabla_{\beta\dot\beta}$ and $\underline\nabla_{\gamma\dot\gamma}$ can be written as
\begin{subequations}
\begin{eqnarray}
\unabla_{a\alpha}\uF_{\alpha\beta}&=&-\epsilon_{\alpha(\beta}\unabla^{\dot\alpha}_{\gamma)}\utpsi_{\dot\alpha a}\,,\label{eq:F_tpsi_jacobi}\\
\unabla_{a\alpha}\utF_{\dot\alpha\dot\beta}&=&\unabla_{\alpha(\dot\alpha}\utpsi_{\dot\beta)a}\,,
\end{eqnarray}
\end{subequations}
so that \eqref{eq:F_B_lagrange_multiplier} and \eqref{eq:fermionic_derivative_B} directly give \eqref{eq:super_tpsi_eom}
\begin{equation}
    \unabla_\alpha^{\dot\alpha}\utpsi_{a\dot\alpha}=2\lambda[\upsi^b_\alpha,\uphi_{ab}]\,.
\end{equation}
Similarly, acting with $\unabla^{\alpha\dot\alpha}$ on \eqref{eq:phi_tpsi_tpsi_jacobi} and using \eqref{eq:bf_constraint} we obtain
\begin{equation}
    \usquare\phi_{ab}=\{\utpsi_{\dot\alpha[b},\utpsi^{\dot\alpha}_{a]}\}+\frac{1}{2}\unabla_{\alpha a}\unabla^{\alpha\dot\alpha}\tilde\psi_{\dot\alpha b}\,,\label{eq:super_phi_eom_partial}
\end{equation}
so that \eqref{eq:fermionic_derivative_psi} gives
\begin{equation}
    \usquare\uphi_{ab}=\{\utpsi_{\dot\alpha[b},\utpsi^{\dot\alpha}_{a]}\}+\lambda\epsilon_{abcd}\{\upsi^c_\alpha,\upsi^{d\alpha}\}+2\lambda[\uphi_{c[a},[\uphi^{cd},\uphi_{b]d}]]\,.
\end{equation}

To prove the remaining implications, we adopt essentially the strategy developed in \cite{Harnad:1985bc,Devchand:1996gv}, to which we refer for more details. We partially fix the gauge on chiral superspace by going to the radial gauge
\begin{equation}
    \rD \uA_{\alpha a}=0\,,
\end{equation}
where $\rD$ is the Euler vector field along the fermionic directions
\begin{equation}
    \mathcal{D}\coloneqq\theta^{a\alpha}\p_{a\alpha}\,.
\end{equation}
The residual gauge invariance corresponds to gauge transformations on chiral superspace that are independent of the fermionic variables, i.e. to ordinary gauge transformations on space-time. Moreover, in this gauge the equations \eqref{eq:fer_der_tF}, \eqref{eq:fer_der_tpsi}, \eqref{eq:fer_der_phi}, \eqref{eq:fer_der_psi}, and \eqref{eq:fer_der_B} readily imply the recursion relations
\begin{subequations}
    \begin{eqnarray}
        \rD\utF_{\dot\alpha\dot\beta}&=&\theta^{a\alpha}\unabla_{\alpha(\dot\alpha}\utpsi_{\dot\beta)a}\,,\label{eq:rec_rel_tF}\\
        \rD\utpsi_{a\dot\alpha}&=&-2\theta^{b\alpha}\unabla_{\alpha\dot\alpha}\uphi_{ab}\,,\label{eq:rec_rel_tpsi}\\
        \rD\uphi_{ab}&=&\epsilon_{abcd}\theta^c_\alpha\upsi^{d\alpha}\,,\label{eq:rec_rel_phi}\\
        \rD\upsi_\alpha^a&=&\theta^{a\beta}\uB_{\alpha\beta}-\theta^{b}_{\alpha}[\uphi_{bc},\uphi^{ac}]\,,\label{eq:rec_rel_psi}\\
        \rD\uB_{\alpha\beta}&=&-2\theta^a_{(\alpha}[\upsi^b_{\beta)},\uphi_{ab}]\,,\label{eq:rec_rel_B}\\
        \rD\uF_{\alpha\beta}&=&\lambda\rD\uB_{\alpha\beta}\,,\label{eq:rec_rel_F}\\
        \rD\uA_{\alpha\dot\alpha}&=&\theta^a_\alpha\utpsi_{a\dot\alpha}\,,\label{eq:rec_rel_Ab}\\
        (1+\rD)\uA_{a\alpha}&=&2\theta^b_\alpha\uphi_{ba}\,.\label{eq:rec_rel_Af}
    \end{eqnarray}
\end{subequations}
These relations defines uniquely the super-fields in terms of their lowest component fields, since the RHSs are all linear in the fermionic variables and since a homogeneous polynomial in the fermionic variables is an eigenstate of $\rD$, the eigenvalue being the degree of the polynomial. In particular, in this gauge the lowest terms in the $\theta$ expansion are 
\begin{subequations}
    \begin{eqnarray}
        \utF_{\dot\alpha\dot\beta}&=&\tilde{F}_{\dot\alpha\dot\beta}+\theta^{a\alpha}\nabla_{\alpha(\dot\alpha}\tilde\psi_{\dot\beta)a}\nonumber\\&&+\frac{1}{2}\theta^{a\alpha}\theta^{b\beta}(\nabla_{\alpha(\dot\alpha}\nabla_{\dot\beta)\beta}\phi_{ab})+\varepsilon_{\beta\alpha}\{\tilde\psi_{b(\dot\alpha},\tilde\psi_{\dot\beta)a}\})+\mathcal{O}(\theta^3)\,,\\
        \utpsi_{a\dot\alpha}&=&\tilde\psi_{a\dot\alpha}-2\theta^{b\alpha}\nabla_{\alpha\dot\alpha}\phi_{ab}-\theta^{b\beta}\theta^{c\gamma}(\epsilon_{abcd}\nabla_{\beta\dot\alpha}\psi^{d\beta}+\varepsilon_{\gamma\beta}[\tilde\psi_{c\dot\alpha},\phi_{ab}])+\mathcal{O}(\theta^3)\,,\\
        \uphi_{ab}&=&\phi_{ab}+\epsilon_{abcd}\theta^c_\alpha\psi^{d\alpha}+\frac{1}{2}\epsilon_{abcd}\theta^c_\alpha\theta^{e\beta}(\delta^d_eB^\alpha_\beta-\delta^\alpha_\beta[\phi_{ef},\phi^{df}])+\mathcal{O}(\theta^3)\,,\\
        \upsi^a_\alpha&=&\psi^a_\alpha+\theta^{b\beta}(\delta_b^a B_{\alpha\beta}-\varepsilon_{\beta\alpha}[\phi_{bc},\phi^{ac}])-\frac{1}{2}\theta^{a\beta}\theta^b_{(\alpha}[\psi^c_{\beta)},\phi_{bc}]\nonumber\\&&-\frac{1}{2}\theta^b_\alpha\theta^f_\beta(2[\phi_{bc},\delta^{[a}_f\psi^{c]\beta}]-\epsilon_{bcfe}[\phi^{ac},\psi^{e\beta}])+\mathcal{O}(\theta^3)\,,\\
        \uB_{\alpha\beta}&=&B_{\alpha\beta}-2\theta^a_{(\alpha}[\psi^b_{\beta)},\phi_{ab}]\nonumber\\&&-\theta^a_{(\alpha}\theta^{c\gamma}(\delta^b_c[B_{\gamma\beta},\phi_{ab}]-\varepsilon_{\gamma\beta}[[\phi_{cd},\phi^{bc}],\phi_{ab}]-\{\psi^b_{\beta)},\psi^d_\gamma\})+\mathcal{O}(\theta^3)\,,\\
        \uA_{\alpha\dot\alpha}&=&A_{\alpha\dot\alpha}+\theta^a_\alpha\tilde\psi_{a\dot\alpha}-\theta^a_\alpha\theta^{b\beta}\nabla_{\beta\dot\alpha}\phi_{ab}+\mathcal{O}(\theta^3)\,,\\
        \uA_{a\alpha}&=&\theta^b_\alpha\phi_{ba}+\frac{1}{3}\theta^b_\alpha\theta^c_\beta\epsilon_{abcd}\psi^{d\beta}+\mathcal{O}(\theta^3)\,.\\
    \end{eqnarray}
\end{subequations}
Conversely, if we assume that we have super-fields defined by the recursion relations and the lowest-component fields, and if we assume that the $\N=4$ equations of motion holds for the lowest-component fields, then the super-fields relations \eqref{eq:fer_der_tF}, \eqref{eq:fer_der_tpsi}, \eqref{eq:fer_der_phi}, \eqref{eq:fer_der_psi}, \eqref{eq:fer_der_B}, \eqref{eq:F_B_lagrangemultiplier2} can be obtained by induction on the fermionic degree via the recursion relations
\begin{subequations}
    \begin{eqnarray}
        (1+\rD)(\unabla_{a\alpha}\utF_{\dot\alpha\dot\beta}-\unabla_{\alpha(\dot\alpha}\utpsi_{\dot\beta)a}&=&-\theta^{b\beta}\unabla_{a\alpha}\unabla_{\beta(\dot\alpha}\utpsi_{\dot\beta)a}+2\theta^b_\alpha[\uphi_{ba},\utF_{\dot\alpha\dot\beta}]\nonumber\\&&-\unabla_{\alpha(\dot\alpha}(\theta^{b\beta}\unabla_{\dot\beta)\beta}\uphi_{ab})+\theta^b_\alpha\{\utpsi_{b(\dot\alpha},\utpsi_{\dot\beta)a}\}\,,\\
        (1+\rD)(\unabla_{a\alpha}\utpsi_{\dot\alpha b}-2\unabla_{\alpha\dot\alpha}\uphi_{ab})&=&2\theta^{c\beta}\unabla_{a\alpha}\unabla_{\beta\dot\alpha}\uphi_{bc}+2\theta^b_\alpha[\uphi_{ba},\utpsi_{\dot\alpha b}]\nonumber\\&&-2\epsilon_{abcd}\theta^{c}_\beta\unabla_{\alpha\dot\alpha}\upsi^{d\beta}+2\theta^c_\alpha[\utpsi_{c\dot\alpha},\uphi_{ab}]\,,\\
        (1+\rD)(\unabla_{a\alpha}\uphi_{bc}-\epsilon_{abcd}\upsi^d_\alpha)&=&-\epsilon_{bcde}\theta^d_\beta\unabla_{a\alpha}\upsi^{e\beta}+2\theta^d_\alpha[\uphi_{da},\uphi_{bc}]\nonumber\\&&-\epsilon_{abcd}(\theta^{d\beta}\uB_{\alpha\beta}-\theta^e_\alpha[\uphi_{ef},\uphi^{df}])\,,\\
        (1+\rD)(\unabla_{a\alpha}\upsi^b_\beta+\varepsilon_{\alpha\beta}[\uphi_{ac},\uphi^{bc}]&=&\delta^b_a(1+\rD)\uB_{\alpha\beta}-\theta^{b\gamma}\unabla_{a\alpha}\uB_{\beta\gamma}\nonumber\\&&+\theta^c_\beta\unabla_{a\alpha}[\uphi_{cd},\uphi^{bd}]+2\theta^c_\alpha[\uphi_{ca},\upsi^b_\beta]\nonumber\\&&+\varepsilon_{\alpha\beta}(\epsilon_{acde}\theta^d_\gamma[\upsi^{e\gamma},\uphi^{bc}]+\theta^{[b}_\gamma\delta^{c]}_g[\uphi_{ac},\upsi^{g\gamma}])\nonumber\\&&+2\delta^b_a\theta^c_{(\alpha}[\upsi^d_{\beta)},\uphi_{cd}],,\\
        (1+\rD)(\unabla_{a\alpha}\uB_{\alpha\beta}+2\varepsilon_{\alpha(\beta}[\upsi^b_{\gamma)},\uphi_{ab}])&=&2\theta^b_{(\beta}\unabla_{|a\alpha|}[\upsi^c_{\gamma)},\uphi_{bc}]+2\theta^b_\alpha[\uphi_{ba},\uB_{\beta\gamma}]\nonumber\\&&+2\varepsilon_{\alpha(\beta}([\theta^{b\delta}\uB_{\gamma)\delta}-\theta^c_{\gamma)}[\uphi_{cd},\uphi^{bd}],\uphi_{ab}])\nonumber\\&&+2\epsilon_{abcd}\varepsilon_{\alpha(\beta}[\upsi^b_{\gamma)},\theta^c_\delta\upsi^{d\delta}]\,,\\
    \end{eqnarray}
\end{subequations}
which follow directly from \eqref{eq:rec_rel_tF}, \eqref{eq:rec_rel_tpsi}, \eqref{eq:rec_rel_phi}, \eqref{eq:rec_rel_psi}, \eqref{eq:rec_rel_B}, \eqref{eq:rec_rel_Ab}, and \eqref{eq:rec_rel_Af}.

Similarly, the constraint equations \eqref{eq:chiral_constraint} follows from induction on the fermionic degree on the recursion relations
\begin{subequations}
    \begin{eqnarray}
        (1+\rD)([\unabla_{a\alpha},\unabla_{\beta\dot\alpha}]-\varepsilon_{\alpha\beta}\utpsi_{a\dot\alpha})&=&-\theta^b_\beta\unabla_{a\alpha}\utpsi_{b\dot\beta}+2\theta^b_\alpha\unabla_{\beta\dot\beta}\uphi_{ba}\nonumber\\&&+2\varepsilon_{\alpha\beta}\theta^{b\gamma}\unabla_{\gamma\dot\beta}\uphi_{ab}\,,\\
        (2+\rD)(\{\unabla_{a\alpha},\unabla_{b\beta}\}-2\varepsilon_{\alpha\beta}\uphi_{ab})&=&-2\theta^c_\beta\unabla_{a\alpha}\uphi_{cb}-2\theta^c_\alpha\unabla_{b\beta}\uphi_{ca}\nonumber\\&&-2\varepsilon_{\alpha\beta}\epsilon_{abcd}\theta^c_\gamma\upsi^{d\gamma}\,,
    \end{eqnarray}
\end{subequations}
thus establishing the equivalence.

\section{Computational details}\label{app:computation}
In this brief appendix, we show in more detail how to derive \eqref{eq:curvedformfactor} from the generating functional \eqref{eq:generatingfunctional}
\begin{equation}
    \int_\M\d^4x\,e^{-\im q\cdot x}\iota_\alpha q^{\alpha\dot\alpha}\iota_\beta q^{\beta\dot\beta}\int_{X^2}\frac{\D\lambda_1\D\lambda_2}{\langle \iota\lambda_1\rangle\langle \iota\lambda_2\rangle}\langle\lambda_1\lambda_2\rangle^2\tr\left(\left.\frac{\partial\twista}{\partial\mu^{\dot\alpha}_1}\right|_X\K_X(\lambda_1,\lambda_2)\left.\frac{\partial\twista}{\partial\mu^{\dot\beta}_2}\right|_X\K_X(\lambda_2,\lambda_1)\right)\,,\label{eq:generatingfunctionalextended}
\end{equation}
here rewritten for the sake of clarity. Its expansion at $n$ points evaluated on momentum eigenstates reads
\begin{equation}
	\begin{aligned}
	&\left(\frac{-1}{2\pi \im}\right)^{n-2}\int_\M\d^4x\,e^{\im(Q-q)\cdot x}\iota_\alpha q^{\alpha\dot\alpha}\iota_\beta q^{\beta\dot\beta}\sum_{i,j}\Biggl\{\frac{\tr \hat T_{{p_1}}\ldots \hat T_{{p_n}}}{\langle p_1p_2\rangle\ldots\langle p_np_1\rangle}\frac{\langle ij\rangle^2}{\langle \iota i\rangle \langle \iota j\rangle}\tilde \kappa_{i\,\dot\alpha}\tilde\kappa_{j\,\dot\beta}\\&-2\int_X\frac{\D\lambda
	}{2\pi \im}\frac{\tr \hat T_{{p_1}}\ldots \hat T_{{p_{i-1}}}\H^{-1}(x,\lambda)\,\partial_{{\dot\alpha}}\twista\, \H(x,\lambda)\hat T_{{p_i}}\ldots \hat T_{{p_n}}}{\langle p_1p_2\rangle\ldots \langle p_{i-1}\lambda\rangle\langle\lambda p_i\rangle\ldots\langle p_np_1\rangle}\frac{\langle\lambda j\rangle^2}{\langle \iota\lambda\rangle\langle \iota j\rangle}\tilde{\kappa}_{j\,\dot\beta}\\&+\int_{X^2}\frac{\D\lambda\D\lambda'}{(2\pi \im)^2}\frac{1}{\langle p_1p_2\rangle\ldots\langle p_{i-1}\lambda\rangle\langle\lambda p_i\rangle\ldots\langle p_{j-1}\lambda'\rangle\langle\lambda' p_j\rangle\ldots\langle p_np_1\rangle}\frac{\langle\lambda\lambda'\rangle^2}{\langle \iota\lambda\rangle\langle \iota\lambda'\rangle}\times\\&\times\tr \hat T_{{p_1}}\ldots \hat T_{{p_{i-1}}}\H^{-1}(x,\lambda)\,\partial_{{\dot\alpha}}\twista\, \H(x,\lambda)\hat T_{{p_i}}\ldots \hat T_{{p_{j-1}}}\H^{-1}(x,\lambda')\,\partial_{{\dot\beta}}\twista\, \H(x,\lambda')\hat T_{{p_j}}\ldots \hat T_{{p_n}}\\&+\text{perms.}\Biggr\}\,,
	\end{aligned}
\end{equation}
where $\hat T_{j}\coloneqq\H^{-1}(x,\kappa_j)T_{j}\H(x,\kappa_j)$, and $+\text{perms.}$ is a sum over the permutations of $\{p_1,\ldots,p_n\}$. Around a Cartan-valued background, the conjugation on the colour factors reduces to $\hat{T}^{\mathtt a_j}=T_{j}\exp(e_jg(x,\kappa_j))$, so that the colour-ordered form factor is 
\begin{equation}
    \begin{aligned}\mathscr{F}=&\frac{1}{\langle 12\rangle\ldots\langle n1\rangle}\int_\M\d^4x\,e^{\im(Q-q)\cdot x+\sum_je_jg(x,\kappa_j)}\iota_\alpha q^{\alpha\dot\alpha}\iota_\beta q^{\beta\dot\beta}\sum_{i,j}\Biggl\{\frac{\langle ij\rangle^2}{\langle \iota i\rangle\langle \iota j\rangle}\tilde\kappa_{i\dot\alpha}\tilde\kappa_{j\dot\beta}\\&-2\int_X\frac{\D\lambda}{2\pi \im}\frac{\langle i-1,i\rangle\langle \lambda j\rangle^2\tilde\kappa_{j\dot\beta}}{\langle i-1,\lambda\rangle\langle\lambda i\rangle\langle \iota\lambda\rangle\langle \iota j\rangle}\left.\frac{\partial\twista}{\partial\mu^{\dot\alpha}}\right|_X\\&+\int_{X^2}\frac{\D\lambda\D\lambda'}{(2\pi \im)^2}\frac{\langle i-1,i\rangle\langle j-1,j\rangle\langle\lambda\lambda'\rangle^2}{\langle i-1,\lambda\rangle\langle\lambda i\rangle\langle j-1,\lambda\rangle\langle \lambda j\rangle\langle \iota\lambda\rangle\langle \iota\lambda'\rangle}\left.\frac{\partial\twista}{\partial\mu^{\dot\alpha}}\right|_X\left.\frac{\partial\twista}{\partial{\mu'}^{\dot\beta}}\right|_X\Biggr\}\,.\end{aligned}\label{eq:pre-formfactor-app}
\end{equation}
We now consider the integral
\begin{equation}
    \mathcal{J}_{\alpha\beta\dot\alpha}(\ell)\coloneqq\int_X\frac{\D\lambda}{2\pi \im}\frac{\lambda_\alpha\lambda_\beta}{\langle \iota\lambda\rangle\langle \ell-1,\lambda\rangle\langle\lambda \ell\rangle}\left.\frac{\partial\twista}{\partial\mu^{\dot\alpha}}\right|_X\,.
\end{equation}
Using the identity
\begin{equation}    	\begin{aligned}\frac{\lambda_\alpha\lambda_\beta}{\langle \iota\lambda\rangle\langle \ell-1\lambda\rangle\langle \ell\lambda\rangle}=&\frac{\kappa_{\ell-1\,\alpha}\kappa_{\ell-1\,\beta}}{\langle \ell-1,\ell\rangle^2\langle \iota,\ell-1\rangle}\left(\frac{\langle \iota\ell\rangle}{\langle \iota\lambda\rangle}-\frac{\langle \ell-1,\ell\rangle}{\langle \ell-1,\lambda\rangle}\right)\\&+\frac{\kappa_{\ell\,\alpha} \kappa_{\ell\,\beta}}{\langle \ell-1,\ell\rangle^2\langle \iota\ell\rangle}\left(\frac{\langle \iota,\ell-1\rangle}{\langle \iota\lambda\rangle}-\frac{\langle \ell,\ell-1\rangle}{\langle \ell\lambda\rangle}\right)\\&-\frac{\kappa_{\ell-1\,\alpha} \kappa_{\ell\,\beta}+\kappa_{\ell-1\,\beta}\kappa_{\ell\,\alpha}}{\langle \ell-1,\ell\rangle^2\langle \iota\lambda\rangle}\,,
	\end{aligned}
\end{equation}
the integral can be evaluated explicitly and reads
\begin{equation}
    \begin{aligned}
    \mathcal{J}_{\alpha\beta\dot\alpha}(\ell)=-\im&\left(\frac{\kappa_{\ell-1\,\alpha} \kappa_{\ell-1\,\beta}}{\langle \ell-1,\iota\rangle\langle \ell-1,\ell\rangle}G_{\dot\alpha}(x,\kappa_{\ell-1})+\frac{\kappa_{\ell\,\alpha}\kappa_{\ell\,\beta}}{\langle \ell \iota\rangle\langle \ell,\ell-1\rangle}G_{\dot\alpha}(x,\kappa_\ell)\right.\\&\left.+\frac{\iota_\alpha \iota_\beta}{\langle \iota,\ell-1\rangle\langle \iota\ell\rangle}G_{\dot\alpha}(x,\iota)\right)\,.
    \end{aligned}
\end{equation}
We then see that the terms with residual integrals over $X$ and $X^2$ don't contribute to the form factor in \eqref{eq:pre-formfactor-app}. For the term with a single integral over $X$, the sum over $i$ is
\begin{equation}
	\sum_{i}\left(\frac{\langle i-1,j\rangle^2}{\langle i-1,\iota\rangle}G_{\dot\alpha}(x,\kappa_{i-1})-\frac{\langle ij\rangle^2}{\langle i\iota\rangle}G_{\dot\alpha}(x,\kappa_i)+\frac{\langle \iota j\rangle^2\langle i-1,i\rangle}{\langle \iota,i-1\rangle\langle \iota i\rangle}G_{\dot\alpha}(x,\iota)\right)\,.
\end{equation}
The first two terms obviously cancel in the sum. The third one is telescopic in $i$ as well, once we complete $\iota_\alpha$ to a basis $\{\iota_\alpha,o_\alpha\}$ of undotted spinors, introducing the spinor $o_\alpha$ normalized such that $\langle \iota o\rangle=1$. The third term then reduces to
\begin{equation}
	\sum_i\frac{\langle i-1,1\rangle}{\langle \iota,i-1\rangle\langle \iota i\rangle}=\sum_i\left(\frac{\langle o i\rangle}{\langle \iota i\rangle}-\frac{\langle o,i-1\rangle}{\langle \iota,i-1\rangle}\right)=0\,.
\end{equation}
Overall, the form factor reads
\begin{equation}
    \mathscr{F}=\frac{1}{\langle 12\rangle\ldots\langle n1\rangle}\sum_{i,j}\frac{\langle ij\rangle^2\langle \iota|q|i]\langle \iota|q|j]}{\langle \iota i\rangle\langle \iota j\rangle}\int_\M\d^4x\,e^{\im(Q-q)\cdot x+\sum_je_jg(x,\kappa_j)}\,.
\end{equation}
Let $\mathcal{S}$ denote the sum over $i,j$
\begin{equation}
	\mathcal{S}\coloneqq q^{\alpha\dot\alpha}q^{\beta\dot\beta}\sum_{i,j}\frac{\langle ij\rangle^2}{\langle \iota i\rangle\langle \iota j\rangle}\iota_\alpha \iota_\beta \tilde\kappa_{i\dot\alpha}\tilde\kappa_{j\dot\beta}\,.
\end{equation}
Using the Schouten identity and performing one of the sums, $\mathcal{S}$ reduces to
\begin{equation}
	\mathcal{S}=q^{\alpha\dot\alpha}q^{\beta\dot\beta}\left(Q_{\gamma\dot\alpha}\sum_i\frac{\iota_\beta\tilde\kappa_{i\dot\beta}\kappa_{i\alpha}\kappa^\gamma_i}{\langle \iota i\rangle}+Q_{\gamma\dot\beta}\sum_i\frac{\iota_\beta\tilde\kappa_{i\dot\alpha}\kappa_{i\alpha}\kappa^\gamma_i}{\langle \iota i\rangle}\right)\,,
\end{equation}
Finally, noticing the identity
\begin{equation}
    \begin{aligned}
	q^{\alpha\dot\alpha}Q_{\gamma\dot\alpha}=\frac{1}{2}(q^{\alpha\dot\alpha}Q_{\gamma\dot\alpha}-q_{\gamma\dot\alpha}Q^{\alpha\dot\alpha})+\frac{1}{2}\delta^\alpha_\gamma q\cdot Q\,,
	\end{aligned}
\end{equation}
we can further simplify $\mathcal{S}$ down to
\begin{equation}
    \mathcal{S}=-(q\cdot Q)^2\,,
\end{equation}
and the form factor is finally (up to an overall numerical factor)
\begin{equation}
	\mathscr{F}=\frac{(q\cdot Q)^2}{\langle 12\rangle\ldots\langle n1\rangle}\int_\M\d^4x\,e^{\im(Q-q)\cdot x+\sum_je_jg(x,\kappa_j)}\,.
\end{equation}

The same procedure can be used to obtain \eqref{eq:trf^3formfactor} from the lifting of $\tr\tilde F^3$ to twistor space. Such lifting reads
\begin{multline}
    \tr\tilde F^3=\int\D\lambda_{123}\,\langle\lambda_1\lambda_2\rangle\langle\lambda_2\lambda_3\rangle\langle\lambda_3\lambda_1\rangle\tr \partial_{{\dot\alpha}}\partial^{{\dot\beta}}\twista_1 \K_{12}\partial_{{\dot\beta}}\partial^{{\dot\gamma}}\twista_2\K_{23}\partial_{{\dot\gamma}}\partial^{{\dot\alpha}}\twista_3\K_{31}\\
    +6\int\D\lambda_{1234}\,\langle\lambda_2\lambda_3\rangle\langle\lambda_3\lambda_4\rangle\langle\lambda_4\lambda_1\rangle\tr\partial_{{(\dot\alpha}}\twista_1\K_{12}\partial_{{\dot\beta)}}\twista_2\K_{23}\partial^{{\dot\beta}}\partial^{{\dot\gamma}}\twista_3\K_{34}\partial_{{\dot\gamma}}\partial^{{\dot\alpha}}\twista_4\K_{41}\\
    -12\int\D\lambda_{12345}\,\langle\lambda_2\lambda_3\rangle\langle\lambda_4\lambda_5\rangle\langle\lambda_5\lambda_1\rangle\tr\partial_{{(\dot\alpha}}\twista_1\K_{12}\partial_{{\dot\beta)}}\twista_2\K_{23}\partial^{{(\dot\beta}}\twista_3\K_{34}\p^{{\dot\gamma)}}\twista_4\K_{45}\p_{{\dot\gamma}}\p^{{\dot\alpha}}\twista_5\K_{51}\\+8\,\varepsilon^{\dot\alpha\dot\delta}\int\D\lambda_{123456}\,\langle\lambda_{2}\lambda_3\rangle\langle\lambda_4\lambda_5\rangle\langle\lambda_6\lambda_1\rangle\tr\partial_{{(\dot\alpha}}\twista_1\K_{12}\partial_{{\dot\beta)}}\twista_2\K_{23}\partial^{{(\dot\beta}}\twista_3\K_{34}\partial^{{\dot\gamma)}}\twista_4\K_{45}\partial_{{(\dot\gamma}}\twista_5\K_{56}\partial_{{\dot\delta)}}\twista_6\K_{61}\,,\label{eq:trf^3generatingfunctional}
\end{multline}

In the case of super form factors in $\N=4$ SYM, the same arguments of Section \ref{sec:genericformfactors} hold around gluonic backgrounds, so we briefly comment only on the derivation of \eqref{eq:super_form_factor_around_fermions} from \eqref{eq:super_form_factor_generating_functional}. We denote the background super-connection as $\A^\text{(B)}=\twista+\twista_a\chi^a$ and a generic external state \eqref{eq:n=4_on_shell_state} as $\A_i$. We first notice that \eqref{eq:super_form_factor_generating_functional} is obtained from the super-field $\underline\phi_{ab}$ using the supersymmetrized $K$-matrix
\begin{equation}
    \frac{1}{2}\tr\phi_{ab}\phi_{ab}=\frac{1}{2}\iota^\alpha \iota^\beta \iota^\gamma \iota^\delta\int\d^8\theta\,\prod_{\gamma,c}\theta^{\gamma c}\,\tr(\p_{\alpha a}\p_{\beta b}\underline K\,\p_{\gamma a}\p_{\delta b}\underline K)\,,
\end{equation}
and integrating by parts twice. The super form factor is obtained by taking the Fourier transform of
\begin{multline}
\iota^\alpha \iota^\beta\left.\Bigg(\sum_{i,j}\int\frac{\D\lambda_{1}\ldots\D\lambda_n}{\langle \iota\lambda_i\rangle\langle \iota \lambda_j\rangle}\frac{\langle \lambda_i\lambda_j\rangle^2}{\langle \lambda_1\lambda_2\rangle\ldots\langle\lambda_n\lambda_1\rangle}\right.\\\left.\tr(\underline\H_1^{-1}\underline{a}_1\underline\H_1\ldots \underline\H_i^{-1}\p_a\underline{a}_i\underline\H_i\ldots \underline\H_j^{-1}\p_b\underline{a}_j\underline\H_j\ldots\underline\H_n^{-1}\underline{a}_n\underline\H_n)\Bigg)\right|_{\theta^{\alpha a}\theta^{\beta b}}\,,
\end{multline}
where we are extracting the $\theta^{\alpha a}\theta^{\beta b}$ component, evaluating this expression on the on-shell state \eqref{eq:n=4_on_shell_state} and summing over permutations. Around a Cartan-valued background, the holomorphic frame is $\underline\H=\exp(-\underline g)$, with
\begin{equation}
    \underline g(x,\theta,\lambda)=\frac{1}{2\pi\im}\int_{X}\frac{\D\lambda'}{\langle\lambda\lambda'\rangle}\frac{\langle \iota\lambda\rangle}{\langle \iota\lambda'\rangle}\left.\A^{(\text{B})}\right|_{X}=g(x,\lambda)+\theta^{\alpha a}\mathcal{G}_{\alpha a}(x,\lambda)\,,
\end{equation}
where
\begin{equation}
    \mathcal{G}_{\alpha a}(x,\lambda)=\frac{1}{2\pi\im}\int_{X}\frac{\D\lambda'}{\langle \lambda\lambda'\rangle}\frac{\langle \iota\lambda\rangle}{\langle \iota\lambda'\rangle}\lambda'_\alpha\left.\twista_a\right|_X\,.
\end{equation}
Note that this function is related to \eqref{eq:g_function_fermionic} as
\begin{equation}
    \iota^\alpha\mathcal{G}_{\alpha a}(x,\lambda)=\langle \iota\lambda\rangle g^{(f)}_a(x,\lambda)\,,
\end{equation}
so that the colour-ordered super form factor is finally given by
\begin{multline}
    \frac{1}{2}\frac{1}{\langle 12\rangle\ldots\langle n1\rangle}\int_\M\d^4x\,e^{\im(Q-q)\cdot x+\sum_je_jg(x,\kappa_j)}\\\sum_{i,j,k,\ell}\frac{\langle ij\rangle^2}{\langle \iota i\rangle\langle \iota j\rangle}\eta_{i\,a}\eta_{j\,b}\iota^\alpha \iota^\beta(\kappa_{k\,\alpha}\eta_{k\,a}+e_k\mathcal{G}_{\alpha a}(x,\kappa_k))(\kappa_{\ell\,\beta}\eta_{\ell\,b}+e_\ell\mathcal{G}_{\beta b}(x,\kappa_{\ell}))\,.
\end{multline}
With a computation completely analogous to the one carried out for the form factor for $\tr\tilde F^2$, the last sums can be shown to be independent of $o^\alpha$ and they coincide precisely with $(\mathcal{Q}^\alpha_a\tilde{\mathcal{Q}}_{\alpha a})^2$.

\bibliography{sd_form_factors}

\providecommand{\href}[2]{#2}\begingroup\raggedright\begin{thebibliography}{100}

\bibitem{Penrose:1976js}
R.~Penrose, \emph{{Nonlinear gravitons and curved twistor theory}},
  \href{http://dx.doi.org/10.1007/BF00762011}{\emph{Gen. Rel. Grav.} {\bfseries
  7} (1976) 31--52}.

\bibitem{Ward:1977ta}
R.~S. Ward, \emph{{On self-dual gauge fields}},
  \href{http://dx.doi.org/10.1016/0375-9601(77)90842-8}{\emph{Phys. Lett.}
  {\bfseries A61} (1977) 81--82}.

\bibitem{Mason:1991rf}
L.~J. Mason and N.~M.~J. Woodhouse, \emph{{Integrability, selfduality, and
  twistor theory}}.
\newblock Oxford University Press, 1996.

\bibitem{Ward:1990vs}
R.~S. Ward and R.~O. Wells, \emph{{Twistor geometry and field theory}}.
\newblock Cambridge Monographs on Mathematical Physics. Cambridge University
  Press, 8, 1991,
  \href{http://dx.doi.org/10.1017/CBO9780511524493}{10.1017/CBO9780511524493}.

\bibitem{Parke:1986gb}
S.~J. Parke and T.~R. Taylor, \emph{{An Amplitude for $n$ Gluon Scattering}},
  \href{http://dx.doi.org/10.1103/PhysRevLett.56.2459}{\emph{Phys. Rev. Lett.}
  {\bfseries 56} (1986) 2459}.

\bibitem{Nair:1988bq}
V.~P. Nair, \emph{{A Current Algebra for Some Gauge Theory Amplitudes}},
  \href{http://dx.doi.org/10.1016/0370-2693(88)91471-2}{\emph{Phys. Lett. B}
  {\bfseries 214} (1988) 215--218}.

\bibitem{Witten:2003nn}
E.~Witten, \emph{{Perturbative gauge theory as a string theory in twistor
  space}}, \href{http://dx.doi.org/10.1007/s00220-004-1187-3}{\emph{Commun.
  Math. Phys.} {\bfseries 252} (2004) 189--258},
  [\href{https://arxiv.org/abs/hep-th/0312171}{{\ttfamily hep-th/0312171}}].

\bibitem{Berkovits:2004hg}
N.~Berkovits, \emph{{An Alternative string theory in twistor space for N=4
  superYang-Mills}},
  \href{http://dx.doi.org/10.1103/PhysRevLett.93.011601}{\emph{Phys. Rev.
  Lett.} {\bfseries 93} (2004) 011601},
  [\href{https://arxiv.org/abs/hep-th/0402045}{{\ttfamily hep-th/0402045}}].

\bibitem{Roiban:2004yf}
R.~Roiban, M.~Spradlin and A.~Volovich, \emph{{On the tree level S matrix of
  Yang-Mills theory}},
  \href{http://dx.doi.org/10.1103/PhysRevD.70.026009}{\emph{Phys. Rev.}
  {\bfseries D70} (2004) 026009},
  [\href{https://arxiv.org/abs/hep-th/0403190}{{\ttfamily hep-th/0403190}}].

\bibitem{Cachazo:2004kj}
F.~Cachazo, P.~Svrcek and E.~Witten, \emph{{MHV vertices and tree amplitudes in
  gauge theory}},
  \href{http://dx.doi.org/10.1088/1126-6708/2004/09/006}{\emph{JHEP} {\bfseries
  09} (2004) 006}, [\href{https://arxiv.org/abs/hep-th/0403047}{{\ttfamily
  hep-th/0403047}}].

\bibitem{Britto:2005fq}
R.~Britto, F.~Cachazo, B.~Feng and E.~Witten, \emph{{Direct proof of tree-level
  recursion relation in Yang-Mills theory}},
  \href{http://dx.doi.org/10.1103/PhysRevLett.94.181602}{\emph{Phys. Rev.
  Lett.} {\bfseries 94} (2005) 181602},
  [\href{https://arxiv.org/abs/hep-th/0501052}{{\ttfamily hep-th/0501052}}].

\bibitem{Brandhuber:2008pf}
A.~Brandhuber, P.~Heslop and G.~Travaglini, \emph{{A Note on dual
  superconformal symmetry of the N=4 super Yang-Mills S-matrix}},
  \href{http://dx.doi.org/10.1103/PhysRevD.78.125005}{\emph{Phys. Rev. D}
  {\bfseries 78} (2008) 125005},
  [\href{https://arxiv.org/abs/0807.4097}{{\ttfamily 0807.4097}}].

\bibitem{Bern:2010ue}
Z.~Bern, J.~J.~M. Carrasco and H.~Johansson, \emph{{Perturbative Quantum
  Gravity as a Double Copy of Gauge Theory}},
  \href{http://dx.doi.org/10.1103/PhysRevLett.105.061602}{\emph{Phys. Rev.
  Lett.} {\bfseries 105} (2010) 061602},
  [\href{https://arxiv.org/abs/1004.0476}{{\ttfamily 1004.0476}}].

\bibitem{ArkaniHamed:2010kv}
N.~Arkani-Hamed, J.~L. Bourjaily, F.~Cachazo, S.~Caron-Huot and J.~Trnka,
  \emph{{The All-Loop Integrand For Scattering Amplitudes in Planar N=4 SYM}},
  \href{http://dx.doi.org/10.1007/JHEP01(2011)041}{\emph{JHEP} {\bfseries 01}
  (2011) 041}, [\href{https://arxiv.org/abs/1008.2958}{{\ttfamily 1008.2958}}].

\bibitem{Cachazo:2013hca}
F.~Cachazo, S.~He and E.~Y. Yuan, \emph{{Scattering of Massless Particles in
  Arbitrary Dimensions}},
  \href{http://dx.doi.org/10.1103/PhysRevLett.113.171601}{\emph{Phys. Rev.
  Lett.} {\bfseries 113} (2014) 171601},
  [\href{https://arxiv.org/abs/1307.2199}{{\ttfamily 1307.2199}}].

\bibitem{Cachazo:2014xea}
F.~Cachazo, S.~He and E.~Y. Yuan, \emph{{Scattering Equations and Matrices:
  From Einstein To Yang-Mills, DBI and NLSM}},
  \href{http://dx.doi.org/10.1007/JHEP07(2015)149}{\emph{JHEP} {\bfseries 07}
  (2015) 149}, [\href{https://arxiv.org/abs/1412.3479}{{\ttfamily 1412.3479}}].

\bibitem{Geyer:2014fka}
Y.~Geyer, A.~E. Lipstein and L.~J. Mason, \emph{{Ambitwistor Strings in Four
  Dimensions}},
  \href{http://dx.doi.org/10.1103/PhysRevLett.113.081602}{\emph{Phys. Rev.
  Lett.} {\bfseries 113} (2014) 081602},
  [\href{https://arxiv.org/abs/1404.6219}{{\ttfamily 1404.6219}}].

\bibitem{Dixon:1996wi}
L.~J. Dixon, \emph{{Calculating scattering amplitudes efficiently}},  in
  \emph{{Theoretical Advanced Study Institute in Elementary Particle Physics
  (TASI 95): QCD and Beyond}}, pp.~539--584, 1, 1996,
  \href{https://arxiv.org/abs/hep-ph/9601359}{{\ttfamily hep-ph/9601359}}.

\bibitem{Bern:2007dw}
Z.~Bern, L.~J. Dixon and D.~A. Kosower, \emph{{On-Shell Methods in Perturbative
  QCD}}, \href{http://dx.doi.org/10.1016/j.aop.2007.04.014}{\emph{Annals Phys.}
  {\bfseries 322} (2007) 1587--1634},
  [\href{https://arxiv.org/abs/0704.2798}{{\ttfamily 0704.2798}}].

\bibitem{Elvang:2013cua}
H.~Elvang and Y.-t. Huang, \emph{{Scattering Amplitudes}},
  \href{https://arxiv.org/abs/1308.1697}{{\ttfamily 1308.1697}}.

\bibitem{Travaglini:2022uwo}
G.~Travaglini et~al., \emph{{The SAGEX review on scattering amplitudes}},
  \href{http://dx.doi.org/10.1088/1751-8121/ac8380}{\emph{J. Phys. A}
  {\bfseries 55} (2022) 443001},
  [\href{https://arxiv.org/abs/2203.13011}{{\ttfamily 2203.13011}}].

\bibitem{Geyer:2022cey}
Y.~Geyer and L.~Mason, \emph{{The SAGEX review on scattering amplitudes Chapter
  6: Ambitwistor Strings and Amplitudes from the Worldsheet}},
  \href{http://dx.doi.org/10.1088/1751-8121/ac8190}{\emph{J. Phys. A}
  {\bfseries 55} (2022) 443007},
  [\href{https://arxiv.org/abs/2203.13017}{{\ttfamily 2203.13017}}].

\bibitem{Chalmers:1996rq}
G.~Chalmers and W.~Siegel, \emph{{The self-dual sector of QCD amplitudes}},
  \href{http://dx.doi.org/10.1103/PhysRevD.54.7628}{\emph{Phys. Rev. D}
  {\bfseries 54} (1996) 7628--7633},
  [\href{https://arxiv.org/abs/hep-th/9606061}{{\ttfamily hep-th/9606061}}].

\bibitem{Mason:2005zm}
L.~J. Mason, \emph{{Twistor actions for non-self-dual fields: A New Foundation
  for twistor-string theory}},
  \href{http://dx.doi.org/10.1088/1126-6708/2005/10/009}{\emph{JHEP} {\bfseries
  10} (2005) 009}, [\href{https://arxiv.org/abs/hep-th/0507269}{{\ttfamily
  hep-th/0507269}}].

\bibitem{Boels:2006ir}
R.~Boels, L.~J. Mason and D.~Skinner, \emph{{Supersymmetric Gauge Theories in
  Twistor Space}},
  \href{http://dx.doi.org/10.1088/1126-6708/2007/02/014}{\emph{JHEP} {\bfseries
  02} (2007) 014}, [\href{https://arxiv.org/abs/hep-th/0604040}{{\ttfamily
  hep-th/0604040}}].

\bibitem{Boels:2007qn}
R.~Boels, L.~J. Mason and D.~Skinner, \emph{{From twistor actions to MHV
  diagrams}},
  \href{http://dx.doi.org/10.1016/j.physletb.2007.02.058}{\emph{Phys. Lett.}
  {\bfseries B648} (2007) 90--96},
  [\href{https://arxiv.org/abs/hep-th/0702035}{{\ttfamily hep-th/0702035}}].

\bibitem{Mason:2010yk}
L.~J. Mason and D.~Skinner, \emph{{The Complete Planar S-matrix of N=4 SYM as a
  Wilson Loop in Twistor Space}},
  \href{http://dx.doi.org/10.1007/JHEP12(2010)018}{\emph{JHEP} {\bfseries 12}
  (2010) 018}, [\href{https://arxiv.org/abs/1009.2225}{{\ttfamily 1009.2225}}].

\bibitem{Adamo:2011cb}
T.~Adamo and L.~Mason, \emph{{MHV diagrams in twistor space and the twistor
  action}}, \href{http://dx.doi.org/10.1103/PhysRevD.86.065019}{\emph{Phys.
  Rev. D} {\bfseries 86} (2012) 065019},
  [\href{https://arxiv.org/abs/1103.1352}{{\ttfamily 1103.1352}}].

\bibitem{Adamo:2011pv}
T.~Adamo, M.~Bullimore, L.~Mason and D.~Skinner, \emph{{Scattering Amplitudes
  and Wilson Loops in Twistor Space}},
  \href{http://dx.doi.org/10.1088/1751-8113/44/45/454008}{\emph{J. Phys. A}
  {\bfseries 44} (2011) 454008},
  [\href{https://arxiv.org/abs/1104.2890}{{\ttfamily 1104.2890}}].

\bibitem{Adamo:2013cra}
T.~Adamo, \emph{{Twistor actions for gauge theory and gravity}}, Ph.D. thesis,
  University of Oxford, 2013.
\newblock \href{https://arxiv.org/abs/1308.2820}{{\ttfamily 1308.2820}}.

\bibitem{Strominger:2017zoo}
A.~Strominger, \emph{{Lectures on the Infrared Structure of Gravity and Gauge
  Theory}},  \href{https://arxiv.org/abs/1703.05448}{{\ttfamily 1703.05448}}.

\bibitem{Costello:2022wso}
K.~Costello and N.~M. Paquette, \emph{{Celestial holography meets twisted
  holography: 4d amplitudes from chiral correlators}},
  \href{https://arxiv.org/abs/2201.02595}{{\ttfamily 2201.02595}}.

\bibitem{Adamo:2019ipt}
T.~Adamo, L.~Mason and A.~Sharma, \emph{{Celestial amplitudes and conformal
  soft theorems}},
  \href{http://dx.doi.org/10.1088/1361-6382/ab42ce}{\emph{Class. Quant. Grav.}
  {\bfseries 36} (2019) 205018},
  [\href{https://arxiv.org/abs/1905.09224}{{\ttfamily 1905.09224}}].

\bibitem{Strominger:2021lvk}
A.~Strominger, \emph{{w(1+infinity) and the Celestial Sphere}},
  \href{https://arxiv.org/abs/2105.14346}{{\ttfamily 2105.14346}}.

\bibitem{Ball:2021tmb}
A.~Ball, S.~A. Narayanan, J.~Salzer and A.~Strominger, \emph{{Perturbatively
  exact $w_{1+\infty}$ asymptotic symmetry of quantum self-dual gravity}},
  \href{http://dx.doi.org/10.1007/JHEP01(2022)114}{\emph{JHEP} {\bfseries 01}
  (2022) 114}, [\href{https://arxiv.org/abs/2111.10392}{{\ttfamily
  2111.10392}}].

\bibitem{Strominger:2021mtt}
A.~Strominger, \emph{{$w_{1+\infty}$ Algebra and the Celestial Sphere: Infinite
  Towers of Soft Graviton, Photon, and Gluon Symmetries}},
  \href{http://dx.doi.org/10.1103/PhysRevLett.127.221601}{\emph{Phys. Rev.
  Lett.} {\bfseries 127} (2021) 221601}.

\bibitem{Newman:1976gc}
E.~T. Newman, \emph{{Heaven and Its Properties}},
  \href{http://dx.doi.org/10.1007/BF00762018}{\emph{Gen. Rel. Grav.} {\bfseries
  7} (1976) 107--111}.

\bibitem{Hansen:1978jz}
R.~O. Hansen, E.~T. Newman, R.~Penrose and K.~P. Tod, \emph{{The Metric and
  Curvature Properties of H Space}},
  \href{http://dx.doi.org/10.1098/rspa.1978.0177}{\emph{Proc. Roy. Soc. Lond.}
  {\bfseries A363} (1978) 445--468}.

\bibitem{Sparling:1990}
G.~A.~J. Sparling, \emph{{Dynamically Broken Symmetry and Global Yang-Mills in
  Minkowski Space}},  in \emph{{Further Advances in Twistor Theory}} (L.~J.
  Mason and L.~P. Hughston, eds.), vol.~231, ch.~1.4.2.
\newblock Pitman Research Notes in Mathematics, 1990.

\bibitem{Newman:1978ze}
E.~T. Newman, \emph{{Source-Free Yang-Mills Theories}},
  \href{http://dx.doi.org/10.1103/PhysRevD.18.2901}{\emph{Phys. Rev. D}
  {\bfseries 18} (1978) 2901--2908}.

\bibitem{Adamo:2021lrv}
T.~Adamo, L.~Mason and A.~Sharma, \emph{{Celestial $w_{1+\infty}$ Symmetries
  from Twistor Space}},
  \href{http://dx.doi.org/10.3842/SIGMA.2022.016}{\emph{SIGMA} {\bfseries 18}
  (2022) 016}, [\href{https://arxiv.org/abs/2110.06066}{{\ttfamily
  2110.06066}}].

\bibitem{Adamo:2022wjo}
T.~Adamo, W.~Bu, E.~Casali and A.~Sharma, \emph{{All-order celestial OPE in the
  MHV sector}}, \href{http://dx.doi.org/10.1007/JHEP03(2023)252}{\emph{JHEP}
  {\bfseries 03} (2023) 252},
  [\href{https://arxiv.org/abs/2211.17124}{{\ttfamily 2211.17124}}].

\bibitem{Brandhuber:2010ad}
A.~Brandhuber, B.~Spence, G.~Travaglini and G.~Yang, \emph{{Form Factors in N=4
  Super Yang-Mills and Periodic Wilson Loops}},
  \href{http://dx.doi.org/10.1007/JHEP01(2011)134}{\emph{JHEP} {\bfseries 01}
  (2011) 134}, [\href{https://arxiv.org/abs/1011.1899}{{\ttfamily 1011.1899}}].

\bibitem{Brandhuber:2011tv}
A.~Brandhuber, O.~Gurdogan, R.~Mooney, G.~Travaglini and G.~Yang,
  \emph{{Harmony of Super Form Factors}},
  \href{http://dx.doi.org/10.1007/JHEP10(2011)046}{\emph{JHEP} {\bfseries 10}
  (2011) 046}, [\href{https://arxiv.org/abs/1107.5067}{{\ttfamily 1107.5067}}].

\bibitem{Brandhuber:2012vm}
A.~Brandhuber, G.~Travaglini and G.~Yang, \emph{{Analytic two-loop form factors
  in N=4 SYM}}, \href{http://dx.doi.org/10.1007/JHEP05(2012)082}{\emph{JHEP}
  {\bfseries 05} (2012) 082},
  [\href{https://arxiv.org/abs/1201.4170}{{\ttfamily 1201.4170}}].

\bibitem{Penante:2014sza}
B.~Penante, B.~Spence, G.~Travaglini and C.~Wen, \emph{{On super form factors
  of half-BPS operators in N=4 super Yang-Mills}},
  \href{http://dx.doi.org/10.1007/JHEP04(2014)083}{\emph{JHEP} {\bfseries 04}
  (2014) 083}, [\href{https://arxiv.org/abs/1402.1300}{{\ttfamily 1402.1300}}].

\bibitem{Brandhuber:2016xue}
A.~Brandhuber, E.~Hughes, R.~Panerai, B.~Spence and G.~Travaglini, \emph{{The
  connected prescription for form factors in twistor space}},
  \href{http://dx.doi.org/10.1007/JHEP11(2016)143}{\emph{JHEP} {\bfseries 11}
  (2016) 143}, [\href{https://arxiv.org/abs/1608.03277}{{\ttfamily
  1608.03277}}].

\bibitem{He:2016jdg}
S.~He and Z.~Liu, \emph{{A note on connected formula for form factors}},
  \href{http://dx.doi.org/10.1007/JHEP12(2016)006}{\emph{JHEP} {\bfseries 12}
  (2016) 006}, [\href{https://arxiv.org/abs/1608.04306}{{\ttfamily
  1608.04306}}].

\bibitem{Brandhuber:2017bkg}
A.~Brandhuber, M.~Kostacinska, B.~Penante and G.~Travaglini, \emph{{Higgs
  amplitudes from $\mathcal{N}=4$ super Yang-Mills theory}},
  \href{http://dx.doi.org/10.1103/PhysRevLett.119.161601}{\emph{Phys. Rev.
  Lett.} {\bfseries 119} (2017) 161601},
  [\href{https://arxiv.org/abs/1707.09897}{{\ttfamily 1707.09897}}].

\bibitem{Brandhuber:2018kqb}
A.~Brandhuber, M.~Kostacinska, B.~Penante and G.~Travaglini,
  \emph{{$\text{Tr}(F^3)$ supersymmetric form factors and maximal
  transcendentality Part II: $0<\mathcal{N}<4$ super Yang-Mills}},
  \href{http://dx.doi.org/10.1007/JHEP12(2018)077}{\emph{JHEP} {\bfseries 12}
  (2018) 077}, [\href{https://arxiv.org/abs/1804.05828}{{\ttfamily
  1804.05828}}].

\bibitem{Brandhuber:2018xzk}
A.~Brandhuber, M.~Kostacinska, B.~Penante and G.~Travaglini,
  \emph{{$\text{Tr}(F^3)$ supersymmetric form factors and maximal
  transcendentality Part I: $\mathcal{N}=4$ super Yang-Mills}},
  \href{http://dx.doi.org/10.1007/JHEP12(2018)076}{\emph{JHEP} {\bfseries 12}
  (2018) 076}, [\href{https://arxiv.org/abs/1804.05703}{{\ttfamily
  1804.05703}}].

\bibitem{Bianchi:2018peu}
L.~Bianchi, A.~Brandhuber, R.~Panerai and G.~Travaglini, \emph{{Form factor
  recursion relations at loop level}},
  \href{http://dx.doi.org/10.1007/JHEP02(2019)182}{\emph{JHEP} {\bfseries 02}
  (2019) 182}, [\href{https://arxiv.org/abs/1812.09001}{{\ttfamily
  1812.09001}}].

\bibitem{Bianchi:2018rrj}
L.~Bianchi, A.~Brandhuber, R.~Panerai and G.~Travaglini, \emph{{Dual conformal
  invariance for form factors}},
  \href{http://dx.doi.org/10.1007/JHEP02(2019)134}{\emph{JHEP} {\bfseries 02}
  (2019) 134}, [\href{https://arxiv.org/abs/1812.10468}{{\ttfamily
  1812.10468}}].

\bibitem{AccettulliHuber:2019abj}
M.~Accettulli~Huber, A.~Brandhuber, S.~De~Angelis and G.~Travaglini,
  \emph{{Complete Form Factors in Yang-Mills from Unitarity and Spinor Helicity
  in Six Dimensions}},
  \href{http://dx.doi.org/10.1103/PhysRevD.101.026004}{\emph{Phys. Rev. D}
  {\bfseries 101} (2020) 026004},
  [\href{https://arxiv.org/abs/1910.04772}{{\ttfamily 1910.04772}}].

\bibitem{Koster:2016ebi}
L.~Koster, V.~Mitev, M.~Staudacher and M.~Wilhelm, \emph{{Composite Operators
  in the Twistor Formulation of N=4 Supersymmetric Yang-Mills Theory}},
  \href{http://dx.doi.org/10.1103/PhysRevLett.117.011601}{\emph{Phys. Rev.
  Lett.} {\bfseries 117} (2016) 011601},
  [\href{https://arxiv.org/abs/1603.04471}{{\ttfamily 1603.04471}}].

\bibitem{Koster:2016loo}
L.~Koster, V.~Mitev, M.~Staudacher and M.~Wilhelm, \emph{{All tree-level MHV
  form factors in $ \mathcal{N} $ = 4 SYM from twistor space}},
  \href{http://dx.doi.org/10.1007/JHEP06(2016)162}{\emph{JHEP} {\bfseries 06}
  (2016) 162}, [\href{https://arxiv.org/abs/1604.00012}{{\ttfamily
  1604.00012}}].

\bibitem{Koster:2016fna}
L.~Koster, V.~Mitev, M.~Staudacher and M.~Wilhelm, \emph{{On Form Factors and
  Correlation Functions in Twistor Space}},
  \href{http://dx.doi.org/10.1007/JHEP03(2017)131}{\emph{JHEP} {\bfseries 03}
  (2017) 131}, [\href{https://arxiv.org/abs/1611.08599}{{\ttfamily
  1611.08599}}].

\bibitem{Chicherin:2016soh}
D.~Chicherin and E.~Sokatchev, \emph{{Demystifying the twistor construction of
  composite operators in ${\mathcal N}=4$ super-Yang\textendash{}Mills
  theory}}, \href{http://dx.doi.org/10.1088/1751-8121/aa6b95}{\emph{J. Phys. A}
  {\bfseries 50} (2017) 205402},
  [\href{https://arxiv.org/abs/1603.08478}{{\ttfamily 1603.08478}}].

\bibitem{Chicherin:2016ybl}
D.~Chicherin, P.~Heslop, G.~P. Korchemsky and E.~Sokatchev, \emph{{Wilson Loop
  Form Factors: A New Duality}},
  \href{http://dx.doi.org/10.1007/JHEP04(2018)029}{\emph{JHEP} {\bfseries 04}
  (2018) 029}, [\href{https://arxiv.org/abs/1612.05197}{{\ttfamily
  1612.05197}}].

\bibitem{Chicherin:2016fac}
D.~Chicherin and E.~Sokatchev, \emph{{$ \mathcal{N} $ = 4 super-Yang-Mills in
  LHC superspace part I: classical and quantum theory}},
  \href{http://dx.doi.org/10.1007/JHEP02(2017)062}{\emph{JHEP} {\bfseries 02}
  (2017) 062}, [\href{https://arxiv.org/abs/1601.06803}{{\ttfamily
  1601.06803}}].

\bibitem{Chicherin:2016fbj}
D.~Chicherin and E.~Sokatchev, \emph{{$ \mathcal{N} $ = 4 super-Yang-Mills in
  LHC superspace part II: non-chiral correlation functions of the stress-tensor
  multiplet}}, \href{http://dx.doi.org/10.1007/JHEP03(2017)048}{\emph{JHEP}
  {\bfseries 03} (2017) 048},
  [\href{https://arxiv.org/abs/1601.06804}{{\ttfamily 1601.06804}}].

\bibitem{Chicherin:2014uca}
D.~Chicherin, R.~Doobary, B.~Eden, P.~Heslop, G.~P. Korchemsky, L.~Mason
  et~al., \emph{{Correlation functions of the chiral stress-tensor multiplet in
  $ \mathcal{N}=4 $ SYM}},
  \href{http://dx.doi.org/10.1007/JHEP06(2015)198}{\emph{JHEP} {\bfseries 06}
  (2015) 198}, [\href{https://arxiv.org/abs/1412.8718}{{\ttfamily 1412.8718}}].

\bibitem{Adamo:2011dq}
T.~Adamo, M.~Bullimore, L.~Mason and D.~Skinner, \emph{{A Proof of the
  Supersymmetric Correlation Function / Wilson Loop Correspondence}},
  \href{http://dx.doi.org/10.1007/JHEP08(2011)076}{\emph{JHEP} {\bfseries 08}
  (2011) 076}, [\href{https://arxiv.org/abs/1103.4119}{{\ttfamily 1103.4119}}].

\bibitem{Eden:2017fow}
B.~Eden, P.~Heslop and L.~Mason, \emph{{The Correlahedron}},
  \href{http://dx.doi.org/10.1007/JHEP09(2017)156}{\emph{JHEP} {\bfseries 09}
  (2017) 156}, [\href{https://arxiv.org/abs/1701.00453}{{\ttfamily
  1701.00453}}].

\bibitem{Casali:2022fro}
E.~Casali, W.~Melton and A.~Strominger, \emph{{Celestial Amplitudes as
  AdS-Witten Diagrams}},  \href{https://arxiv.org/abs/2204.10249}{{\ttfamily
  2204.10249}}.

\bibitem{Melton:2022fsf}
W.~Melton, S.~A. Narayanan and A.~Strominger, \emph{{Deforming Soft Algebras
  for Gauge Theory}},  \href{https://arxiv.org/abs/2212.08643}{{\ttfamily
  2212.08643}}.

\bibitem{Furry:1951zz}
W.~H. Furry, \emph{{On Bound States and Scattering in Positron Theory}},
  \href{http://dx.doi.org/10.1103/PhysRev.81.915}{\emph{Phys. Rev.} {\bfseries
  81} (1951) 115--124}.

\bibitem{DeWitt:1967ub}
B.~S. DeWitt, \emph{{Quantum Theory of Gravity. 2. The Manifestly Covariant
  Theory}}, \href{http://dx.doi.org/10.1103/PhysRev.162.1195}{\emph{Phys. Rev.}
  {\bfseries 162} (1967) 1195--1239}.

\bibitem{tHooft:1975uxh}
G.~'t~Hooft, \emph{{The Background Field Method in Gauge Field Theories}},  in
  \emph{{Functional and Probabilistic Methods in Quantum Field Theory. 1.
  Proceedings, 12th Winter School of Theoretical Physics, Karpacz, Feb 17-March
  2, 1975}}, pp.~345--369, 1975.

\bibitem{Abbott:1981ke}
L.~F. Abbott, \emph{{Introduction to the Background Field Method}}, {\emph{Acta
  Phys. Polon.} {\bfseries B13} (1982) 33}.

\bibitem{Adamo:2017nia}
T.~Adamo, E.~Casali, L.~Mason and S.~Nekovar, \emph{{Scattering on plane waves
  and the double copy}},
  \href{http://dx.doi.org/10.1088/1361-6382/aa9961}{\emph{Class. Quant. Grav.}
  {\bfseries 35} (2018) 015004},
  [\href{https://arxiv.org/abs/1706.08925}{{\ttfamily 1706.08925}}].

\bibitem{Adamo:2017sze}
T.~Adamo, E.~Casali, L.~Mason and S.~Nekovar, \emph{{Amplitudes on plane waves
  from ambitwistor strings}},
  \href{http://dx.doi.org/10.1007/JHEP11(2017)160}{\emph{JHEP} {\bfseries 11}
  (2017) 160}, [\href{https://arxiv.org/abs/1708.09249}{{\ttfamily
  1708.09249}}].

\bibitem{Adamo:2018mpq}
T.~Adamo, E.~Casali, L.~Mason and S.~Nekovar, \emph{{Plane wave backgrounds and
  colour-kinematics duality}},
  \href{http://dx.doi.org/10.1007/JHEP02(2019)198}{\emph{JHEP} {\bfseries 02}
  (2019) 198}, [\href{https://arxiv.org/abs/1810.05115}{{\ttfamily
  1810.05115}}].

\bibitem{Adamo:2020qru}
T.~Adamo and A.~Ilderton, \emph{{Classical and quantum double copy of
  back-reaction}}, \href{http://dx.doi.org/10.1007/JHEP09(2020)200}{\emph{JHEP}
  {\bfseries 09} (2020) 200},
  [\href{https://arxiv.org/abs/2005.05807}{{\ttfamily 2005.05807}}].

\bibitem{Adamo:2021hno}
T.~Adamo, A.~Ilderton and A.~J. MacLeod, \emph{{One-loop multicollinear limits
  from 2-point amplitudes on self-dual backgrounds}},
  \href{http://dx.doi.org/10.1007/JHEP12(2021)207}{\emph{JHEP} {\bfseries 12}
  (2021) 207}, [\href{https://arxiv.org/abs/2103.12850}{{\ttfamily
  2103.12850}}].

\bibitem{Adamo:2020syc}
T.~Adamo, L.~Mason and A.~Sharma, \emph{{MHV scattering of gluons and gravitons
  in chiral strong fields}},
  \href{http://dx.doi.org/10.1103/PhysRevLett.125.041602}{\emph{Phys. Rev.
  Lett.} {\bfseries 125} (2020) 041602},
  [\href{https://arxiv.org/abs/2003.13501}{{\ttfamily 2003.13501}}].

\bibitem{Adamo:2020yzi}
T.~Adamo, L.~Mason and A.~Sharma, \emph{{Gluon scattering on self-dual
  radiative gauge fields}},  \href{https://arxiv.org/abs/2010.14996}{{\ttfamily
  2010.14996}}.

\bibitem{Adamo:2022mev}
T.~Adamo, L.~Mason and A.~Sharma, \emph{{Graviton scattering in self-dual
  radiative space-times}},  \href{https://arxiv.org/abs/2203.02238}{{\ttfamily
  2203.02238}}.

\bibitem{Chalmers:1997sg}
G.~Chalmers and W.~Siegel, \emph{{Dual formulations of Yang-Mills theory}},
  \href{https://arxiv.org/abs/hep-th/9712191}{{\ttfamily hep-th/9712191}}.

\bibitem{Costello:2022upu}
K.~Costello and N.~M. Paquette, \emph{{On the associativity of one-loop
  corrections to the celestial OPE}},
  \href{https://arxiv.org/abs/2204.05301}{{\ttfamily 2204.05301}}.

\bibitem{Bu:2022dis}
W.~Bu and E.~Casali, \emph{{The 4d/2d correspondence in twistor space and
  holomorphic Wilson lines}},
  \href{https://arxiv.org/abs/2208.06334}{{\ttfamily 2208.06334}}.

\bibitem{Costello:2022jpg}
K.~Costello, N.~M. Paquette and A.~Sharma, \emph{{Top-Down Holography in an
  Asymptotically Flat Spacetime}},
  \href{http://dx.doi.org/10.1103/PhysRevLett.130.061602}{\emph{Phys. Rev.
  Lett.} {\bfseries 130} (2023) 061602},
  [\href{https://arxiv.org/abs/2208.14233}{{\ttfamily 2208.14233}}].

\bibitem{Wolkow:1935zz}
D.~M. Wolkow, \emph{{\"Uber eine Klasse von Losungen der Diracschen
  Gleichung}}, \href{http://dx.doi.org/10.1007/BF01331022}{\emph{Z. Phys.}
  {\bfseries 94} (1935) 250--260}.

\bibitem{Seipt:2017ckc}
D.~Seipt, \emph{{Volkov States and Non-linear Compton Scattering in Short and
  Intense Laser Pulses}},  in \emph{{Proceedings, Quantum Field Theory at the
  Limits: from Strong Fields to Heavy Quarks (HQ 2016): Dubna, Russia, July
  18-30, 2016}}, pp.~24--43, 2017,
  \href{https://arxiv.org/abs/1701.03692}{{\ttfamily 1701.03692}},
  \href{http://dx.doi.org/10.3204/DESY-PROC-2016-04/Seipt}{DOI}.

\bibitem{Dixon:2004za}
L.~J. Dixon, E.~W.~N. Glover and V.~V. Khoze, \emph{{MHV rules for Higgs plus
  multi-gluon amplitudes}},
  \href{http://dx.doi.org/10.1088/1126-6708/2004/12/015}{\emph{JHEP} {\bfseries
  12} (2004) 015}, [\href{https://arxiv.org/abs/hep-th/0411092}{{\ttfamily
  hep-th/0411092}}].

\bibitem{Henn:2019mvc}
J.~Henn, B.~Power and S.~Zoia, \emph{{Conformal Invariance of the One-Loop
  All-Plus Helicity Scattering Amplitudes}},
  \href{http://dx.doi.org/10.1007/JHEP02(2020)019}{\emph{JHEP} {\bfseries 02}
  (2020) 019}, [\href{https://arxiv.org/abs/1911.12142}{{\ttfamily
  1911.12142}}].

\bibitem{Chicherin:2022bov}
D.~Chicherin and J.~M. Henn, \emph{{Symmetry properties of Wilson loops with a
  Lagrangian insertion}},
  \href{http://dx.doi.org/10.1007/JHEP07(2022)057}{\emph{JHEP} {\bfseries 07}
  (2022) 057}, [\href{https://arxiv.org/abs/2202.05596}{{\ttfamily
  2202.05596}}].

\bibitem{Penrose:1986uia}
R.~Penrose and W.~Rindler, \emph{{Spinors and Space-Time}}, vol.~2 of
  \emph{Cambridge Monographs on Mathematical Physics}.
\newblock Cambridge Univ. Press, Cambridge, UK, 1986,
  \href{http://dx.doi.org/10.1017/CBO9780511564048}{10.1017/CBO9780511564048}.

\bibitem{Adamo:2017qyl}
T.~Adamo, \emph{{Lectures on twistor theory}},
  \href{http://dx.doi.org/10.22323/1.323.0003}{\emph{PoS} {\bfseries
  Modave2017} (2018) 003}, [\href{https://arxiv.org/abs/1712.02196}{{\ttfamily
  1712.02196}}].

\bibitem{vanderBurg:1969}
M.~G.~T. van~der Burg, \emph{{Gravitational Waves in General Relativity 10.
  Asymptotic expansions for the Einstein-Maxwell field}},
  \href{http://dx.doi.org/10.1098/rspa.1969.0072}{\emph{Proc. Roy. Soc. Lond.}
  {\bfseries A310} (1969) 221--230}.

\bibitem{Strominger:2013lka}
A.~Strominger, \emph{{Asymptotic Symmetries of Yang-Mills Theory}},
  \href{http://dx.doi.org/10.1007/JHEP07(2014)151}{\emph{JHEP} {\bfseries 07}
  (2014) 151}, [\href{https://arxiv.org/abs/1308.0589}{{\ttfamily 1308.0589}}].

\bibitem{Barnich:2013sxa}
G.~Barnich and P.-H. Lambert, \emph{{Einstein-Yang-Mills theory: Asymptotic
  symmetries}}, \href{http://dx.doi.org/10.1103/PhysRevD.88.103006}{\emph{Phys.
  Rev.} {\bfseries D88} (2013) 103006},
  [\href{https://arxiv.org/abs/1310.2698}{{\ttfamily 1310.2698}}].

\bibitem{Penrose:1962ij}
R.~Penrose, \emph{{Asymptotic properties of fields and space-times}},
  \href{http://dx.doi.org/10.1103/PhysRevLett.10.66}{\emph{Phys. Rev. Lett.}
  {\bfseries 10} (1963) 66--68}.

\bibitem{Penrose:1980yx}
R.~Penrose, \emph{{Null hypersurface initial data for classical fields of
  arbitrary spin and for general relativity}},
  \href{http://dx.doi.org/10.1007/BF00756234}{\emph{Gen. Rel. Grav.} {\bfseries
  12} (1980) 225--264}.

\bibitem{Penrose:1984uia}
R.~Penrose and W.~Rindler, \emph{{Spinors and Space-Time}}, vol.~1 of
  \emph{Cambridge Monographs on Mathematical Physics}.
\newblock Cambridge Univ. Press, Cambridge, UK, 1984,
  \href{http://dx.doi.org/10.1017/CBO9780511564048}{10.1017/CBO9780511564048}.

\bibitem{Penrose:1967wn}
R.~Penrose, \emph{{Twistor algebra}},
  \href{http://dx.doi.org/10.1063/1.1705200}{\emph{J. Math. Phys.} {\bfseries
  8} (1967) 345}.

\bibitem{Mason:1986}
L.~J. Mason, \emph{{Dolbeault representative from characteristic initial data
  at null infinity}},  in \emph{{Further Advances in Twistor Theory}} (L.~J.
  Mason and L.~P. Hughston, eds.), vol.~231, ch.~1.2.16.
\newblock Pitman Research Notes in Mathematics, 1990.

\bibitem{Penrose:1969ae}
R.~Penrose, \emph{{Solutions of the zero-rest-mass equations}},
  \href{http://dx.doi.org/10.1063/1.1664756}{\emph{J. Math. Phys.} {\bfseries
  10} (1969) 38--39}.

\bibitem{Eastwood:1981jy}
M.~G. Eastwood, R.~Penrose and R.~O. Wells, \emph{{Cohomology and Massless
  Fields}}, \href{http://dx.doi.org/10.1007/BF01942327}{\emph{Commun. Math.
  Phys.} {\bfseries 78} (1981) 305--351}.

\bibitem{Newman:1980fr}
E.~T. Newman, \emph{{Selfdual gauge fields}},
  \href{http://dx.doi.org/10.1103/PhysRevD.22.3023}{\emph{Phys. Rev.}
  {\bfseries D22} (1980) 3023--3033}.

\bibitem{Witten:1978xx}
E.~Witten, \emph{{An Interpretation of Classical Yang-Mills Theory}},
  \href{http://dx.doi.org/10.1016/0370-2693(78)90585-3}{\emph{Phys. Lett. B}
  {\bfseries 77} (1978) 394--398}.

\bibitem{Harnad:1985bc}
J.~P. Harnad and S.~Shnider, \emph{{CONSTRAINTS AND FIELD EQUATIONS FOR
  TEN-DIMENSIONAL SUPERYANG-MILLS THEORY}},
  \href{http://dx.doi.org/10.1007/BF01454971}{\emph{Commun. Math. Phys.}
  {\bfseries 106} (1986) 183}.

\bibitem{Witten:1985nt}
E.~Witten, \emph{{Twistor - Like Transform in Ten-Dimensions}},
  \href{http://dx.doi.org/10.1016/0550-3213(86)90090-8}{\emph{Nucl. Phys. B}
  {\bfseries 266} (1986) 245--264}.

\bibitem{Devchand:1996gv}
C.~Devchand and V.~Ogievetsky, \emph{{Interacting fields of arbitrary spin and
  N \ensuremath{>} 4 supersymmetric selfdual Yang-Mills equations}},
  \href{http://dx.doi.org/10.1016/S0550-3213(96)90129-7}{\emph{Nucl. Phys. B}
  {\bfseries 481} (1996) 188--214},
  [\href{https://arxiv.org/abs/hep-th/9606027}{{\ttfamily hep-th/9606027}}].

\bibitem{Ferber:1977qx}
A.~Ferber, \emph{{Supertwistors and Conformal Supersymmetry}},
  \href{http://dx.doi.org/10.1016/0550-3213(78)90257-2}{\emph{Nucl. Phys. B}
  {\bfseries 132} (1978) 55--64}.

\bibitem{Volovich:1983aa}
I.~V. Volovich, \emph{{Supersymmetric Yang-Mills Theories and Twistors}},
  \href{http://dx.doi.org/10.1016/0370-2693(83)90133-8}{\emph{Phys. Lett. B}
  {\bfseries 129} (1983) 429--431}.

\bibitem{Volovich:1983ii}
I.~V. Volovich, \emph{{Superselfduality for Supersymmetric Yang-Mills Theory}},
  \href{http://dx.doi.org/10.1016/0370-2693(83)91211-X}{\emph{Phys. Lett. B}
  {\bfseries 123} (1983) 329--331}.

\bibitem{Adamo:2021rfq}
T.~Adamo, A.~Cristofoli and P.~Tourkine, \emph{{Eikonal amplitudes from curved
  backgrounds}},  \href{https://arxiv.org/abs/2112.09113}{{\ttfamily
  2112.09113}}.

\bibitem{Mason:2008jy}
L.~J. Mason and D.~Skinner, \emph{{Gravity, Twistors and the MHV Formalism}},
  \href{http://dx.doi.org/10.1007/s00220-009-0972-4}{\emph{Commun. Math. Phys.}
  {\bfseries 294} (2010) 827--862},
  [\href{https://arxiv.org/abs/0808.3907}{{\ttfamily 0808.3907}}].

\bibitem{Boels:2013bi}
R.~H. Boels, R.~S. Isermann, R.~Monteiro and D.~O'Connell,
  \emph{{Colour-Kinematics Duality for One-Loop Rational Amplitudes}},
  \href{http://dx.doi.org/10.1007/JHEP04(2013)107}{\emph{JHEP} {\bfseries 04}
  (2013) 107}, [\href{https://arxiv.org/abs/1301.4165}{{\ttfamily 1301.4165}}].

\bibitem{Rosly:1996vr}
A.~A. Rosly and K.~G. Selivanov, \emph{{On amplitudes in selfdual sector of
  Yang-Mills theory}},
  \href{http://dx.doi.org/10.1016/S0370-2693(97)00268-2}{\emph{Phys. Lett. B}
  {\bfseries 399} (1997) 135--140},
  [\href{https://arxiv.org/abs/hep-th/9611101}{{\ttfamily hep-th/9611101}}].

\bibitem{Selivanov:1998hn}
K.~G. Selivanov, \emph{{On tree form-factors in (supersymmetric) Yang-Mills
  theory}}, \href{http://dx.doi.org/10.1007/s002200050006}{\emph{Commun. Math.
  Phys.} {\bfseries 208} (2000) 671--687},
  [\href{https://arxiv.org/abs/hep-th/9809046}{{\ttfamily hep-th/9809046}}].

\bibitem{Mason:2009afn}
L.~J. Mason and D.~Skinner, \emph{{Gravity, Twistors and the MHV Formalism}},
  \href{http://dx.doi.org/10.1007/s00220-009-0972-4}{\emph{Commun. Math. Phys.}
  {\bfseries 294} (2010) 827--862},
  [\href{https://arxiv.org/abs/0808.3907}{{\ttfamily 0808.3907}}].

\bibitem{Adamo:2019zmk}
T.~Adamo and A.~Ilderton, \emph{{Gluon helicity flip in a plane wave
  background}}, \href{http://dx.doi.org/10.1007/JHEP06(2019)015}{\emph{JHEP}
  {\bfseries 06} (2019) 015},
  [\href{https://arxiv.org/abs/1903.01491}{{\ttfamily 1903.01491}}].

\bibitem{Atiyah:1981ey}
M.~F. Atiyah, \emph{{Green's functions for selfdual four manifolds}},
  {\emph{Adv. Math. Suppl. Stud.} {\bfseries 7} (1981) 129--158}.

\bibitem{Lipstein:2012vs}
A.~E. Lipstein and L.~Mason, \emph{{From the holomorphic Wilson loop to `d log'
  loop-integrands for super-Yang-Mills amplitudes}},
  \href{http://dx.doi.org/10.1007/JHEP05(2013)106}{\emph{JHEP} {\bfseries 05}
  (2013) 106}, [\href{https://arxiv.org/abs/1212.6228}{{\ttfamily 1212.6228}}].

\bibitem{Lipstein:2013xra}
A.~E. Lipstein and L.~Mason, \emph{{From $d$ logs to dilogs the super
  Yang-Mills MHV amplitude revisited}},
  \href{http://dx.doi.org/10.1007/JHEP01(2014)169}{\emph{JHEP} {\bfseries 01}
  (2014) 169}, [\href{https://arxiv.org/abs/1307.1443}{{\ttfamily 1307.1443}}].

\bibitem{Bern:1993qk}
Z.~Bern, G.~Chalmers, L.~J. Dixon and D.~A. Kosower, \emph{{One loop N gluon
  amplitudes with maximal helicity violation via collinear limits}},
  \href{http://dx.doi.org/10.1103/PhysRevLett.72.2134}{\emph{Phys. Rev. Lett.}
  {\bfseries 72} (1994) 2134--2137},
  [\href{https://arxiv.org/abs/hep-ph/9312333}{{\ttfamily hep-ph/9312333}}].

\bibitem{Mahlon:1993si}
G.~Mahlon, \emph{{Multi - gluon helicity amplitudes involving a quark loop}},
  \href{http://dx.doi.org/10.1103/PhysRevD.49.4438}{\emph{Phys. Rev. D}
  {\bfseries 49} (1994) 4438--4453},
  [\href{https://arxiv.org/abs/hep-ph/9312276}{{\ttfamily hep-ph/9312276}}].

\bibitem{Bern:1994ju}
Z.~Bern, L.~J. Dixon, D.~C. Dunbar and D.~A. Kosower, \emph{{One loop gauge
  theory amplitudes with an arbitrary number of external legs}},  in
  \emph{{Workshop on Continuous Advances in QCD}}, 2, 1994,
  \href{https://arxiv.org/abs/hep-ph/9405248}{{\ttfamily hep-ph/9405248}}.

\bibitem{Boels:2007gv}
R.~Boels, \emph{{A Quantization of twistor Yang-Mills theory through the
  background field method}},
  \href{http://dx.doi.org/10.1103/PhysRevD.76.105027}{\emph{Phys. Rev. D}
  {\bfseries 76} (2007) 105027},
  [\href{https://arxiv.org/abs/hep-th/0703080}{{\ttfamily hep-th/0703080}}].

\bibitem{Bullimore:2010pj}
M.~Bullimore, L.~J. Mason and D.~Skinner, \emph{{MHV Diagrams in Momentum
  Twistor Space}}, \href{http://dx.doi.org/10.1007/JHEP12(2010)032}{\emph{JHEP}
  {\bfseries 12} (2010) 032},
  [\href{https://arxiv.org/abs/1009.1854}{{\ttfamily 1009.1854}}].

\bibitem{Bern:1994zx}
Z.~Bern, L.~J. Dixon, D.~C. Dunbar and D.~A. Kosower, \emph{{One loop n point
  gauge theory amplitudes, unitarity and collinear limits}},
  \href{http://dx.doi.org/10.1016/0550-3213(94)90179-1}{\emph{Nucl. Phys. B}
  {\bfseries 425} (1994) 217--260},
  [\href{https://arxiv.org/abs/hep-ph/9403226}{{\ttfamily hep-ph/9403226}}].

\bibitem{Brandhuber:2004yw}
A.~Brandhuber, B.~J. Spence and G.~Travaglini, \emph{{One-loop gauge theory
  amplitudes in N=4 super Yang-Mills from MHV vertices}},
  \href{http://dx.doi.org/10.1016/j.nuclphysb.2004.11.023}{\emph{Nucl. Phys. B}
  {\bfseries 706} (2005) 150--180},
  [\href{https://arxiv.org/abs/hep-th/0407214}{{\ttfamily hep-th/0407214}}].

\bibitem{Brown:1978yj}
L.~S. Brown and D.~B. Creamer, \emph{{Vacuum Polarization about Instantons}},
  \href{http://dx.doi.org/10.1103/PhysRevD.18.3695}{\emph{Phys. Rev. D}
  {\bfseries 18} (1978) 3695}.

\bibitem{Corrigan:1979di}
E.~Corrigan, P.~Goddard, H.~Osborn and S.~Templeton, \emph{{Zeta Function
  Regularization and Multi - Instanton Determinants}},
  \href{http://dx.doi.org/10.1016/0550-3213(79)90346-8}{\emph{Nucl. Phys. B}
  {\bfseries 159} (1979) 469--496}.

\bibitem{Bardeen:1995gk}
W.~A. Bardeen, \emph{{Selfdual Yang-Mills theory, integrability and multiparton
  amplitudes}}, \href{http://dx.doi.org/10.1143/PTPS.123.1}{\emph{Prog. Theor.
  Phys. Suppl.} {\bfseries 123} (1996) 1--8}.

\bibitem{Costello:2021bah}
K.~J. Costello, \emph{{Quantizing local holomorphic field theories on twistor
  space}},  \href{https://arxiv.org/abs/2111.08879}{{\ttfamily 2111.08879}}.

\bibitem{Monteiro:2022nqt}
R.~Monteiro, R.~Stark-Much\~ao and S.~Wikeley, \emph{{Anomaly and double copy
  in quantum self-dual Yang-Mills and gravity}},
  \href{https://arxiv.org/abs/2211.12407}{{\ttfamily 2211.12407}}.

\end{thebibliography}\endgroup
\bibliographystyle{JHEP}

\end{document}